\documentclass[twocolumn]{aastex63}
\usepackage[utf8]{inputenc}
\usepackage{graphicx}
\usepackage{amssymb}
\usepackage{amsmath}
\usepackage{float}
\usepackage{multirow}
\usepackage{textcomp}
\usepackage{gensymb}
\usepackage{enumitem}  
\usepackage{hyperref}
\usepackage{natbib}
\usepackage{comment}
\usepackage{systeme}
\usepackage[Symbol]{upgreek}
\usepackage{xcolor}
\definecolor{xlinkcolor}{cmyk}{1,1,0,0}

\newcommand{\chandra}{\emph{Chandra}}
\newcommand{\xmm}{\emph{XMM-Newton}}

\shorttitle{Galaxy group NGC\,741-742}
\shortauthors{Rajpurohit et al.}

\begin{document}

\title{A Deep Dive into the NGC~741 Galaxy Group: Insights into a Spectacular Head-Tail Radio Galaxy from VLA, MeerKAT, uGMRT and LOFAR }

\correspondingauthor{Kamlesh Laxmi Rajpurohit}
\email{kamlesh.rajpurohit@cfa.harvard.edu}

\author[0000-0001-7509-2972]{K. Rajpurohit} 
\affil{Center for Astrophysics $|$ Harvard \& Smithsonian, 60 Garden Street, Cambridge, MA 02138, USA}
\affil{Th\"{u}ringer Landessternwarte, Sternwarte 5, 07778 Tautenburg, Germany}

\author[0000-0002-5671-6900]{E. O'Sullivan}
\affil{Center for Astrophysics $|$ Harvard \& Smithsonian, 60 Garden Street, Cambridge, MA 02138, USA}

\author[0000-0002-4962-0740]{G. Schellenberger}
\affil{Center for Astrophysics $|$ Harvard \& Smithsonian, 60 Garden Street, Cambridge, MA 02138, USA}

\author[0000-0003-4120-9970]{M. Brienza}
\affil{INAF-IRA, via Gobetti 101, 40129 Bologna, Italy} 

\author[0009-0007-0318-2814]{J. M. Vrtilek}
\affil{Center for Astrophysics $|$ Harvard \& Smithsonian, 60 Garden Street, Cambridge, MA 02138, USA}

\author[0000-0002-9478-1682]{W. Forman}
\affil{Center for Astrophysics $|$ Harvard \& Smithsonian, 60 Garden Street, Cambridge, MA 02138, USA}

\author{L. P. David }
\affil{Center for Astrophysics $|$ Harvard \& Smithsonian, 60 Garden Street, Cambridge, MA 02138, USA}

\author[0000-0001-6812-7938]{T. Clarke}
\affil{Naval Research Laboratory, 4555 Overlook Avenue SW, Code 7213, Washington, DC 20375, US}

\author[0000-0002-9325-1567]{A. Botteon}
\affil{INAF-IRA, via Gobetti 101, 40129 Bologna, Italy} 

\author[0000-0002-2821-7928]{F. Vazza}
\affil{Dipartimento di Fisica e Astronomia, Universit\`a di Bologna, via P. Gobetti 93/2, 40129, Bologna, Italy}
\affil{INAF-IRA, via Gobetti 101, 40129 Bologna, Italy} 

\author[0000-0002-1634-9886]{S. Giacintucci}
\affil{Naval Research Laboratory, 4555 Overlook Avenue SW, Code 7213, Washington, DC 20375, US}

\author[0000-0003-2206-4243]{C. Jones}
\affil{Center for Astrophysics $|$ Harvard \& Smithsonian, 60 Garden Street, Cambridge, MA 02138, USA}

\author[0000-0002-3369-7735]{M. Br\"uggen}
\affil{Hamburger Sternwarte, Universit\"at Hamburg, Gojenbergsweg 112, 21029, Hamburg, Germany}

\author[0000-0001-5648-9069]{T. W. Shimwell}
\affil{ASTRON, Netherlands Institute for Radio Astronomy, Oude Hoogeveensedijk 4, 7991 PD, Dwingeloo, The Netherlands}
\affil{Leiden Observatory, Leiden University, PO Box 9513, NL-2300 RA Leiden, The Netherlands} 

\author[0000-0003-2792-1793]{A. Drabent}
\affil{Th\"{u}ringer Landessternwarte, Sternwarte 5, 07778 Tautenburg, Germany}

\author{F. Loi}
\affil{INAF-Osservatorio Astronomico di Cagliari, Via della Scienza 5, 09047 Selargius, CA, Italy}

\author[0000-0002-3937-7126]{S. I. Loubser}
\affil{Centre for Space Research, North-West University, Potchefstroom 2520, South Africa}

\author[0000-0002-3104-6154]{K. Kolokythas}
\affil{Centre for Radio Astronomy Techniques and Technologies, Department of Physics and Electronics, Rhodes University, P.O. Box 94, Makhanda 6140, South Africa}
\affiliation{South African Radio Astronomy Observatory, Black River Park North, 2 Fir St, Cape Town, 7925, South Africa}

\author[0000-0003-3165-9804]{I. Babyk}
\affil{Center for Astrophysics $|$ Harvard \& Smithsonian, 60 Garden Street, Cambridge, MA 02138, USA}

\author{H. J. A. R\"ottgering}
\affil{Leiden Observatory, Leiden University, PO Box 9513, NL-2300 RA Leiden, The Netherlands}


\begin{abstract}
We present deep, wideband multifrequency radio observations (144\,MHz{$-$}8\,GHz) of the  remarkable galaxy group NGC\,741, which yield crucial insights into the interaction between the infalling head-tail radio galaxy (NGC\,742) and the main group.  Our new data provide an unprecedentedly detailed view of the NGC\,741-742 system, including the shock cone, disrupted jets from NGC\,742, the long ($\sim \rm 255\,kpc$) braided southern radio tail, and eastern lobe-like structure ($\sim100\,\rm kpc$), and reveal, for the first time, complex radio filaments throughout the tail and lobe, and a likely vortex ring behind the shock cone. The cone traces the bow shock caused by the supersonic ($\mathcal{M}\sim2$) interaction between the head-tail radio galaxy NGC~742 and the intragroup medium (IGrM) while the ring may have been formed by interaction between the NGC~742 shock and a previously existing lobe associated with NGC~741. This interaction plausibly compressed and re-accelerated the radio plasma. We estimate that shock-heating by NGC~742  has likely injected $\sim$2-5$\times$10$^{57}$~erg of thermal energy into the central 10~kpc cooling region of the IGrM, potentially affecting the cooling and feedback cycle of NGC~741. A comparison with {\chandra} X-ray images shows that some of the previously detected thermal filaments align with radio edges, suggesting compression of the IGrM as the relativistic plasma of the NGC\,742 tail interacts  with the surrounding medium. Our results highlight that multi-frequency observations are key to disentangling the complex, intertwined origins of the variety of radio features seen in the galaxy group NGC\,741, and the need for simulations to reproduce all the detected features.
\end{abstract}


\section{Introduction}
 \label{sec:intro}

Mergers are an important process for the growth of galaxy groups and galaxy clusters. They are a common subject of study in clusters, with numerous examples of shock heating of the intracluster medium (ICM) \citep[e.g.,][]{McNamara2004,Bourdin2013,Ubertosi2023,vanWeeren2017a,Botteon2018}. In a growing number of merging clusters, very asymmetric and filamentary morphologies of tailed radio galaxies have been observed \citep[e.g., ][]{Wilber2018,Gendron2020,Rudnick2022,Botteon2022a,Rajpurohit2022b,Koribalski2024}. The properties of such radio galaxies, in some cases coincident with X-ray shocks, suggest an interaction between the radio galaxy, its surrounding medium, and the shock. However, the lower density of the intragroup medium (IGrM) and the lower velocities of the galaxy population have led to fewer studies of group-scale mergers. The lower mass and IGrM temperature of groups mean that the infall of individual large galaxies may have a greater impact than is the case for clusters. The heating effect of group-central galaxies may affect the cooling and feedback cycle.  

\begin{figure*}[!thbp]
    \centering
    \includegraphics[width=0.90\textwidth]{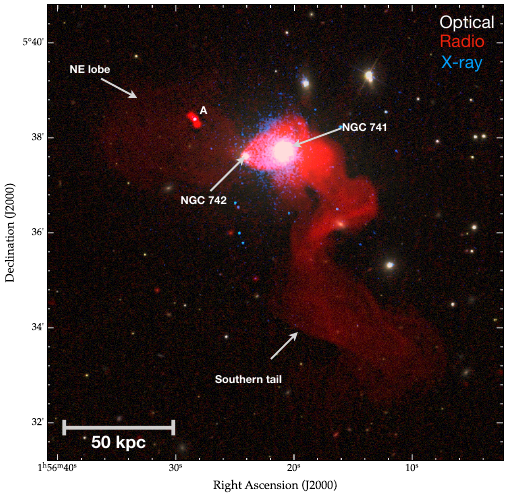}
    \vspace{-0.2cm}
 \caption{Radio, X-ray, and optical overlay of NGC\,741-742. The intensity in red shows the radio emission observed with uGMRT and MeerKAT. The radio image properties are given in Table\,\ref{imaging} (IM8 and IM13). The intensity in blue shows \chandra~ X-ray emission in the $0.5{-}2.0$ keV band, and in the background is the color composite optical image created using Sloan Digital Sky Survey data with $r$, $g$, and $i$ intensities represented in blue, green, and red, respectively. Source A is a background radio galaxy.}
      \label{overlay}
\end{figure*}

Only a handful of group-group mergers with high enough velocities to drive shocks \citep[e.g.,][]{Randalletal09,Russelletal14,O'Sullivan2019} are known. In this paper, we focus on a spectacular example of a high-velocity head tail radio galaxy NGC\,742, driving a shock front in the galaxy group around NGC\,741, (Figure\,\ref{overlay}).

\setlength{\tabcolsep}{5pt}
\begin{table*}[!htbp]
\caption{Observational overview: VLA, MeerKAT, uGMRT, and LOFAR observations.}
\begin{center}
\begin{tabular}{ l  c  c c c  c c c c c}
  \hline  \hline  
\multirow{1}{*}{}&  \multicolumn{4}{c}{ VLA$^{\ast}$} & \multirow{1}{*}{MeerKAT} & \multicolumn{2}{c}{ uGMRT}  &\multirow{1}{*}{LOFAR}   \\  
 \cline{2-5} \cline{7-8} 
&\multicolumn{2}{c}{C-band} &\multicolumn{2}{c}{S-band}& L-band & \multirow{1}{*}{Band\,4} & \multirow{1}{*}{Band\,3} &  HBA\\
 \cline{2-3} \cline{4-5} 
& B array &C-array & B-array &C-array&  &  & \\
\hline
Frequency &4-8\,GHz  &4-8\,GHz &2-4\,GHz &2-4\,GHz&0.9-1.7\,GHz &550-850\,MHz&300-500\,MHz\ &120-169\,MHz\\ 
Channel width  &27\,MHz & 27\,MHz&2\,MHz&2\,MHz &23.3\,kHz&97.7\,kHz & 42\,kHz&12.2\,kHz\\ 
No of  IF$^{\ast\ast} $ & 34&34&16&16& 1&1&1& 1 &\\ 
No of channels per IF & 64&64&64&64& 4096&4028&4028& 64 &\\ 
On source time &4\,hrs &4\,hrs &4\,hrs&4\,hrs&8\,hrs&8\,hrs &8\,hrs &8\,hrs \\
\hline 
\end{tabular}
\end{center}
{Notes.$^{\ast}$ VLA S-band observations have a total of 1024 channels at each array while C-band has 2048 channels;$^{\ast\ast}$ IF stands for intermediate frequency, i.e., the number of spectral windows. The VLA, MeerKAT, and uGMRT Band\,4 observations were performed in full polarization mode. Detailed polarization and Faraday analysis will be presented in a later paper.}
\label{Tabel:obs}
\end{table*} 

NGC\,741 is a nearby group located at a redshift of $z$=0.018. The giant elliptical NGC\,741 is surrounded by $\sim$40 fainter neighbors with a velocity dispersion $\sigma$=432$^{+50}_{-46}$~km~s$^{-1}$ \citep{ZabludoffMulchaey98} and lies at the center of a diffuse, X-ray-emitting IGrM (Figure\,\ref{overlay}). The discovery of a bright, extended radio source centered on NGC\,741 \citep[4C~05.10][]{Birkinshaw1985} led to its classification as a somewhat distorted double-lobed radio galaxy. However, higher resolution Very Large Array \citep{Venkatesan1994,Jetha2008} and Giant Metrewave Radio Telescope (GMRT) images \citep{Giacintucci2011} show compact cores associated with NGC\,741 and a smaller neighboring elliptical, NGC~742. 

A high resolution ~4.9~GHz radio image shows two jets emerging from NGC~742, both bending sharply westward and blending together to form a  130\,kpc long tail  to the south-west \citep{Schellenberger2017}. Despite crossing NGC\,741, the entire tail is interpreted as originating from the AGN of NGC~742. Essentially all diffuse radio emission west of NGC~742 is considered to be part of this head-tail source. Low frequency 235\,MHz and 610\,MHz radio observations revealed fainter extended radio emission to the east of NGC\,741 \citep{Schellenberger2017}.

Deep {\chandra} and {\xmm} X-ray observations reveal several gaseous filaments, one of them linking NGC\,741 and NGC~742. They are apparently part of a larger V-shaped structure pointing east with NGC~742 at its apex. \cite{Schellenberger2017} interpret this as a bow shock driven into the IGrM by NGC~742 as it passes through the group core. NGC\,741 and NGC~742 are separated by $\sim$45\arcsec\ ($\sim$15~kpc) in projection, but the  sharply bent jets of NGC~742 indicate that it has a large velocity in the plane of the sky, and its recession velocity \citep[5969~km~s$^{-1}$,][]{MahdaviGeller04} is $\sim$480 km~s$^{-1}$ greater than that of NGC\,741 \citep[5485~km~s$^{-1}$,][]{VandenBoschetal15}. Based on the length and radiative age of the south-west tail, \cite{Schellenberger2017} estimated true velocities in the range $\sim$1100-1400~km~s$^{-1}$, highly supersonic in the IGrM (kT$\sim$1-2~keV, sound speed $\sim$520-730 km~s$^{-1}$). X-ray observations also  show a probable cavity in the IGrM about 16 kpc to the west of NGC\,741. 

In this paper, we present a deep, multifrequency radio continuum and spectral analysis of the spectacular galaxy group NGC\,741.  It has several unusual features that provide special insight into the evolution of the group and the interaction process between the infalling galaxy NGC\,742 and the main group NGC\,741, intriguing narrow X-ray filaments, and gas sloshing features.  We use radio observations conducted with MeerKAT \citep{Jonas2016}, Karl G. Jansky Very Large Array (VLA), the upgraded Giant Metrewave Radio Telescope (uGMRT) and the new LOw-Frequency ARray \citep[LOFAR;][]{Haarlem2013} High-Band Antenna. These observations were mainly undertaken to provide higher resolution radio images, thus allowing us to study this exceptional group with large frequency coverage. 

The paper is structured as follows: Section\,\ref{observations} provides an overview of the observations and details about the data reduction. Section\,\ref{results} presents the newly obtained total images. The analysis is presented and discussed in Section\,\ref{discussion}. Finally, a summary of the findings is provided in Section\,\ref{summary}.

Throughout this paper, we adopt a flat $\Lambda$CDM cosmology with $H_{\rm{ 0}}=69.6$ km s$^{-1}$\,Mpc$^{-1}$, $\Omega_{\rm{ m}}=0.286$, and $\Omega_{\Lambda}=0.714$ \citep{Bennett2014}. At the group's redshift, $1\arcsec$ corresponds to a physical scale of 0.368\,kpc.

\section{Observations and data reduction}
\label{observations}


\subsection{VLA}
NGC\,741-742 was observed by the VLA in the S-band (2-4\,GHz) and C-band (4-8\,GHz) in C and B arrays (Project code: 24A-053; PI: K. Rajpurohit). For observational details, see Table\,\ref{Tabel:obs}. All four polarization products (RR, RL, LR, and LL) were recorded. For both bands and arrays, 3C\,147 was included as the primary calibrator and unpolarized calibrator, observed for 8 minutes at the start of each observing run. J0834$+$5534 was included as a phase calibrator, and 3C\,138  as a polarization calibrator.

The data were calibrated and imaged with the Common Astronomy Software Applications \citep[$\tt{CASA}$;][]{McMullin2007,casa2022} package (version 6.1). Data obtained from different observing runs were calibrated separately but in the same manner. The data reduction steps consisted of Hanning smoothing followed by radio frequency interference (RFI) inspection.  Bad data were flagged using ${\tt tfcrop}$ mode from the ${\tt flagdata}$ task. For target scans, we performed additional flagging using ${\tt AOFlagger}$ \citep{Offringa2010}. After this, we determined and applied elevation-dependent gain tables and antenna offset correction. We corrected the initial bandpass correction using 3C\,147. This prevents the flagging of good data due to the bandpass roll-off at the edges of the spectral windows. 

We used the S-band and C-band 3C\,147 and 3C\,138 models provided by the ${\tt CASA}$ software package and set the flux density scale according to \cite{Perley2013}. An initial phase calibration was performed using both calibrators over a few channels per spectral window. The bandpass response were determined using 3C\,147, followed by the gain calibration. For polarization, the leakage corrections were determined using the unpolarized calibrator 3C\,147 and the absolute position angle using the polarized calibrator 3C\,138.  All calibration solutions were applied to the target field. The resulting calibrated data were averaged by a factor of 4 in frequency per spectral window to perform Rotational Measure synthesis (RM-synthesis). After calibrating each array, we created initial images of the target field.

After initial calibration and flagging, several rounds of self-calibration were performed on each array data to refine the calibration for each individual data set. Imaging was done in the W-projection algorithm in ${\tt CASA}$ \citep{Cornwell2008}. Clean masks were used for each imaging step. These masks were made using the {\tt PyBDSF} source detection package \citep{Mohan2015}. The spectral index and curvature were considered during deconvolution using $\tt{nterms}=3$ \citep{Rau2011}.

After self-calibration, the C and B array data from each band (C and S) were combined. Final imaging of the combined data sets was done in {\tt WSClean} \citep{Offringa2014} with $\tt{multiscale}$ and $\tt{Briggs}$ weighting. The multiscale setting assumes that the emission can be modeled as a collection of components at various spatial scales, making it necessary to account for the extended emission. The images were corrected for the primary beam attenuation.

\subsection{MeerKAT}
We observed the group with MeerKAT in the L-band  (Project code:
SCI-20220822; PI: E. O'Sullivan).  An overview of the MeerKAT observation is given in Table\,\ref{Tabel:obs}. All four polarization products (XX, XY, YX, and YY) were recorded using the 4K correlator mode covering a frequency range of 0.85-1.7~GHz.  J0408-6545 was the primary calibrator used for flux and bandpass calibration, observed at the beginning, middle and end of the observing run. J0059+0006 was included as a polarization calibrator with multiple scans at different parallactic angles. J0108+0134 served as a gain calibrator.

\begin{deluxetable*}{c c c r c c c r}[!htbp]
\tablecaption{Radio image properties}
\tablehead{ & Name & Restoring Beam & Robust  & \textit{uv}-cut & \textit{uv}-taper & RMS noise\\ 
&&&parameter&&&$\upmu\,\rm Jy\,beam^{-1}$}
\startdata
LOFAR HBA&IM1&$7.0\arcsec \times 7.0\arcsec$&$-0.5$&$ \geq\rm0.1\,k\uplambda$&$-$&120\\
(120--168\,MHz)&IM2&$15\arcsec \times 15\arcsec $&$-0.5$&$-$&8\arcsec&130\\
&IM3&$15\arcsec \times 15\arcsec $&$-0.5$&$  \geq\rm0.1\,k\uplambda$&8\arcsec&150\\
\hline   
 &IM4&$6.0\arcsec \times 6.0\arcsec$&0.0&$-$&&50\\
 uGMRT Band\,3 &IM5&$7.0\arcsec \times 7.0\arcsec$&$-0.5$&$ \geq\rm0.1\,k\uplambda$&$2\arcsec$&55\\
 (300--500\,MHz)&IM6&$15\arcsec \times 15\arcsec $&$0.0$&$ -$&8\arcsec&60\\
 &IM7&$15\arcsec \times 15\arcsec $&$-0.5$&$ \geq\rm0.1\,k\uplambda$&8\arcsec&70\\
\hline   
 &IM8&$3.4\arcsec \times 2.0\arcsec$&$-2.0$&$-$&$-$&60\\
 uGMRT Band\,4 &IM9&$7.0\arcsec \times 7.0\arcsec$&$-0.5$&$ \geq\rm0.1\,k\uplambda$&$2\arcsec$&14\\
 (550--850\,MHz)&IM10&$15\arcsec \times 15\arcsec $&$0.0$&$ -$&8\arcsec&28\\
 &IM11&$15\arcsec \times 15\arcsec $&$-0.5$&$ \geq\rm0.1\,k\uplambda$&8\arcsec&30\\
  \hline   
 &IM12&$5.5\arcsec \times 5.5\arcsec$&-0.5&$\geq\rm0.1\,k\uplambda$&$-$&6\\
MeerKAT L-band &IM13&$7.0\arcsec \times 7.0\arcsec$&$$-0.5$$&$\geq\rm0.1\,k\uplambda$&2\arcsec&7 \\
(0.9--1.7\,GHz)&IM14&$15\arcsec \times 15\arcsec $&$0.0$&$-$&10\arcsec& 9\\
&IM15&$15\arcsec \times 15\arcsec $&$-0.5$&$ \geq\rm0.1\,k\uplambda$&10\arcsec&10 \\ 
 \hline   
&IM16&$1.5\arcsec \times 1.4\arcsec$&-0.5&$-$&$-$&4\\
 &IM17&$5.5\arcsec \times 5.5\arcsec$&-0.5&$\geq\rm0.1\,k\uplambda$&$2\arcsec$&5\\
VLA S-band &IM18&$7.0\arcsec \times 7.0\arcsec$&$$-0.5$$&$\geq\rm0.1\,k\uplambda$&4\arcsec& 6\\
(2--4\,GHz)&IM19&$15\arcsec \times 15\arcsec $&$0.0$&$-$&10\arcsec& 6\\
&IM20&$15\arcsec \times 15\arcsec $&$-0.5$&$ \geq\rm0.1\,k\uplambda$&10\arcsec& 7\\
   \hline   
 &IM21&$1.1\arcsec \times 0.7\arcsec$&0.0&$-$&$-$&3\\
  &IM22&$5.5\arcsec \times 5.5\arcsec$&-0.5&$\geq\rm0.1\,k\uplambda$&$2\arcsec$&3\\
VLA C-band &IM23&$7.0\arcsec \times 7.0\arcsec$&$$-0.5$$&$\geq\rm0.1\,k\uplambda$&4\arcsec& 4\\
(4--8\,GHz)&IM24&$15\arcsec \times 15\arcsec $&$0.0$&$-$&10\arcsec& 5\\
&IM25&$15\arcsec \times 15\arcsec $&$-0.5$&$ \geq\rm0.1\,k\uplambda$&10\arcsec& 5\\
\enddata
\tablecomments{Imaging was always performed in {\tt WSCLEAN} using {\tt multiscale} and with {\tt Briggs} weighting scheme. Primary beam correction for the VLA images was performed in {\tt CASA}, and MeerKAT and uGMRT images using   {\tt WSCLEAN}. }
\label{imaging}
\end{deluxetable*}

The data were calibrated using the Containerized Automated Radio Astronomy Calibration \citep[$\tt{CARACal}$;][]{caracal2020} pipeline {\footnote{\url{https://ascl.net/2006.014}}} to perform full-polarization calibration. The first step consists of flagging in {\tt CARACAL}, including shadowed antennas, autocorrelations, known RFI channels, and the tfcrop algorithm.  Thereafter, {\tt AOflagger} \citep{Offringa2010} was used to flag bad data using the {\tt firstpass\_QUV.rfis} strategy.  {\tt CARACal} modeled the primary calibrator J0408-6545 using the MeerKAT local sky models. Following this, cross-calibration was performed to solve for the time-dependent delays and complex gains of each antenna and the bandpass corrections.

The polarization calibration was performed using the {\tt CARACal} ``polcal" strategy as recommended when observing an unknown (non standard) polarized calibrator.  In the end, all the solutions were applied to the target. The resulting calibrated data were averaged by a factor of five in frequency.  We ran {\tt AOflagger} on the averaged calibrated target data as well to flag the weak amplitude RFI.

After initial calibration, we created an initial image of the target field using {\tt WSClean}. We carried out several rounds of phase self-calibration using the ${\tt CASA}$ task {\tt gaincal}, followed by two final rounds of amplitude and phase self-calibration.

\subsection{uGMRT}
We observed the group with the upgraded GMRT in Bands\,3 and 4 (Project code: ddtC284; PI: K. Rajpurohit) using the GMRT Wideband Backend (GWB). In Band\,4, the source was observed in two observing runs, see Table\,\ref{Tabel:obs}, for the observational details. Source 3C\,48 was included as a flux and bandpass calibrator.

The initial calibration of the wideband uGMRT data was performed using the Source Peeling and Atmospheric Modeling \citep[$\tt{SPAM}$;][]{Intema2009} pipeline. For details about the main data reduction steps, we refer to \cite{Rajpurohit2021c}. In summary, both Band\,3 and Band\,4 data were split into six sub-bands. The flux densities of 3C\,48 were set according to \citet{Scaife2012}. After this, the data were averaged, flagged, and corrected for bandpass. Subsequently, we visually inspected the \texttt{SPAM} calibrated data for the presence of RFI, where affected data were subsequently flagged using \texttt{AOFlagger}. Finally, we combined all the data and proceeded to imaging.

The {\tt SPAM} pipeline performs direction-dependent selfcalibration, however it failed for the NGC\,741 uGMRT data. Therefore, the imaging and self calibration were performed in {\tt WSClean} and {\tt CASA} respectively. Several rounds of phase only calibration were carried out in \texttt{CASA} to refine the gain solutions followed by two final rounds of amplitude-phase calibration.

\begin{figure*}[!thbp]
    \centering
    \includegraphics[width=1.0\textwidth]{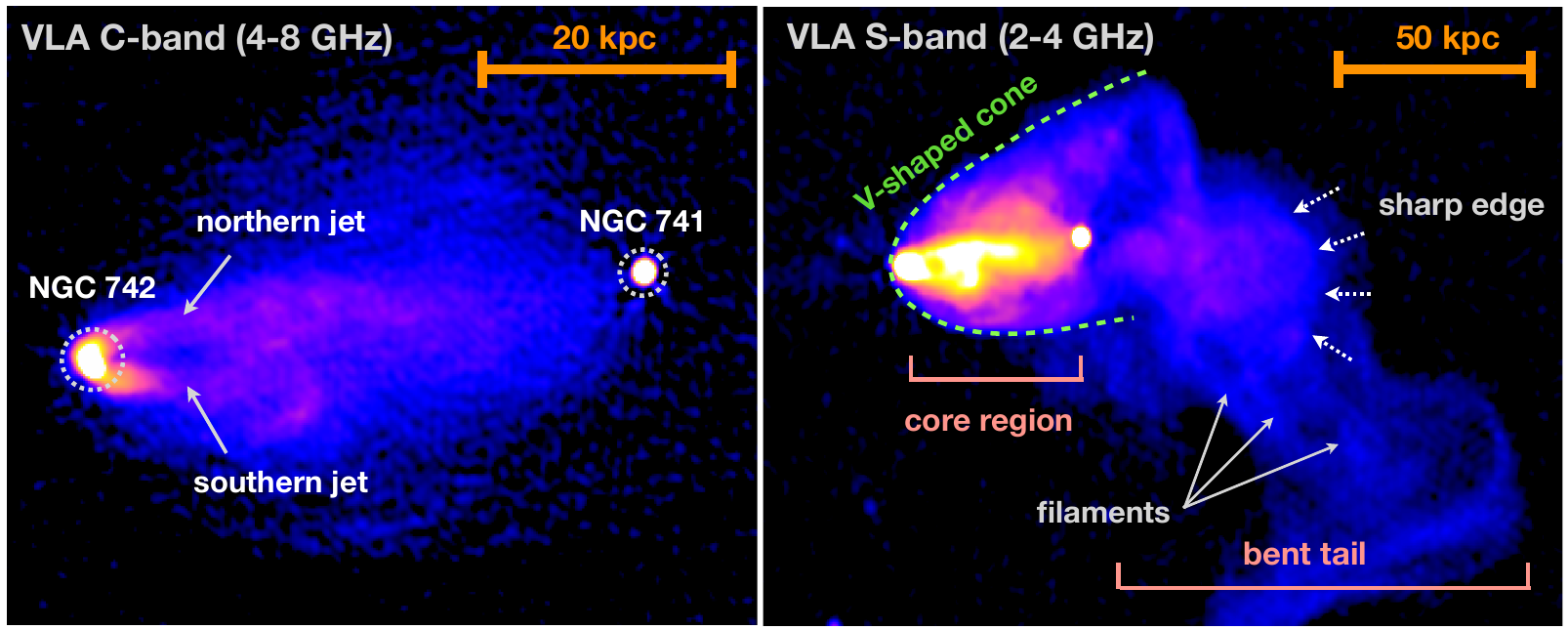}
     \caption{Radio structures in the inner region of the galaxy group NGC\,741-742. \textit{Left:} VLA C-band image showing the core region with two bent jets emerging from NGC~742 surrounded by low surface brightness diffuse emission. \textit{Right:} VLA S-band image revealing the complex morphology. The image properties are given in Table\,\ref{imaging} (IM21 and IM16).}
  \label{structure}
\end{figure*}  

\subsection{LOFAR}
The target was observed with LOFAR HBA as part of the LOFAR Two-metre Sky Survey  \citep[LoTSS;][]{Shimwell2017,Shimwell2019,Shimwell2022}. The observations were conducted in HBA dual inner mode. In particular, we used 12 datasets related to the LoTSS pointings P029+06, P027+06 and P028+04 whose centers lie at 0.68, 2.13, and 2.14 degrees away from the target position, respectively. The target is outside the DR2 survey area and thus the number of datasets is large because at low declination observations are split into multiple runs in order to observe targets at higher elevations.

The data reduction and calibration were performed using the standard LoTSS DR2 pipeline \citep{Tasse2020}, which comprises the {\tt PreFactor} pipeline \citep{vanWeeren2016c,Williams2016} and the {\tt DDF-pipeline} \citep{Tasse2020,Shimwell2019}. The {\tt PreFactor} pipeline corrects for direction-independent effects such as ionospheric Faraday rotation, offsets between XX and YY phases, and clock offsets. The direction-dependent calibration utilizes the {\tt DDF-pipeline}, which corrects for ionospheric distortions. To further improve the calibration and easier re-imaging, the data were processed by the ``extraction+self-calibration" scheme \citep{vanWeeren2020} where all sources outside a square region and centered on the target were subtracted from the visibilities followed by  several loops of self-calibration. The flux density scale from LoTSS data is aligned with the NRAO VLA Sky Survey (NVSS) \citep{Shimwell2022,Hardcastle2016}

\subsection{Flux density scale}
Since we used different flux scales, the overall flux scale for all observations (MeerKAT, LOFAR, uGMRT, and VLA) was checked by comparing the spectra of compact sources in the field of view between 144\,MHz and 6\,GHz. The LOFAR 144~MHz data points were found to be systemically low,  therefore, a correction factor was applied to the LOFAR 144~MHz data. The uncertainty in the flux density measurements was estimated as
\begin{equation}
\Delta S =  \sqrt {(f \cdot S)^{2}+{N}_{{\rm{ beams}}}\ (\sigma_{{\rm{rms}}})^{2}},
\end{equation}
where $f$ is the absolute flux density calibration uncertainty, $S$ is the flux density, $\sigma_{{\rm{ rms}}}$ is the RMS noise, and $N_{{\rm{beams}}}$ is the number of beams. We assumed absolute flux density uncertainties of 10\% for LOFAR HBA data \citep{Shimwell2022}, uGMRT Band\,3 and Band\,4 \citep{Chandra2017} and MeerKAT L-band data, 5\% for VLA S-band and 2.5\% for the VLA  C-band data \citep{Perley2013}.  

The overall extent of the radio emission reported in this paper, unless specified otherwise, in measured where the radio emission is $\geq3\sigma$. All output images are in the J2000 coordinate system and are corrected for primary beam attenuation using either {\tt CASA} and {\tt WSClean}.

\begin{figure*}[!thbp]
    \centering
        \includegraphics[width=0.49\textwidth]{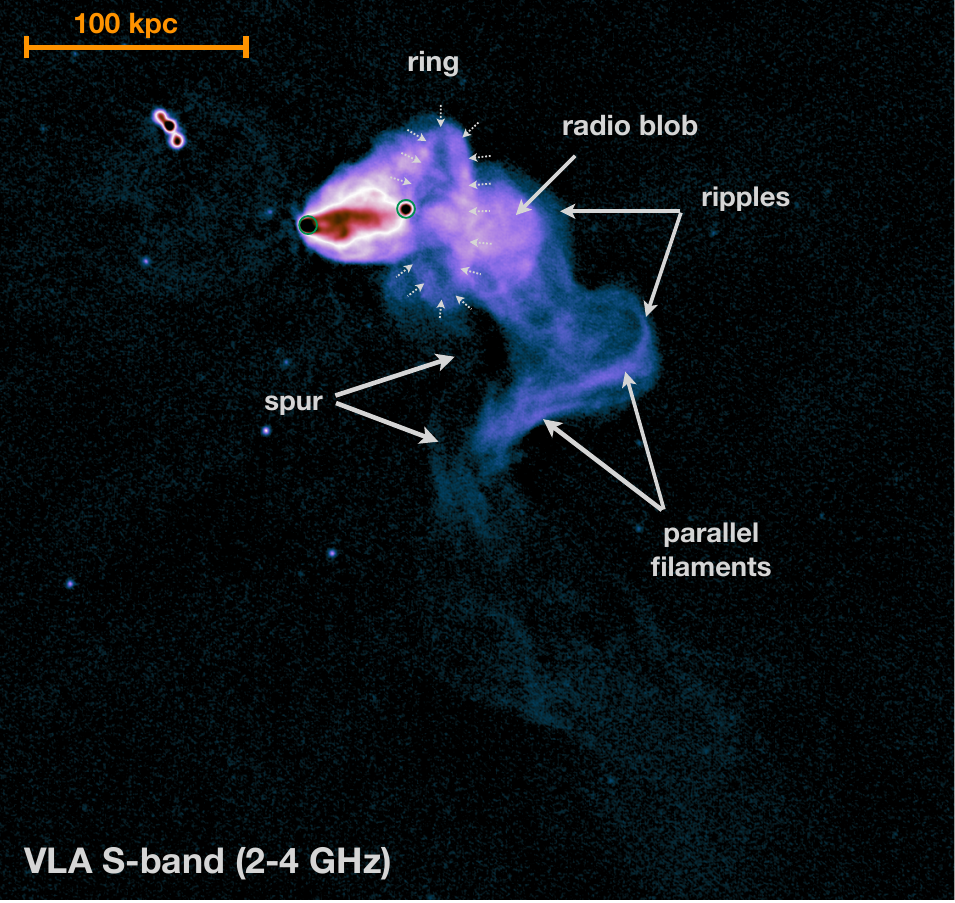}
                \includegraphics[width=0.495\textwidth]{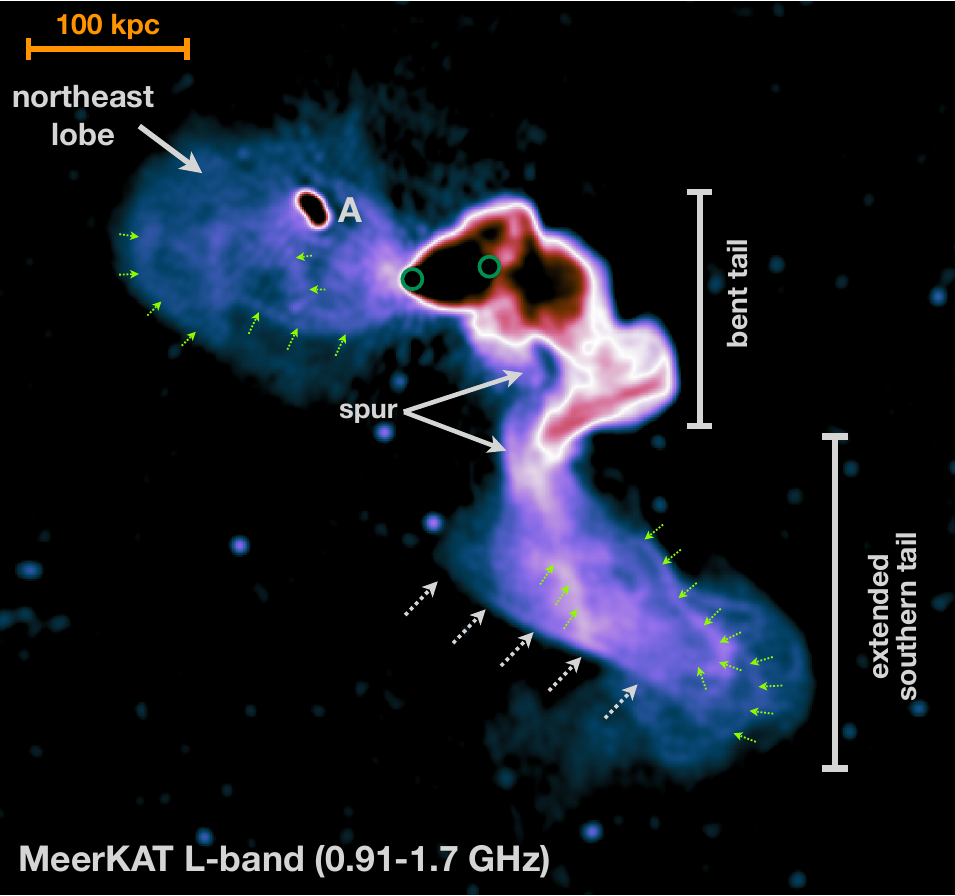}
 \caption{Radio structures in the outer region of the galaxy group NGC\,741-742. \textit{Left:} VLA S-band image revealing new features, namely a ring (also marked with white arrows), radio blob, ripples, parallel filaments, and other small scales features. \textit{Right:} MeerKAT L-band Stokes I map revealing filaments (marked with green arrows), extended tails, and a sharp radio edge (shown with white arrows). The nuclei of NGC~741 and NGC~742 are highlighted with green circles in both images. Note that images do not show the exact same region. The image properties are given in Table\,\ref{imaging} (IM16 and IM13).}
   \label{outer_structure}
\end{figure*}  

\section{Results: continuum emission}
\label{results}
\begin{figure*}[!thbp]
    \centering
    \includegraphics[width=1.0\textwidth]{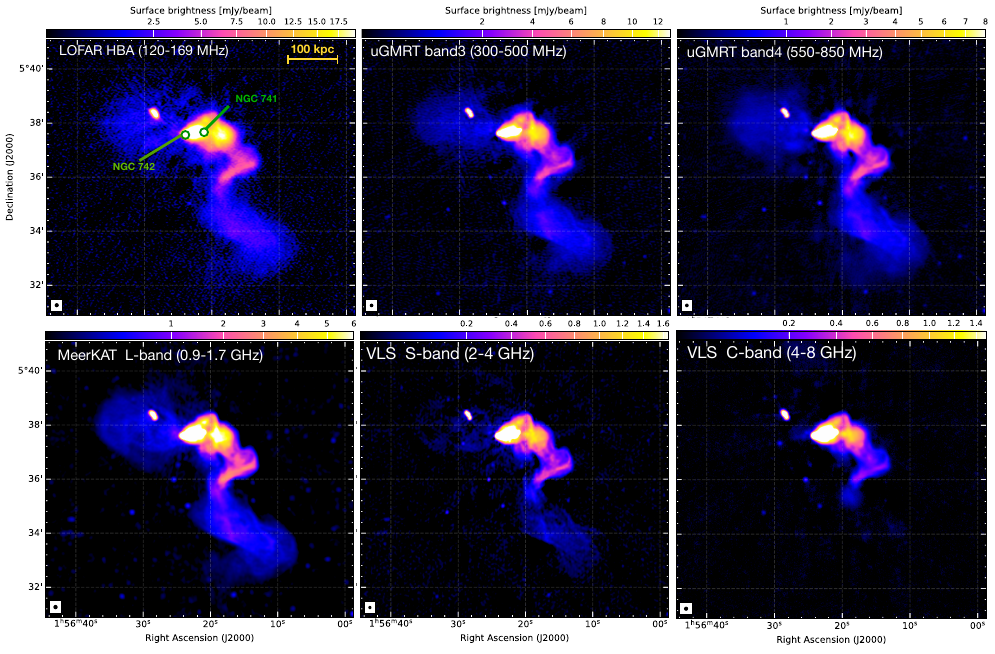}
    
\vspace{-0.25cm}
 \caption{ LOFAR, uGMRT Band\,3  and Band\,4, MeerKAT, and VLA S-band  and C-band radio images at a common resolution of 7\arcsec\ of NGC\,741-742 in square root scale. The beam size is indicated in the bottom left corner of each image. The images reveal filaments and extended tails as a function of observing frequency. From top to bottom, the images correspond to IM1, IM5, IM9, IM13, IM18, and IM23, respectively (for image properties are given in Table\,\ref{imaging}).}
      \label{high_res}
\end{figure*}

\begin{figure*}[!thbp]
    \centering
    \includegraphics[width=1.0\textwidth]{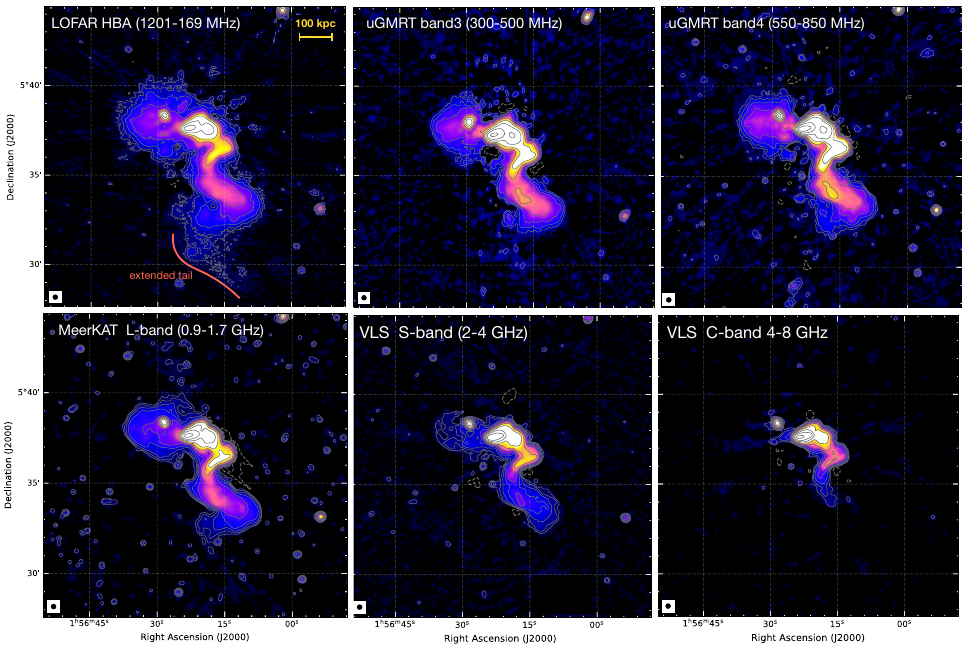}
    
    \vspace{-0.25cm}
 \caption{Medium resolution (15\arcsec) LOFAR, uGMRT Band\,3, Band\,4, MeerKAT, VLA S-band, and C band radio images of NGC\,741-742 in square root scale (in Jy/beam unit). The beam size is indicated in the bottom left corner of each image. The images reveal filaments and extended tails as a function of observing frequency. Here, from top to bottom, the images correspond to IM2, IM6, IM10, IM14, IM19 and IM24, respectively. The image properties are given in Table\,\ref{imaging}. Contours drawn at  $[1, 2, 4, 8 ...]\times 3.0\sigma_{\rm rms}$. Dashed contours depict the $-3.0\sigma_{\rm rms}$ contours.  }
      \label{low_res}
\end{figure*}

Deep VLA, MeerKAT, uGMRT, and LOFAR total intensity images are shown in Figures\,\ref{structure} to \ref{low_res}.  Our new multi-frequency images provide an unprecedented view of the radio emission from the galaxy group NGC\,741 over a broad frequency range. For the first time, the VLA and LOFAR observations allowed us to image the radio emission at high resolution both at high and low frequencies. Our 144\,MHz LOFAR image is about a factor of 25 deeper than the GMRT 150\,MHz image presented by \cite{Schellenberger2017}. Additionally, the uGMRT, MeerKAT, and VLA data are more than 6 times deeper than the published images at the same frequencies. 

To facilitate the discussion we divided the source in three main regions; core (covering NGC\,741 and NGC~742), bent tail and extended southern tail (see Figures\,\ref{structure} and \ref{outer_structure} for regions). The new total intensity images recovered these three regions known from previous radio observations \citep{Schellenberger2017}. However, our new images reveal previously unseen features and additional emission. In this section we discuss the most prominent features in the various regions of the source.

\subsection{Core region}
To resolve the structures in the core region and to examine the jets, we use the highest-resolution VLA images. Figure~\ref{structure} shows two compact cores associated with NGC\,741 and NGC~742. The VLA C-band ($1.0\arcsec\times0.7\arcsec$ beam size) image reveals two bent jets emerging from NGC~742 (Figure~\ref{structure} left panel) also reported by \citet{Schellenberger2017}. However, our new images allow us to trace the jets in detail. In particular, the northern jet appears notably longer than its southern counterpart, which diminishes halfway along its path (Figure~\ref{structure} left panel). The morphology suggests twisting, with the southern jet overlapping its northern counterpart, but the true physical structure is unclear, including whether the northern jet bends toward the south and the one or both jets may lose their collimation and broaden. By contrast, the NGC\,741 core is compact and shows no sign of radio jets in the VLA high resolution images. The Very-long-baseline interferometry observations also do not find any jet from NGC\,741 \citep{Chun2000}. Both NGC\,741 and 742 are surrounded by diffuse radio emission. 

A striking feature in the radio images is a V-shaped cone-like structure with NGC~742 at its apex (Figure~\ref{structure} right panel), also reported by \cite{Schellenberger2017}. However, our new images allow us to study this structure at higher spatial resolution. The connection between the NGC~742 jets and the V-shaped cone is unclear, as is the path of the jets/tail to the east of NGC\,741. Surface brightness discontinuities are often seen in the thermal emission of galaxy groups undergoing mergers, either as merger cold fronts,  sloshing fronts, or shock fronts. Based on its position,  shape, and the presence of a temperature discontinuity, \cite{Schellenberger2017} suggested the presence a shock front coincident with the V-shaped structure seen at the radio band. 

\subsection{Bent tail region}
Figure~\ref{structure} (right panel) shows the VLA S-band (3~GHz) image. While the 6~GHz VLA C-band image reveals the base of NGC~742's jets, the S-band image traces new intricate structures extending tens of kiloparsecs into the bent tail. After the jets, the emission from NGC\,742 undergoes a dramatic change in structure, direction, and surface brightness (Figures~\ref{structure} and \ref{outer_structure}). The radio emission extends across NGC\,741 before bending to the southwest, after which its structure becomes complex, forming a bright, well-defined cloud (labeled as radio blob) at about $80\arcsec$  distance from the NGC\,742 core. The radio blob covers an area of $39\arcsec\times43\arcsec$ and has a sharply defined western edge (marked with arrows in Figure~\ref{structure} right panel).

As depicted in Figure~\ref{outer_structure} left panel, for the first time our new observations reveal a remarkable ring-like feature to the west of NGC\,741 (marked with arrows and labeled as `ring'). It has a major axis diameter of about 1.2\arcmin\ (in physical scale about 26~kpc) and minor axis diameter of about 26\arcmin (about 9.6\,kpc). The width of the ring is approximately 12\arcsec\ (5.2~kpc). The northern edge of the V-shaped cone aligns with the northern perimeter of the ring, while to the south, the ring extends well past the cone's edge (Figure~\ref{outer_structure} left panel). The ring appears to surround the radio blob and NGC~742 tail structures, although it is possible they are only superimposed along the line of sight. There is a sharp decrease in the radio source brightness in a region between the newly discovered ring and radio blob. 

 The bent tail region shows ripples (i.e., sharp turns) along its western edge (Figure~\ref{outer_structure} right panel). To the east of the ripples are thin filaments (see Figures~\ref{structure} left panel). The bent tail from the NGC\,742 stretches out to 1.8\arcmin ($\sim\rm 108\,kpc$) before bending toward the west and southwest, its outer edge marked by two parallel filaments (see Figure\,\ref{outer_structure} left panel). Both filaments are approximately 33 kpc long, but the northern filament is 4.5 kpc wide, significantly wider than the southern one  (1.5 kpc). Similar pairs of filaments are reported in CL2015 \citep{Andreon2019}, NGC~6065 \citep{Candini2023} and Abell 194 \citep{Rudnick2022}

\subsection{Extended southern tail region}
As shown in Figure\,\ref{outer_structure} right panel, on larger scales we see the extended southern tail, also reported by previous observations \citep{Schellenberger2017,Giacintucci2011}. However, our new images reveal additional extent of the tail and for the first time the presence of filaments (some of them are marked with green arrows) with a braided appearance across the entire southern extent of the tail, as if multiple filaments within the tail are twisted around one another. These thin filaments lack any large scale ordering. The eastern edge of the southern tail is edge-sharpened (arrows in Figure \ref{outer_structure}). We observe structure in the tail all the way out to its farthest extent. The large extension of the tail becomes fainter, extending toward the southeast, transitioning from a bright channel of filamentary substructures to fainter, diffuse extended emission. 

The new radio images also show bright spur-like features (Figure\,\ref{outer_structure} left panel, labeled as `spur') to the south of the NGC\,741 core which are apparently linked to the southern tail by a bridge of faint emission. This connection is most apparent in the MeerKAT L-band data and at low frequencies (Figure\,\ref{outer_structure} right panel). The nature of this structure is unclear. One possibility is that it may be a second tail, or secondary channel within the tail, associated with NGC~742.

\subsection{Northeast lobe} 
Our new images confirm the presence of a radio lobe-like emission located to the east of NGC~742 (Figure\,\ref{outer_structure} right panel) detected partially by \citet{Giacintucci2011} and \citet{Schellenberger2017} in the GMRT 235\,GHz image.  In our new images, the northeast lobe (NE) is extended and detected at a high significance level up to 1.28\,GHz. No corresponding lobe is seen to the west of NGC\,741. However, \cite{Schellenberger2017} identified a possible cavity in the X-ray image,  which may mark its location. The NE lobe overlaps another radio source, NVSS~J015628+053820, with core plus lobes morphology (labeled as A in Figure\,\ref{overlay}). No redshift is available for the source, but it is thought to be a distant background galaxy.

\begin{figure*}[!thbp]
    \centering
    \includegraphics[width=0.66\textwidth]{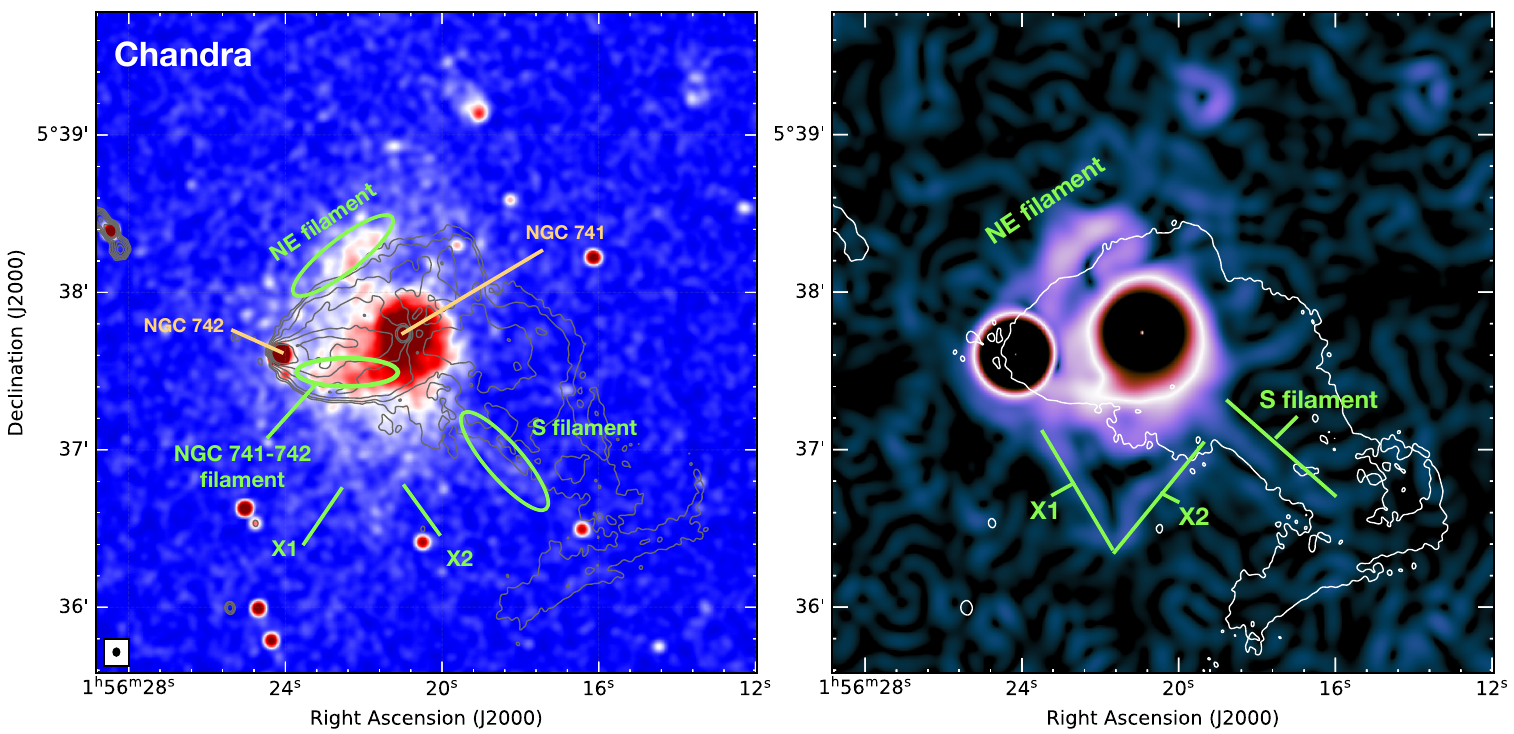}
      \includegraphics[width=0.327\textwidth]{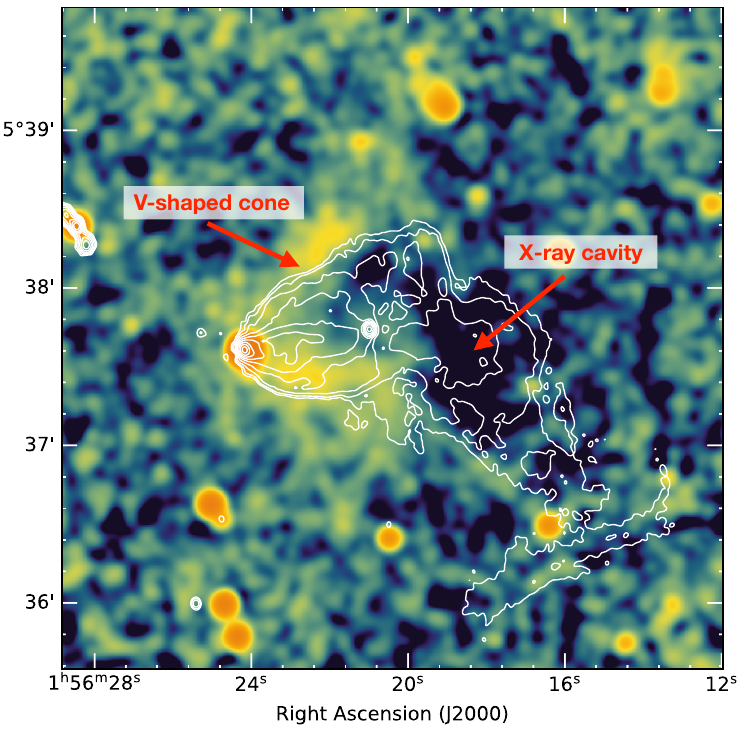}
 \caption{ \textit{Left:} Background subtracted, exposure corrected {\chandra}  0.5-2.0 keV image overlaid with C-band VLA radio contours.  The {\chandra} map is smoothed with a Gaussian of full width at half maximum (FWHM)  3\arcsec. Radio contours are drawn at $[1, 2, 4, 8 ...]\times 4.0\sigma_{\rm rms}$ \textit{Middle:} {\chandra} 0.5-2.0~keV image image processed with the Gaussian Gradient Method with $\sigma= 10$ pixels, highlighting  X-ray filaments. Radio contour is drawn at  $1\times 3.0\sigma_{\rm rms}$. \textit{Right:} Residuals of the best-fit 2D beta model overlaid with the VLA C-band high-resolution contours, revealing a small cavity and a V-shaped structure. Radio contours are same as the left panel.}
      \label{Xray}
\end{figure*}  

The NE lobe and the extended southern tail exhibit distinct morphologies (Figure\,\ref{low_res}). If the emission in these two regions is associated with radio lobes/tails, one likely reason for the difference in morphology can be the environment in which the host radio galaxy resides. In case the thermal density of the gas surrounding the NE lobe and the southern tail differ, we would expect to see different lobe/tail morphologies. In a high-density environment, an elongated tail is expected whereas in a low-density environment, a shorter and wider lobe. The distribution of thermal emission indeed suggests that the NE lobe is located in a low density IGrM \citep{Schellenberger2017} compared to the southern tail, which is consistent with its morphology, i.e., wider lobe.   

For the first time, we also see small-scale filaments embedded within the NE lobe in the  high-resolution MeerKAT image (marked with green arrows in Figure\,\ref{outer_structure}). Since the NGC~742 jets bend to the southwest, this lobe is unlikely to be associated with the infalling galaxy.  

\subsection{Radio emission as a function of frequency} 
To compare the total extent of the radio emission as a function of frequencies, in Figures\,\ref{high_res} and \ref{low_res} we show the LOFAR, MeerKAT, VLA and uGMRT images at a common resolution of 7\arcsec\ and 15\arcsec, respectively. At the highest frequency probed by our observations (i.e., 4-8\,GHz), only the core and bent tail regions are visible (LLS $\sim 60$\,kpc), while at the lowest frequency (144\,GHz), we recovered the largest extent of the radio emission (LLS$\sim$255\,kpc) from the NGC~741-742 system.

The NE lobe is not visible in the C-band (Figure\,\ref{low_res}), and its extent grows towards low frequencies, hinting at the steeper spectral index in the outermost regions. It has an LLS of about 100~kpc, 90~kpc, 70~kpc and 57~kpc  at 144~MHz, 400~MHz, 700~MHz, 1.28~GHz and 3~GHz, respectively. 

The main southern tail is clearly detected from 144~MHz to 3~GHz, with its extent increasing toward lower frequencies (Figure\,\ref{low_res}). The most extended part (labeled as extended tail) is only detected at 144~MHz. The LSS of the extended southern tail is about 150 ~kpc at 144~MHz, while it is about 85~kpc at 700~MHz and 1.28~GHz.

\begin{figure}[!thbp]
    \centering
        \includegraphics[width=0.49\textwidth]{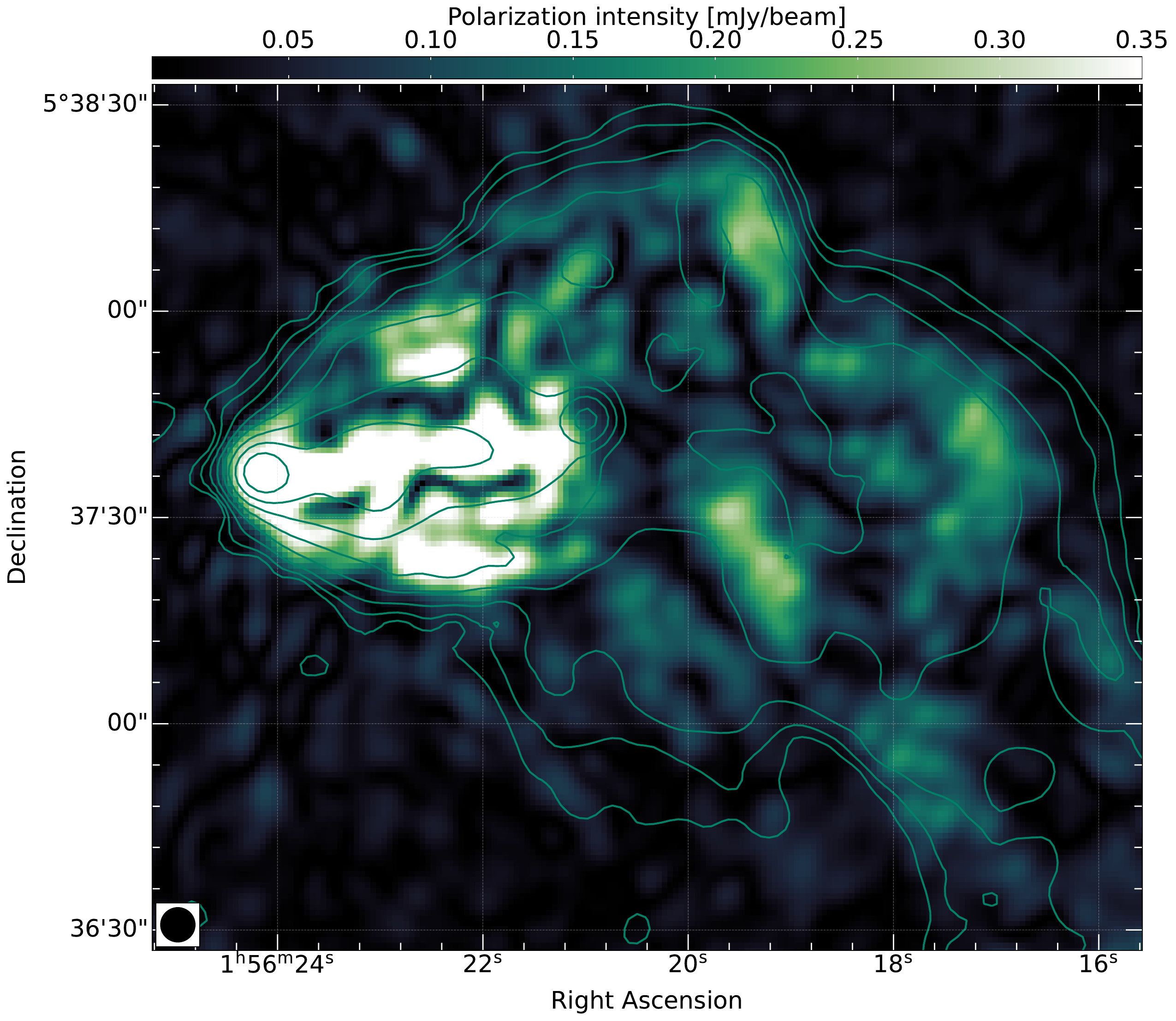}
         \includegraphics[width=0.49\textwidth]{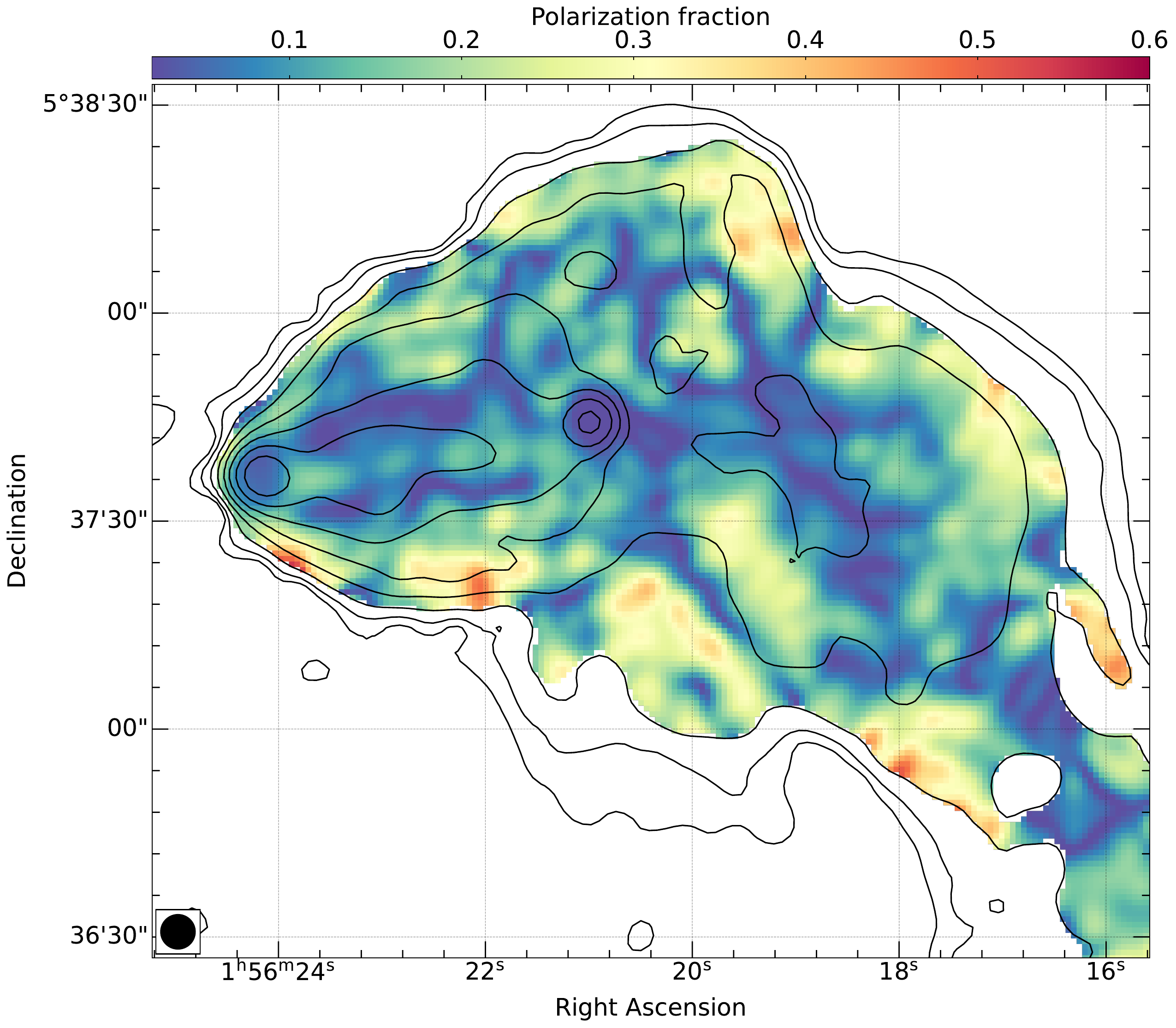}
         
    \vspace{-0.20cm}
 \caption{VLA S-band polarization intensity and fractional polarization maps overlaid with Stokes I contours. The strongest polarization is observed along the V-shaped Mach cone, most likely due to shock compression, and in the X-ray filament connecting NGC\,741-NGC\,742. Stokes I S-band contours levels are drawn at  $[1, 2, 4, 8 ...]\times 7.0\sigma_{\rm rms}$. The beam size is indicated in the bottom left corner of the image.}
      \label{poli}
\end{figure}  

\section{Analysis and Discussion}
\label{discussion}

\subsection{Radio and X-ray comparison}
To study the relationship between X-ray and radio emission, we use the reduced \textit{Chandra} maps presented in \cite{Schellenberger2017}. Figure\,\ref{Xray} shows a {\chandra} 0.5-2~keV X-ray image with VLA C-band radio contours superimposed. In the core region, the brightest X-ray emission originates from the cores of NGC\,741 and NGC~742 (labeled as NGC\,741-NGC\,742 filament). As reported by \cite{Schellenberger2017}, the surrounding halo emission is highly structured, with several filaments:  a northeast filament (NE-filament), a linear feature to the south (S-filament), and two, thin line-shaped filaments (X1 and X2) to the south of the core region. We note that X1 was not reported by \cite{Schellenberger2017}, however it is detected at $\geq3\sigma$ level in Chandra maps. Filament S lies along the edge of the radio tail, and therefore probably consists of gas compressed by the motion of the relativistic tail plasma. The remarkable linearity of X1 and X2, which are not correlated with radio structures, might be explained if magnetic fields stabilize them against tidal shear and dissipation into the surrounding medium.

To highlight the filaments in the X-rays, we also used the Gaussian Gradient Magnitude (GGM) filter method \citep{Sanders2016}. The GGM filter uses Gaussian derivatives to determine the magnitude of surface brightness gradients in the image. We applied the GGM filter to the exposure-corrected {\chandra} images. In Figure\,\ref{Xray} middle panel, we show the resulting map. The X-ray filaments are much more visible in this image. The width of the NE filament is $\sim7~\rm kpc$. The length of X1, X2, and S-filament are 20\,kpc, 19\,kpc, and 18~kpc, respectively, and all have a width of about 3~kpc. 


\setlength{\tabcolsep}{15pt}
\begin{table*}[htp]
\caption{Flux densities of the different regions of NGC\,741-NGC\,742.}
\begin{center} 
\begin{tabular}{*{10}{c}}
\hline \hline
\multirow{1}{*}{Regions} &\multirow{1}{*}{LOFAR} &\multicolumn{2}{c}{uGMRT} &\multirow{1}{*}{MeerKAT}& \multicolumn{2}{c}{VLA} \\
 \cline{3-4}  \cline{6-7} 

& $S_{\rm144\,MHz}$ &${S_{\rm400\,MHz}}$&${S_{\rm700\,MHz}}$ & ${S_{\rm1.28\,GHz}}$&${S_{\rm3.0\,GHz}}$&${S_{\rm6.0\,MHz}}$\\
  & (mJy) & (mJy) & (mJy) & (mJy)  & (mJy) &(mJy) \\

  \hline 
Total emission& $6000\pm800$ &$2500\pm300$&$1650\pm200$&$1050\pm100$&$450\pm30$&$190\pm7$\\
Core region & $2000\pm250$&$1000\pm100$ & $690\pm70$&$470\pm50$ & $250\pm10$& $140\pm6$\\
Bent tail region & $2000\pm250$& $930\pm100$& $640\pm80$&$390\pm40$& $160\pm8$& $50\pm2$\\
S-tail& $1300\pm150$& $430\pm50$& $265\pm28$&$140\pm15$& $30\pm2$& $-$\\
NE lobe & $700\pm80$ &$190\pm20$&$110\pm12$&$60\pm4$&$7.0\pm0.4$&$-$\\
\hline 
\end{tabular}
\end{center} 
{Notes. All reported flux densities were extracted from 15\arcsec\ images created with ${\tt robust}=-0.5$ and a \textit{uv}-cut of $0.1k\lambda$ excluding the core region where we used 7\arcsec\ images to properly subtract the contribution from NGC\,741 and NGC~742 nuclei. The image properties are given in Table\,\ref{imaging}. The regions where the flux densities were extracted are indicated in the Figure\,\ref{spectra} left panel. Absolute flux density scale uncertainties are assumed to be 10\% for LOFAR, uGMRT Band\,3, uGMRT Band\,4 and MeerKAT, 5\% for the VLA S-band and 2.5\% for VLA C-band data.}
\label{Tabel:Tabel2}   
\end{table*}   

The brightest of the X-ray filaments appears to connect NGC\,741 and NGC\,742 (Figure\,\ref{Xray}). In the high-resolution VLA S-band image, we see this filament coinciding with the southern edge of the V-shaped cone (see Figure\,\ref{structure}). Additionally, the radio brightness is enhanced in the same region with a hint of a small-scale, thin feature (Figure\,\ref{structure} right panel) that may be tracing the X-ray filament. This bright filament comprises low-entropy gas, consistent with being stripped from NGC\,742  \citep{Schellenberger2017}. Alternatively, it could consist of gas from the core of NGC\,741, drawn outward during the close encounter. We measure the strongest polarization along this filament (see Section\,\ref{polarization}).

Figure\,\ref{Xray} right panel shows the {\chandra} residual map. As reported by \cite{Schellenberger2017} a V-shaped structure is observed, with a hint of a temperature discontinuity in the same region. This feature appears spatially distinct and coincides with the V-shaped cone detected in radio maps (Figure\,\ref{structure}). This suggests that very likely it arises from compression or shock heating of the IGrM by the passage of NGC~742 through the group core. To the west of NGC\,741, there is a region of negative residuals which \cite{Schellenberger2017} interpret as a small cavity. This region coincides with the bright radio blob observed in the radio images.

The mechanical power of the cavities is found to be correlated with the AGN radio power \citep[e.g.,][]{O'Sullivan2011,Liu2019}. In general, luminous radio galaxies are expected to have a large cavity power. The cavity powers in groups are found to be generally low, even for those radio luminous systems. The radio power of the radio blob is $L_{\rm 1.3\,GHz}=\sim 1\times10^{23}\rm\,W\,Hz\,s^{-1}$. Based on the sound crossing timescale, \cite{Schellenberger2017} reported a cavity power of $P_{\rm cav}=5.1\pm0.5\times10^{42}\rm\,erg\,s^-{1}$. These estimated values agree well (within the scatter) with the known mechanical power versus radio power relations for galaxy groups and clusters. 

We also observe cospatiality between the detected thermal X-ray filaments and the outer edge of the radio emission (see Figure\,\ref{Xray}). The NE-filament traces the leading edge of the V-shaped cone in the radio. This apparent connection is clearer in the {\chandra} residual map, see Figure\,\ref{Xray} right panel. The S-filament is aligned with one of the filaments seen in the bent tail region (Figure\,\ref{structure} bottom left). This likely implies a physical connection between the non-thermal plasma and the surrounding thermal medium. The location of the NE and S filaments at the edges of radio structures suggests compression of the IGrM as the relativistic plasma of the NGC~742 tail interacts with the IGrM. Polarization and Faraday analysis of the radio emission may provide insight into compression and interaction with the IGrM. X1 and X2 have no radio counterpart, but may be related to the perturbation caused by the infalling galaxy \citep{Schellenberger2017}. The non-detection of X1 and X2 at the radio band could be due to absence of relativistic electrons.

\subsection{S-band polarization}
\label{polarization}
The S-band VLA images of the Stokes parameters I, Q, and U were obtained using robust=0 weighting. These images serve only the purpose of studying the polarized emission in the core region of NGC~742-741. A detailed analysis of polarization and Faraday rotation measure (RM), combining all available data from the L, S, and C-bands, will be presented in a future paper. 

We obtained the linear polarized intensity ($p$) and fractional polarization ($\pi$) maps from the Stokes I, Q, and U maps as
\begin{subequations}
\begin{align}
p = \sqrt{Q^2 + U^2},\\
\pi =  \frac{\sqrt{Q^2 + U^2}}{I}.
\end{align}
\end{subequations}
Figure\,\ref{poli} shows the resulting $5.5\arcsec$ resolution 2-4\,GHz polarization intensity and fractional polarization maps.

The polarized emission across the source is detected from the core and bent tail regions.  The nuclei of NGC~742 exhibits a fractional polarization of about 6\% while the nuclei of NGC\,741 is almost unpolarized in the S-band ($<2\%$). The average degree of polarization across the jets of NGC\,742 is 9\%. We emphasize that we did not correct for the effect of Faraday rotation. Therefore it could be that the true polarization fraction is even higher. It is worth noting that between 2-4~GHz, the maximum polarization of about 48\% occurs along the brightest X-ray filament connecting NGC\,741 and NGC~742 (Figure\,\ref{poli} and Figure\,\ref{Xray}).  Strong polarization is also observed along the boundaries of the V-shaped shock cone (including the NGC\,741-742 filament), with an average degree of polarization of 20\%. This is consistent with compression of the magnetic field along the shock front. The ring is also polarized at S-band.


\begin{figure*}[!thbp]
    \centering
            \includegraphics[width=0.49\textwidth]{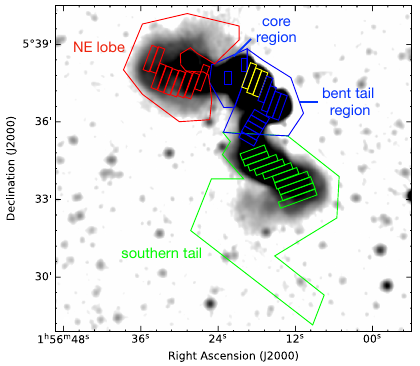}
    \includegraphics[width=0.49\textwidth]{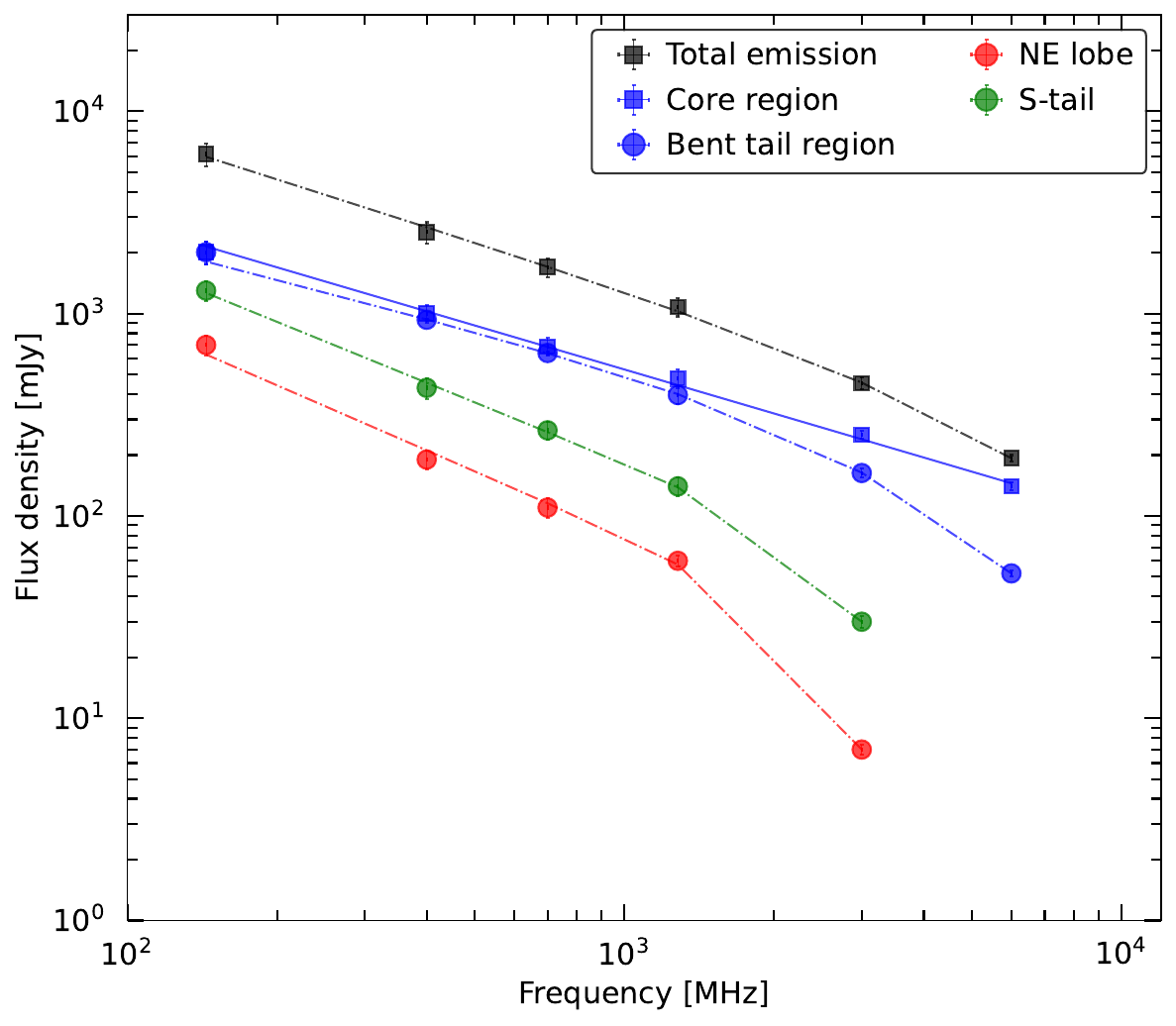}
   \vspace{-0.2cm}
 \caption{\textit{Left}: MeerKAT 1.28~GHz Stokes I image showing regions where the flux densities were extracted to obtain the integrated spectra (bis regions) and radio color-color diagram/global spectra (rectangles). \textit{Right}: Integrated spectra of subregions of NGC\,741-742 between 144\,MHz and 6\,GHz. Dashed lines show the fitted broken power law and the solid line shows the core region spectrum which is a power law spectrum. Except for the core region, all other regions show a high-frequency steepening as expected due to aging of electrons. }
      \label{spectra}
\end{figure*}  


\begin{figure}[!thbp]
    \centering
      \includegraphics[width=0.49\textwidth]{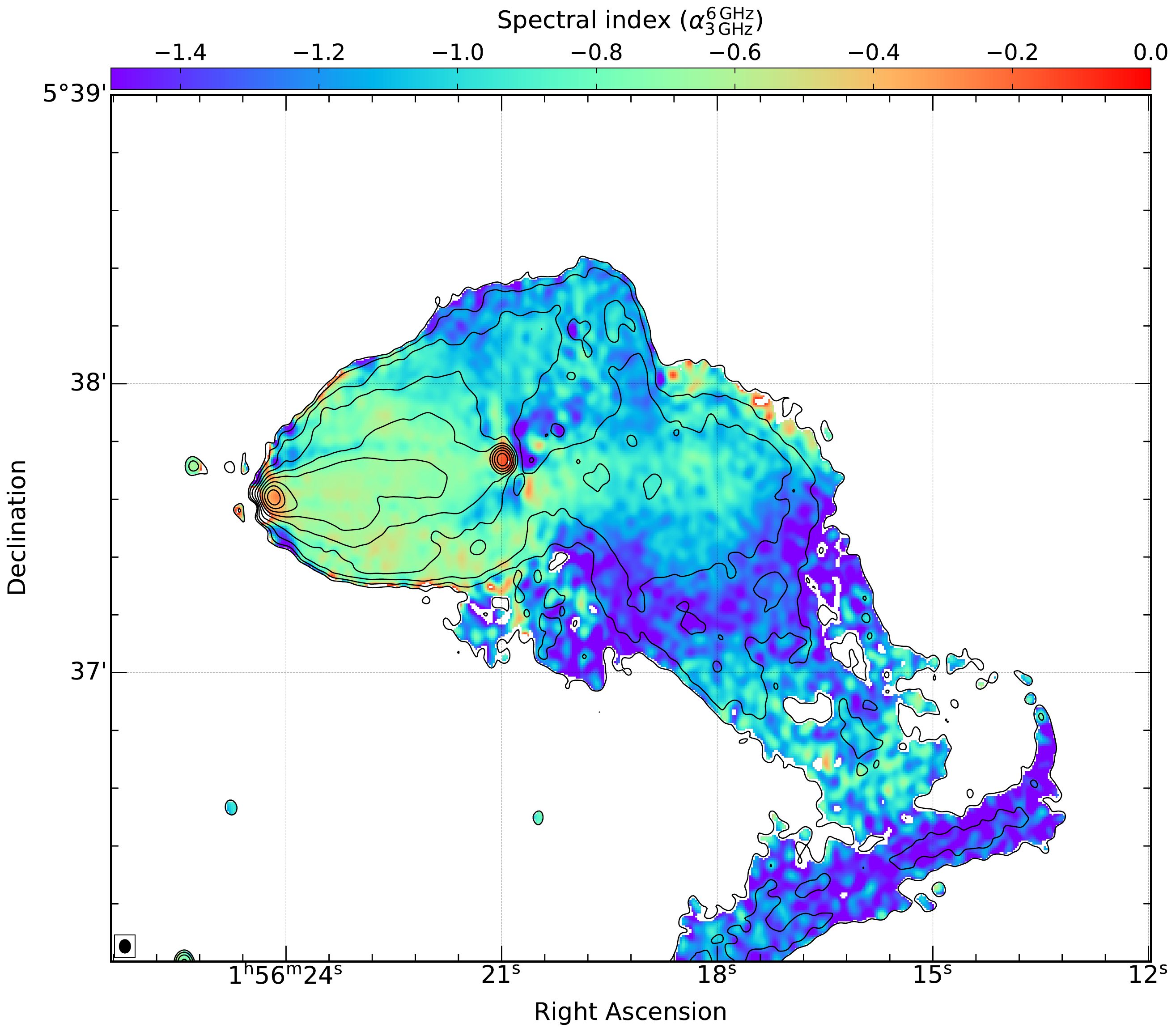}
 \caption{High resolution ($2.8\arcsec\times 2.3\arcsec$) spectral index maps of the inner region created between 3 and 6\,GHz. The core region is the flattest and the ring and blob are comparatively steeper.  The spectral index steepens when moving from the core to the bent regions, suggesting electron aging. Contour levels are drawn at  $[1, 2, 4, 8 ...]\times 3.0\sigma_{\rm rms}$ and are from the VLA C-band image. The beam size is indicated in the bottom left corner.}
      \label{high_res_index}
\end{figure} 

\subsection{Radio spectral index and curvature  analysis}
To study the spectral characteristics{\footnote{We define the radio spectral index, $\alpha$, so that $S_{\nu}\propto\nu^{\alpha}$, where $S$ is the flux density at frequency $\nu$.} }of the radio emission in NGC\,741-742, we use our VLA~C-band (1-2\,GHz), VLA S-band (2-4\,GHz), MeerKAT L-band (0.9-1.7~GHz), uGMRT Band\,3 (300-500\,GHz), uGMRT Band\,4 (550-850\,GHz), and LOFAR (120-169~MHz) observations. 

The radio observations reported here were performed using four different interferometers, each with different uv coverage. This requires careful attention when comparing flux density measurements of extended emission. To derive reliable flux densities and spectral index maps, we created images at 144~MHz, 400~MHz, 700~MHz, 1.28\,GHz, 3~GHz, and 6~GHz with Briggs weighting and a robust parameter of $-0.5$. To ensure that the flux distribution has a similar effect at all the observed frequencies, we create images with a common inner uv-cut at $100\lambda$, which is the well-sampled shortest baseline of the uGMRT data. This uv-cut is applied to all the other telescopes. The full size of the source is about $6.2\arcmin$ and $3\arcmin$ at 3~GHz and 6~GHz, respectively. The largest angular scale recovered by the VLA at S and C bands is 8.2\arcmin\ and 4\arcmin\ , respectively. This implies that we are not missing any significant flux density due to the missing angular scale issue at high frequencies. To reveal the spectral properties of different spatial scales, we tapered the images accordingly. The imaging parameters and the image properties are summarized in Table\,\ref{imaging}.

\begin{figure*}[!thbp]
    \centering
      \includegraphics[width=0.98\textwidth]{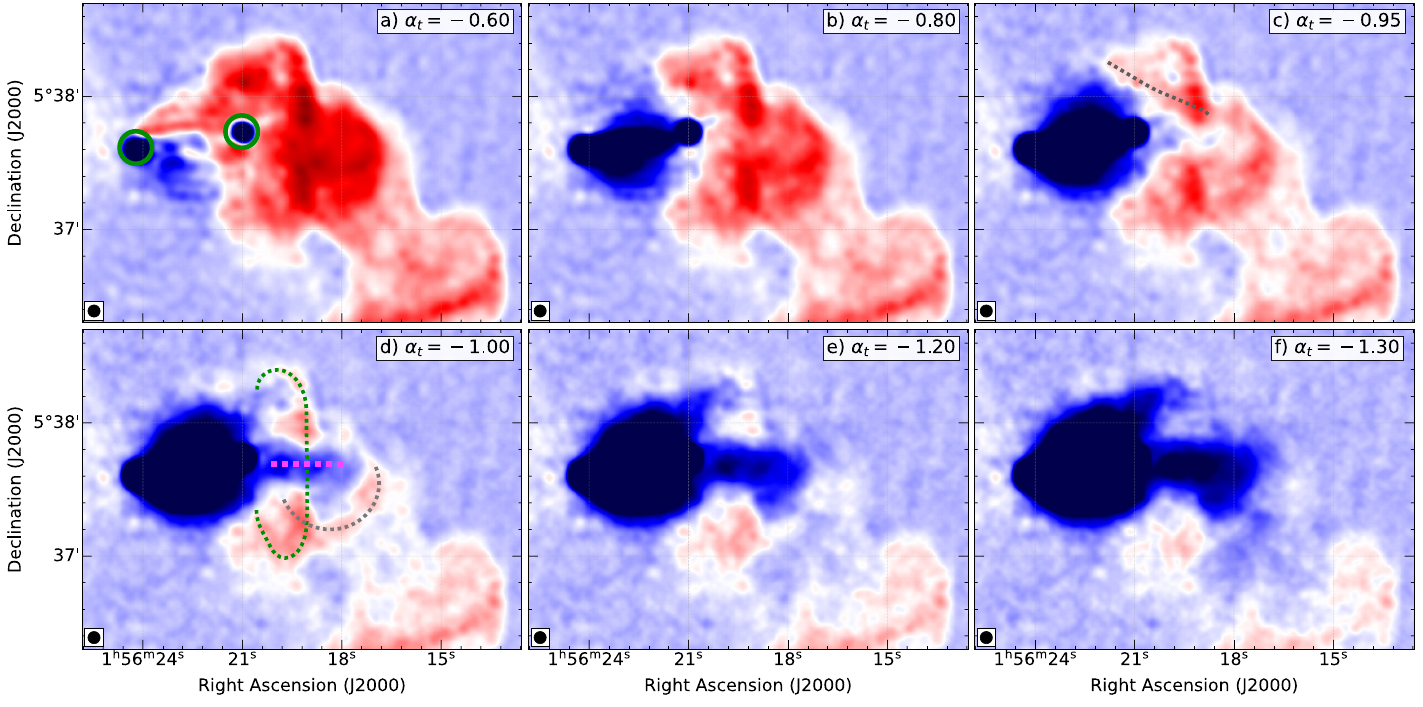}
 \caption{Gallery of spectral tomography maps between 1.28 and 6~GHz at $5.5\arcsec$ resolution. The range of $\alpha_t$ is $-0.60$, $-0.80$,  $-0.95$, $-1.00$, $-1.20$, and  $-1.30$. The regions with a spectrum steeper than $\alpha_t$ appear positive (red-white regions), while regions with a flatter spectrum appear negative (blue). These images demonstrate that there are several features in the inner region with different spectral indices. The main overlapping/distinct features are shown with dotted lines. Green circles mark the location of NGC\,741 and NGC\,742 core. The image properties are given in Table\,\ref{imaging} (IM12 and IM17).}
      \label{tomo}
\end{figure*}  

\subsubsection{Integrated spectrum}
To obtain integrated radio spectra of NGC\,741-742 as a whole, and of the various subregions, we created images at 15\arcsec\ resolution. We chose this resolution  because it provides high signal-to-noise in low-surface-brightness regions (in particular for the NE lobe), while allowing us to identify distinct substructures.  The same maps are used to perform radio color-color and global spectrum analysis. 

The regions used to extract flux densities are indicated in Figure\,\ref{spectra} left. The regions are based on the LOFAR 144~MHz images since the radio emission is most extended at that frequency.  The flux densities are summarized in Table\,\ref{Tabel:Tabel2}. Our new LOFAR 144~MHz and uGMRT 700\,GHz measurements for the total emission are consistent with those reported by \cite{Schellenberger2017} using narrow band GMRT 150~MHz and 610~MHz data. Our 1.28\,GHz MeerKAT flux density is also in line with the VLA DnC-array and NVSS data \citep{Schellenberger2017}. 

We also created an image at 4.8\,GHz. However, our 4.8\,GHz flux density value is about $250\pm10$\,mJy which is significantly lower than the VLA (DnC) 4.8~GHz measurement by \cite{Schellenberger2017}, namely $300\pm20$. We emphasize that our VLA 4.8\,GHz map is more sensitive than \cite{Schellenberger2017} and we used multiscale cleaning. To exactly compare our values with \cite{Schellenberger2017}, we re-imaged the C-band data at 20\arcsec\ resolution without any uv-cut using ${\tt robust}=0$. The resulting image shows a flux density of $280\pm12$ which is comparable to the value reported in \cite{Schellenberger2017}. This implies that the difference in flux density at 4.8\,GHz is simply due to different imaging parameters and resolution.

The resulting spectra from NGC~742-741 and its subregions, obtained by our flux density measurements, are shown in Figure\,\ref{spectra} right. The wideband interferometric observations enabled us to study the integrated spectrum in regions over a large frequency range. The integrated spectrum of the total emission from NGC~742-741 (excluding source A) is significantly curved at frequencies above 1.28~GHz, implying the presence of strong radiative losses. The low frequency spectral index is flat, namely $\alpha_{\rm 144~MHz}^{\rm 1.28\,GHz}=-0.80\pm0.02$. At high frequencies, the spectral index is steep, $\alpha_{\rm 1.28~GHz}^{\rm 6.0\,GHz}=-1.11\pm0.02$.

The subregions show different spectral indices. The core region, comprising the shock cone and the jets, can be characterized by a power-law spectrum and has a spectral index of $\alpha_{\rm 144~MHz}^{\rm 6\,GHz}=-0.71\pm0.02$. Such a flat spectral index is expected, both for active jets and for shock compression of radio plasma.  The bent tail region has a low frequency spectral index of $\alpha_{\rm 144~MHz}^{\rm 1.28\,GHz}=-0.75\pm0.02$, steepening at high frequencies to $\alpha_{\rm 144~MHz}^{\rm 1.28\,GHz}=-1.33\pm0.02$.

The NE lobe and southern tail are not visible at 6~GHz, and, therefore, we exclude 6~GHz data. The NE lobe is very steep at high frequency, $\alpha_{\rm 1.28~GHz}^{\rm 3.0\,GHz}=-2.52\pm0.04$. Between 144~MHz and 1.28~GHz, the spectral index is quite flat $\alpha_{\rm 144~MHz}^{\rm 1.28\,GHz}=-1.13\pm0.04$. In contrast, the low frequency overall spectrum of the southern tail is $\alpha_{\rm 144~MHz}^{\rm 1.28\,GHz}=-1.02\pm0.04$, which is flatter than in the NE lobe. The southern tail also shows a steepening above 1.28\,GHz and has a high frequency spectral index of $\alpha_{\rm 1.28~GHz}^{\rm 3.0\,GHz}=-1.89\pm0.04$, which is flatter than the NE lobe in the same frequency range. The difference in spectral index between the NE lobe and the southern tail suggests that the NE lobe is older or that it has same age but expanded more rapidly.  

\subsubsection{Spatially resolved spectral index and spectral curvature maps}
We created spectral index maps at different resolutions and frequency sets to highlight various structures. First, to resolve the spectral properties in the inner region, we created a spectral index map at the highest possible resolution ($2.8\arcsec\times2.3\arcsec$) using 3~GHz and 6~GHz data. Additionally, we also created a set of low (7\arcsec) and high resolution (5.5\arcsec) spectral index maps. In all cases we included only pixels with a flux density above $3\sigma_{\rm rms}$ detected at the observed frequencies.


\begin{figure*}[!thbp]
    \centering
      \includegraphics[width=1.0\textwidth]{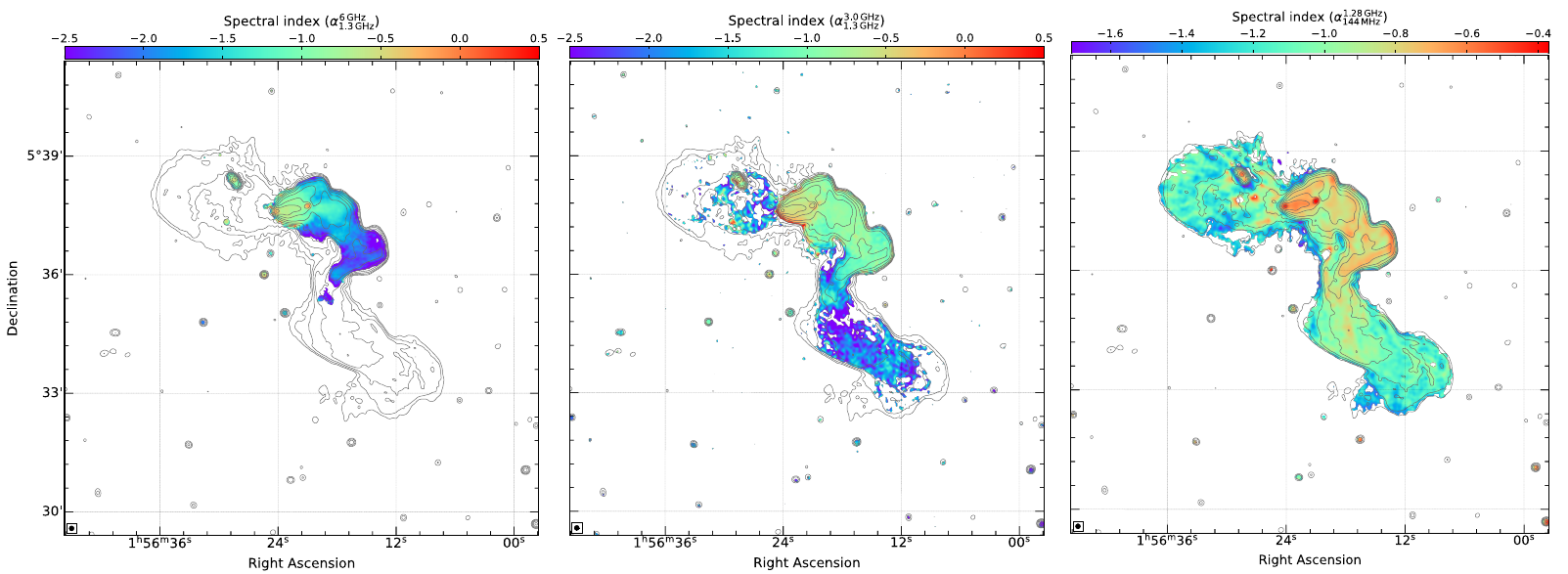}
      
    \vspace{-0.1cm}
 \caption{Spectral index maps of NGC\,741-742 at 5.5\arcsec~ (left and middle) and 7\arcsec~ (right). In all maps, contour levels are drawn at  $[1, 2, 4, 8 ...]\times 3.0\sigma_{\rm rms}$ and are from MeerKAT L-band except bottom right where the uGMRT Band\,3 contours are shown. The beam size is indicated in the bottom left corner of each image. These maps were created using images IM17 and IM22 (left), IM12 and IM17 (middle), and IM13 and IM1 (right), see Table\,\ref{imaging} for image properties.}
      \label{index_maps}
\end{figure*}  


The highest resolution spectral index map is shown in Figure\,\ref{high_res_index}.  There are no clear homogeneous trends in the spectral distribution across NGC\,741-742, although it is evident that the spectral index steepens from the core to the bent tail region. The nuclei of both NGC\,741 and NGC~742 appear flat, with the mean spectral indices of about $-0.5$. The spectral index in the innermost nuclei of the NGC~741 is between $-0.2$ to $-0.3$ The rest of the core region has a spectral index of $-0.75$ including the shock cone. The spectral index in the radio blob at the high frequency is about $-1.0$ surrounded by regions with steep spectral indices of about $-1.4$. This suggests that the region contains a mix of both old and young radio plasma. The filaments (see Figure\,\ref{structure} for labeling) in the bent tail region exhibit a flat spectral index of $-1.0$. At the parallel filaments, again, the spectral index is apparently steep, namely $-1.4$ between 3 and 6~GHz.

To investigate the spectral distribution further in the core and bent tail regions, we employ the ``Spectral Tomography Technique" \citep{Katz1993} which allows tracing fine-scale spectral index changes and identifying and differentiating components with distinct regions that may overlap along the line-of-sight \citep{Rajpurohit2021a,Rajpurohit2022b}. This technique involves subtracting scaled versions of one image from another using a reference spectral index value ($\alpha_t$). Structures with the spectral index $\alpha_t$ disappear, while those with a flatter spectral index than $\alpha_t$ are over-subtracted, resulting in negative residuals relative to the surrounding emission. Conversely, structures with a steeper spectral index than $\alpha_t$ are under-subtracted, leading to positive residuals.

We used 5.5\arcsec\ L and C-band images to create tomography maps, these are shown in Figure\,\ref{tomo}. In these maps, regions with a spectral index flatter than $\alpha_t$ appear blue, while those with a steeper index appear red-white. In the $\alpha_t=-0.6$ map, we observe that the jets from NGC~742 and both cores (NGC\,741 and NGC~742 marked with circles) has a spectral index flatter than $-0.60$. The radio emission to the north and south of  NGC\,742-741 jets pops up at $\alpha_t=-0.80$, implying that in these regions, the spectral index ranges between $-0.61$ and $-0.79$. However, the region to the south of the NGC\,742 jets is flatter than to the north. 

\begin{figure*}[!thbp]
    \centering
      \includegraphics[width=0.465\textwidth]{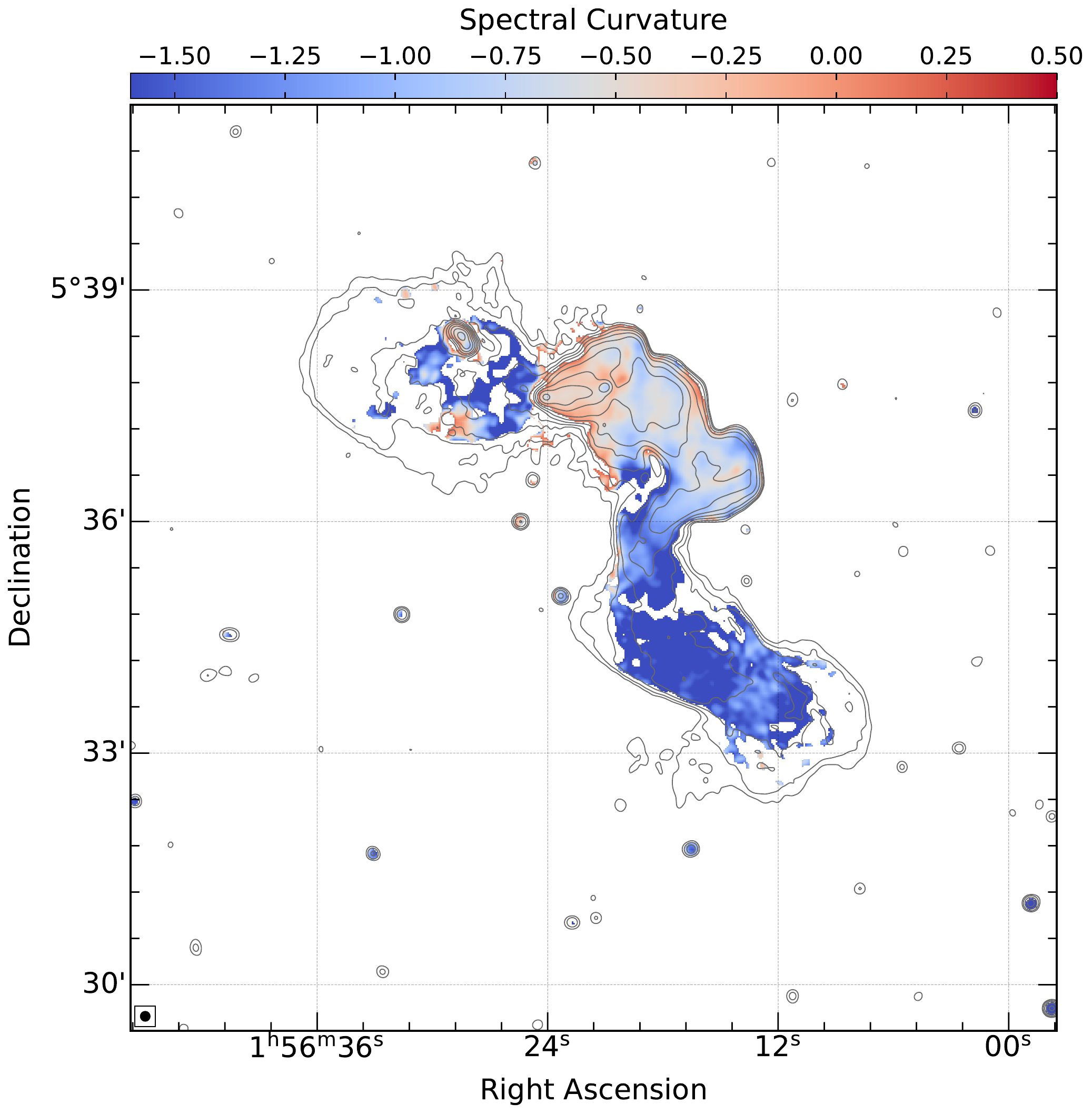}
    \includegraphics[width=0.45\textwidth]{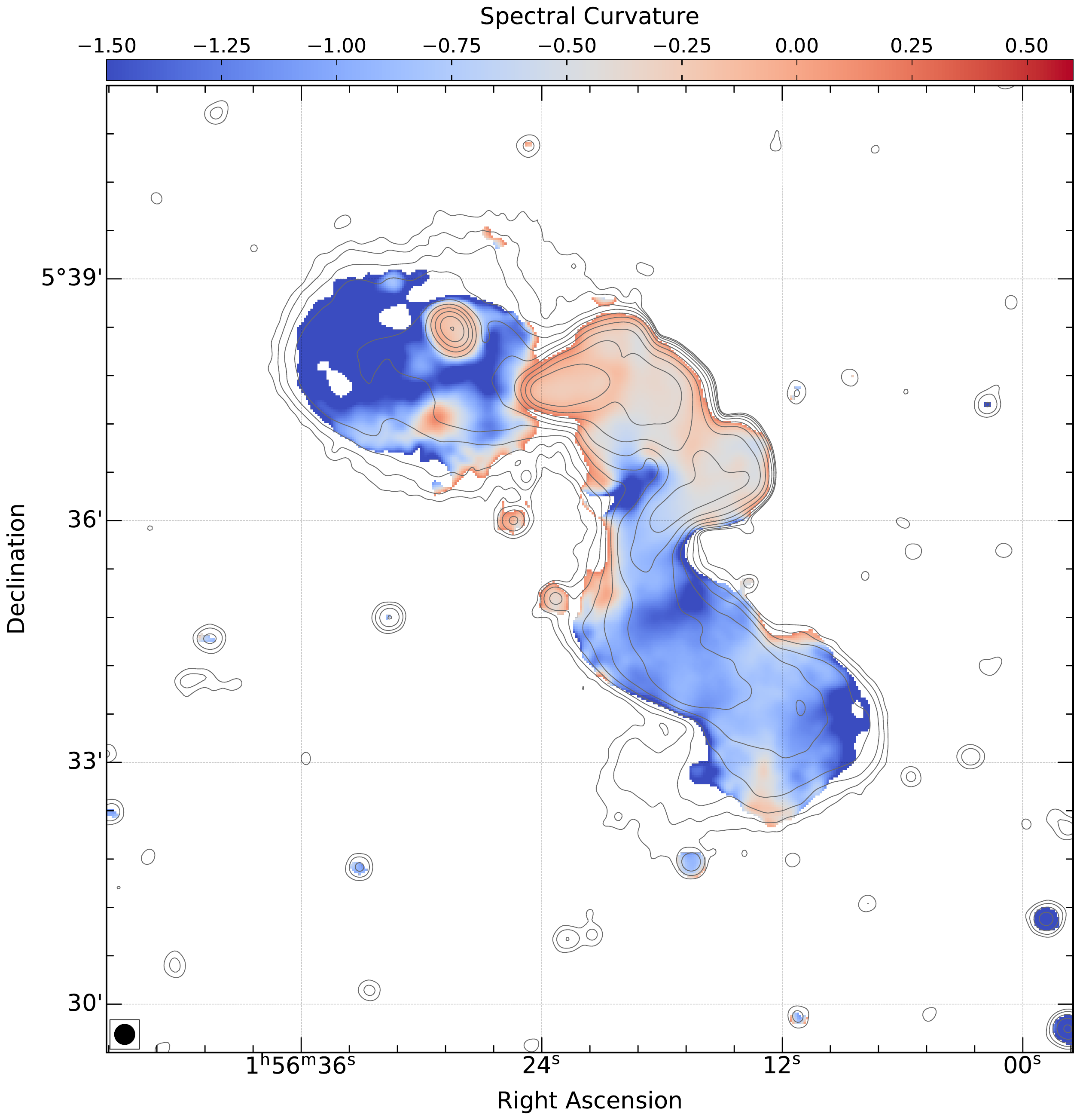}
 \caption{Spectral curvature maps of NGC\,741-742: 7\arcsec resolution high frequency curvature map created using $\alpha_{144\,MHz}^{400\,MHz} -\alpha_{1.28\,GHz}^{3\,GHz}$ images (left) and $15\arcsec$  resolution low frequency curvature map created using $\alpha_{144\,MHz}^{400\,MHz} -\alpha_{700\,GHz}^{1.2\,GHz}$ images (right). These maps indicate that, compared to the southern tail, the NE lobe is significantly curved. In all maps, contour levels are drawn at  $[1, 2, 4, 8 ...]\times 3.0\sigma_{\rm rms}$ and are from the MeerKAT L-band. These maps were created using images IM1, IM5, IM13, IM18, IM3, IM7, IM11, and IM15 (see Table\,\ref{imaging} for image properties).}
      \label{curvature_maps}
\end{figure*}  

There are structures in the bent tail region crossing each other (dashed lines/curves in Figure \ref{tomo}) with spectral indices that are either steeper or flatter than the reference spectral index. This suggests the presence of overlapping structures with different spectral indices. At the location of the ring, there is evidence of two spectral components; at $\alpha_t=-1.00$, we see a flatter component (magenta line) overlapping the steep spectrum ring (green line). The ring almost disappears at $\alpha_t=-1.30$ (Figure \ref{tomo} panel f), suggesting that its spectral index is between $-1.20$ to $-1.30$ between 1.3 and 6\,GHz. This also implies that the ring is a distinct structure. It is worth noting that the northern and the southern edges of the ring are steeper than the V-shaped cone. The radio blob is visible at  $\alpha_t=-1.0$ and $\alpha_t=-1.20$ (Figure \ref{tomo} panels d and e). There also also other overlapping, marked with gray dashed lines in Figure \ref{tomo} panels c and d, again highlighting the complexity.     

In Figure\,\ref{index_maps}, we show spectral index maps for several pairs of frequencies and at different resolutions ($5.5\arcsec$ and 7\arcsec). As seen in the high resolution spectral index map, at both high and low frequencies, the cores of NGC\,741 and NGC~742 are flat. The spectral index in the bent tail region is as steep as $-2.5$ between 1.3 and 6~GHz but flatter at low frequency (144~MHz and 1.28~GHz), implying a high-frequency steepening. In the low frequency spectral index map (Figure\,\ref{index_maps} right panel), along the bent tail and southern tail regions, we see regions with flat spectral indices ($-0.6$ to $-0.8$) narrow regions surrounded by steep spectrum diffuse regions ($-0.9$ to $-1.5$). The flat spectral indices are mainly observed in the parallel filaments, ripples, ring, filaments in the southern tail, and the core region. Such a flat spectral index is a signature of particle re-acceleration, for example, from a shock that adiabatically compresses the fossil plasma and possibly also re-accelerates the radio plasma in the emitting region. From the spectral distribution it is clear that most of the filaments (marked with green arrows in Figure\,\ref{outer_structure}) and the ring exhibit high frequency steepening.  The average spectral index of the southern tail and the NE lobe is about  $-1.03$ and  $-1.13$, respectively, between 144~MHz and 1.28~GHz. These values are consistent with our integrated spectral index estimates.

Using our multifrequency data, we also obtained spectral curvature maps.  In Figure\,\ref{curvature_maps}, we show  maps at 7\arcsec\ and 15\arcsec\ resolutions. These maps are derived as:
\begin{equation}
    SC=-\alpha_{\rm low} + \alpha_{\rm high}, 
\end{equation}
where $\alpha_{\rm low}$ is the low frequency spectral index between 144~MHz and 700~MHz, whereas $\alpha_{\rm high}$ is between 1.28~GHz and 3~GHz. According to the chosen spectral index convention, the curvature is negative for a convex spectrum. A value of $SC=0$ implies no curvature. 

As shown in the high-resolution curvature map (Figure\,\ref{curvature_maps} left panel), in the core region, the $SC$ is $\sim -0.6$. The ring has a constant curvature of about =$-0.75$. The ripples are less curved. The radio blob and filaments in the bent tail region exhibit an SC equal to about $-0.6$, i.e., similar to the core region. The filaments in the southern tail are less curved ($SC \sim-1.10$) than the surrounding diffuse emission. In Figure\,\ref{curvature_maps} right panel, we show the 15\arcsec curvature map. There is a general trend that the curvature increases when moving away from the group center, which is expected due to electron aging. Evidently, the NE lobe is more curved compared to the southern tail.

At the parallel filaments, the spectral index is flat at low frequency (Figure\,\ref{index_maps} bottom right) but steep at high frequency (Figure\,\ref{index_maps} top left). Several individual filaments in the southern tail and NE lobe stand out with flatter spectral indices relative to the surrounding diffuse emission. It is worth noting that  not all filaments have the same spectral indices; the filaments in bent tail region are flatter and less curved than those in the southern tail.

\begin{figure}[!thbp]
    \centering
      \includegraphics[width=0.48\textwidth]{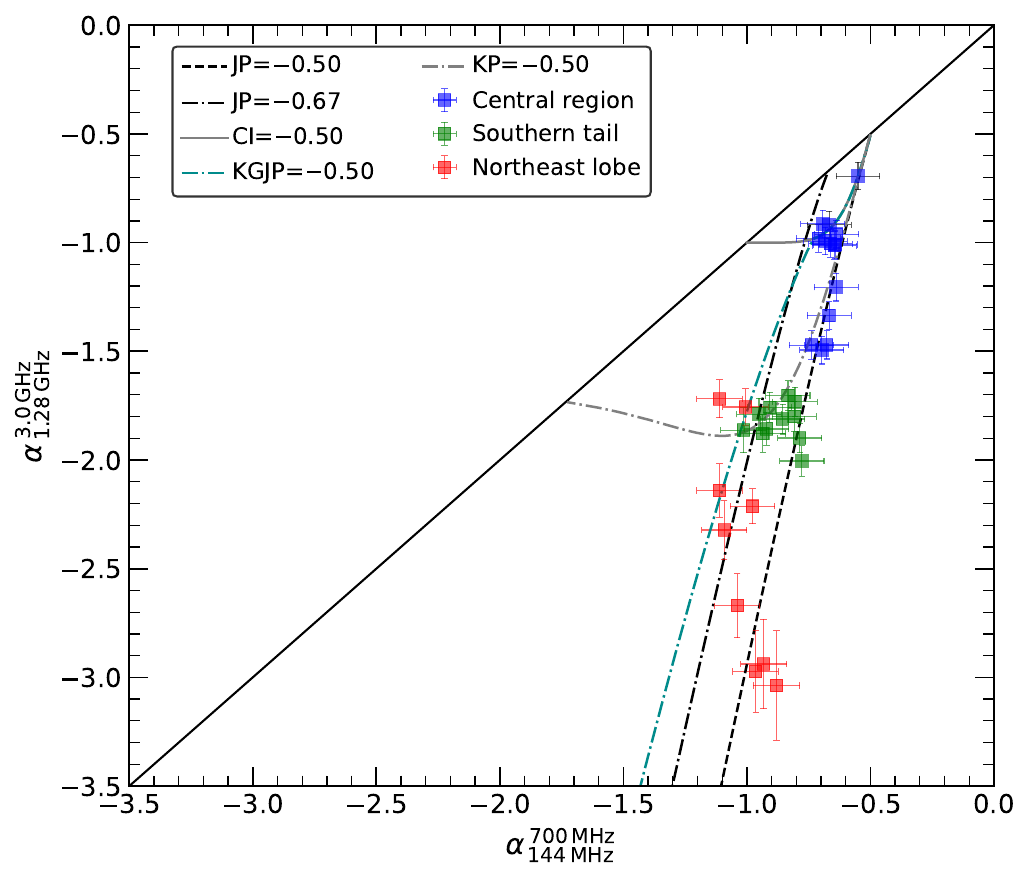}
    
    \vspace{-0.2cm}
 \caption{Radio color-color plot of NGC\,741~742 created using 15\arcsec~ resolution radio maps superimposed with JP, CI, KP, and KGJP spectral aging model trajectories. The plot reveals two distinct trajectories shown by the standard aging models in the color-color plane. The central region (blue circles) and the southern tail (red) reasonably well match the JP model with $\alpha_{\rm inj}=-0.50.$ }
      \label{cc_plot}
\end{figure}  

\begin{figure*}[!thbp]
    \centering
    \includegraphics[width=0.33\textwidth]{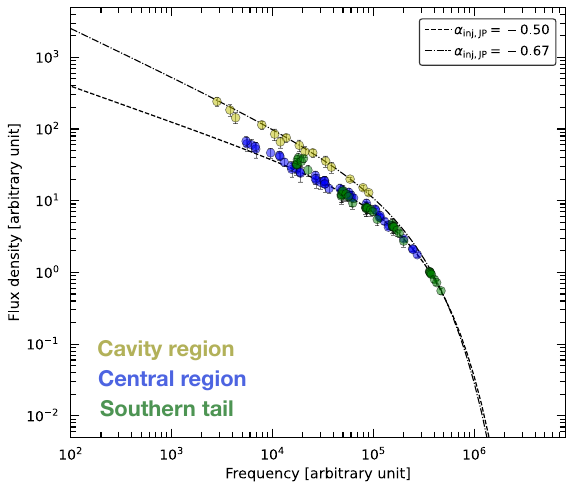}
    \includegraphics[width=0.3265\textwidth]{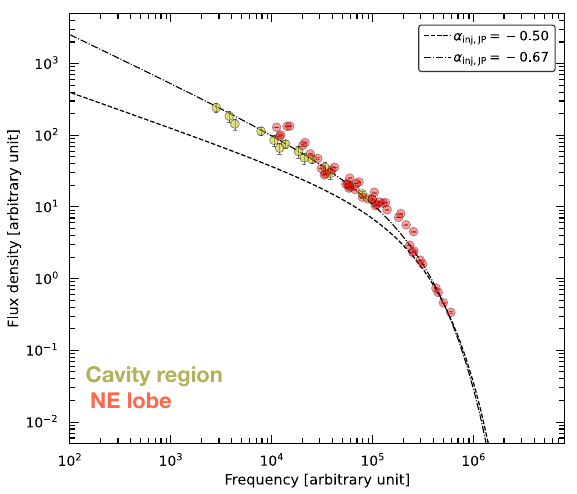}
    \includegraphics[width=0.3255\textwidth]{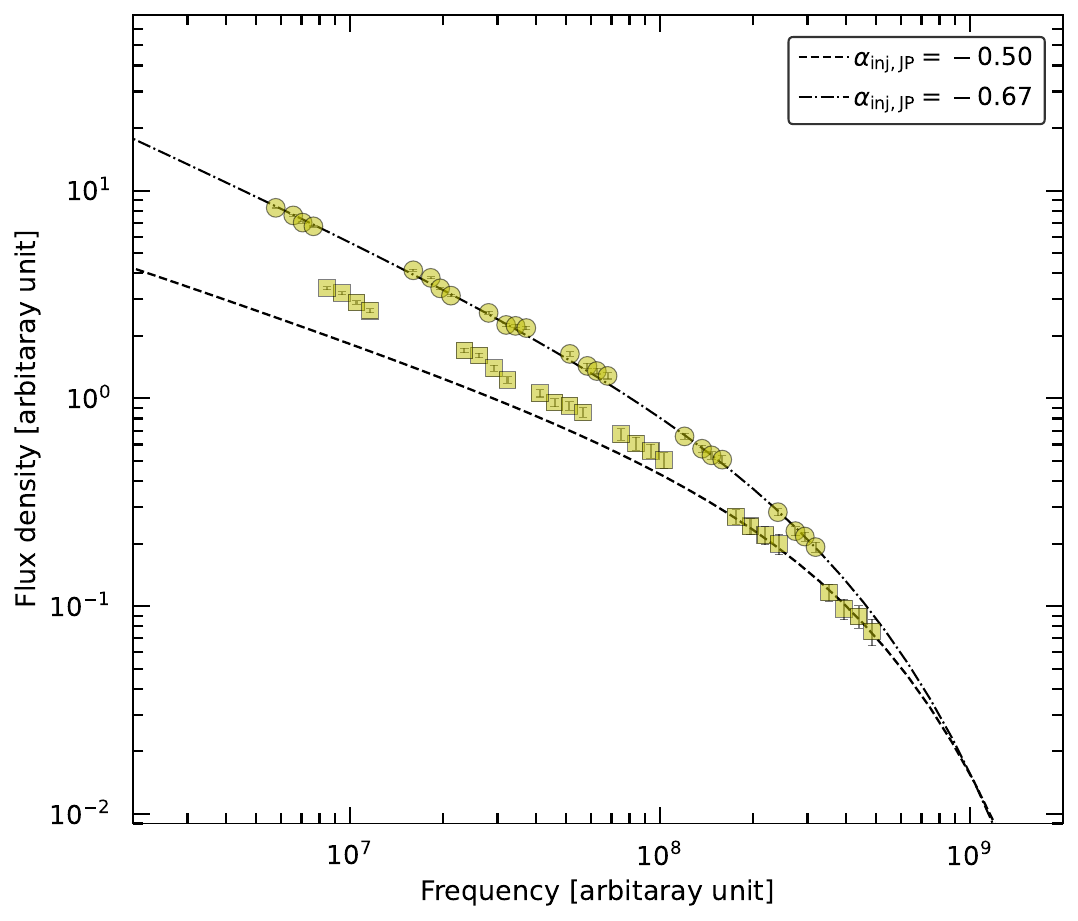}
   \vspace{-0.2cm} 
 \caption{\textit{Left} Global radio spectra of different regions (cavity, central and extended southern tail) created using 15\arcsec\, maps and five frequency data points (excluding C-band). The spectrum of each region has been shifted in $\log(\nu)$ and $\log(I)$ space to create the ``global" spectrum. The JP model lines for $\alpha_{inj}=-0.50$ and $\alpha_{inj}=-0.67$ are shown with dashed and dotted-dashed lines, respectively. \textit{Middle}: Global spectrum for the NE lobe. \textit{Right}: The cavity region (yellow circles/square) global spectrum is created using six frequency data points covering frequency range from 144\,GHz to 6\,GHz using 7\arcsec\, images (right). The yellow circles and squares are fit with the JP model lines for $\alpha_{inj}=-0.67$ and for $\alpha_{inj}=-0.50$, respectively. These plots reveal two different global spectra and suggest that the NE lobe and the cavity regions have the same injection index suggesting that both features possibly have a similar origin.}
      \label{global_spectrum_NE}
\end{figure*} 

A population of non-thermally emitting ``radio filaments" from radio galaxies in a wide variety of physical conditions, lengths (tens of parsecs to kpc scales), and orientation are becoming ubiquitous in radio observations \citep{Brienza2021,Rudnick2022,Ramatsoku2020,Condon2021,Knowles2022,Giacintucci2022,Botteon2022a}. They are observed in isolation and/or bundles and in lobes and/or tails. The origin of these radio filaments is unclear. It has been suggested that they could be related to supersonic MHD turbulence where such intermittent structures are predicted, a boost in the magnetic field, magnetic re-connection, instabilities in the plasma flow or emission from old remnant plasma previously injected by an AGN that accumulates under the influence of buoyancy in the cluster or group environment \citep{Beattie2024,Brienza2021,Rudnick2022}.

Simulations of the interactions between radio galaxies and the external medium with the inclusion of cosmic-ray electrons, show that shocks (with Mach numbers 2-4), propagating through low-density radio cocoons, can generate vorticity, disrupting jets and fine filaments \citep{Nolting2019}. The flatter spectral index, with respect to surrounding emission, observed across individual filaments in NGC\,742-741 may result from lower-energy cosmic-ray electrons being boosted by adiabatic compression and radiating at a given frequency in comparatively stronger magnetic fields. Polarization and Faraday maps may shed light on their origin. 

\subsubsection{Radio color-color diagram}
Radio color-color diagrams are an additional powerful tool for understanding the spectral properties of radio sources. In a radio color-color diagram, a low-frequency spectral index is plotted against a high-frequency spectral index.  We performed the color-color analysis using 15\arcsec~ resolution LOFAR, Band\,4, MeerKAT, and VLA S-band maps. To maximize the area available for studying the curvature, we exclude the VLA C-band data. 

The spectral indices were extracted from a grid of rectangular boxes (Figure\,\ref{spectra}) covering the majority of the source (excluding the cores of NGC~741 and NGC~742, source A, and low surface brightness regions at the edges of the NE lobe and SW tail). For the low-frequency spectral index, we used 144~MHz and 700~MHz maps, while for the high-frequency one, 1.28~MHz and 3~GHz maps. The resulting plot is shown in Figure\,\ref{cc_plot}. The solid black line is a power law line where  ${\rm \alpha}_{\rm low}=\alpha_{\rm high}$. 

Changes in spectral properties over time will result in regions moving along a ``trajectory" in the color-color plot as they age \citep[e.g.,][]{vanWeeren2012a,Rajpurohit2020a}. By examining the trajectory, one can determine whether multiple spectral shapes are present within the source.  If the color-color points from all locations in the source align along a single trajectory, this suggests that there is only one spectral shape throughout the source.  It is worth noting that the trajectory in the color-color diagram is conserved for changes in the magnetic field, adiabatic expansion or compression, and the radiation losses for standard spectral aging models, e.g., Jaffe-Perola \citep[JP;][]{Jaffe1973}, Kardashev-Pacholczyk \citep[KP;][]{Kardashev1962}, continuous injection \citep[CI;][]{Pacholczyk1970}, KGJP \citep{Komissarov1994}. These models follow distinct trajectories in the color-color plot.

In Figure\,\ref{cc_plot}, we show the JP model at injection indices of $-0.50$ and $-0.67$. All data points from the source are below the power-law line (solid black line), implying a negative curvature, as expected for an ageing optically thin plasma. However, the entire emitting region is inconsistent with a single trajectory in the color-color plane. Moreover, except the JP, all other models, namely CI, KP and KGJP  are inconsistent with the observed spectral shape. The central region (blue points) and the southern tail (green points) are better consistent with the JP model with $\alpha_{\rm inj}=-0.50$, but inconsistent with all other models.

There is an apparent difference in the trajectory of the NE lobe and the southern tail in the color-color diagram (Figure\,\ref{cc_plot}). The data points from the NE lobe are most curved at high frequency, while at low frequency, they exhibit a more or less constant spectral index (see Figure\,\ref{cc_plot}). The observed trajectory of of the NE lobe is not expected in any aging models. However, a part of the trajectory (5 data points) seems to be consistent with the JP model with $\alpha_{\rm inj}=-0.67$. Within the central region (ie., core and bent tail regions), there are some areas where the data points deviate from the JP model curve with $\alpha_{\rm inj}=-0.50$ (see discussion in Section\,\ref{global_spectrum}).

\subsubsection{Global spectrum}
\label{global_spectrum}
Since the data points in the radio color-color plot show different spectral shape, we also used the ``shift technique" introduced by \cite{Katz1993,Rudnick1994}, which is based on the idea that the individual radio spectra for different regions within the source trace some part of the ``global spectrum" depending on the energy losses and magnetic fields in those regions. This technique allows to separate regions with different properties, if they exists. If we observe two plasma regions that are identical except for the strength of their magnetic fields or electron energy, their spectra would be related by a shift in $\log(I)$ and $\log(\nu)$ \citep{vanWeeren2012a}. A single global spectrum across the entire emitting region suggests uniform underlying physical parameters throughout. The shifts made in the frequency $log(\nu)$ are related to $\gamma^2B$, where $B$ is the magnetic field and $\gamma$ is the electron energy. The shifts in $log(I)$ are related to $N_TB$, where $N_T$ is the total number of relativistic electrons along the line of sight. 

\begin{figure*}[!thbp]
    \centering
     \includegraphics[width=0.48\textwidth]{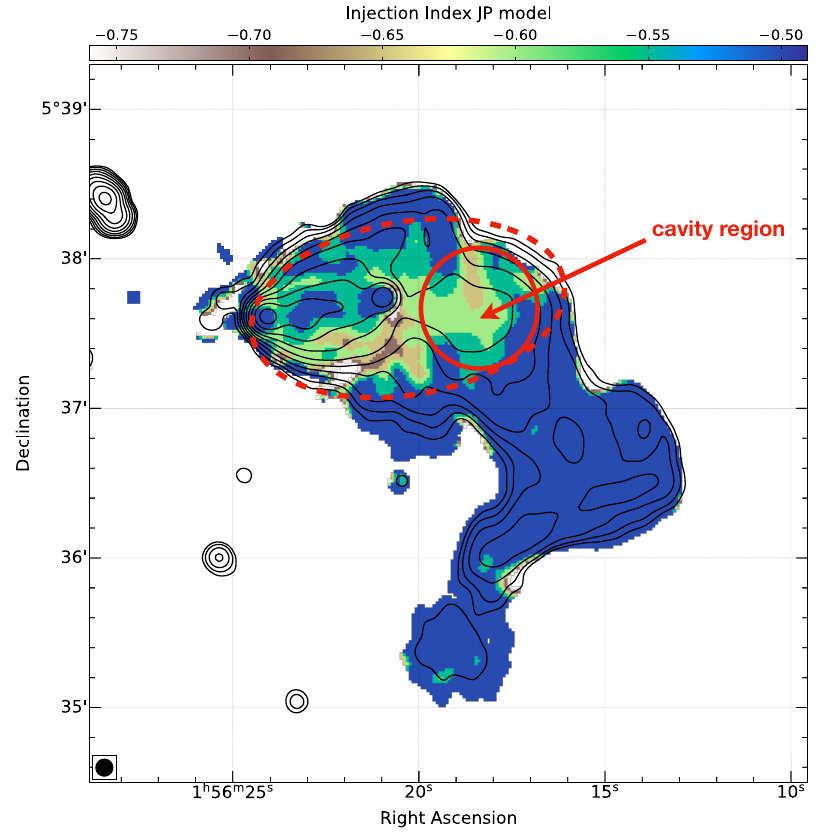} 
     \includegraphics[width=0.48\textwidth]{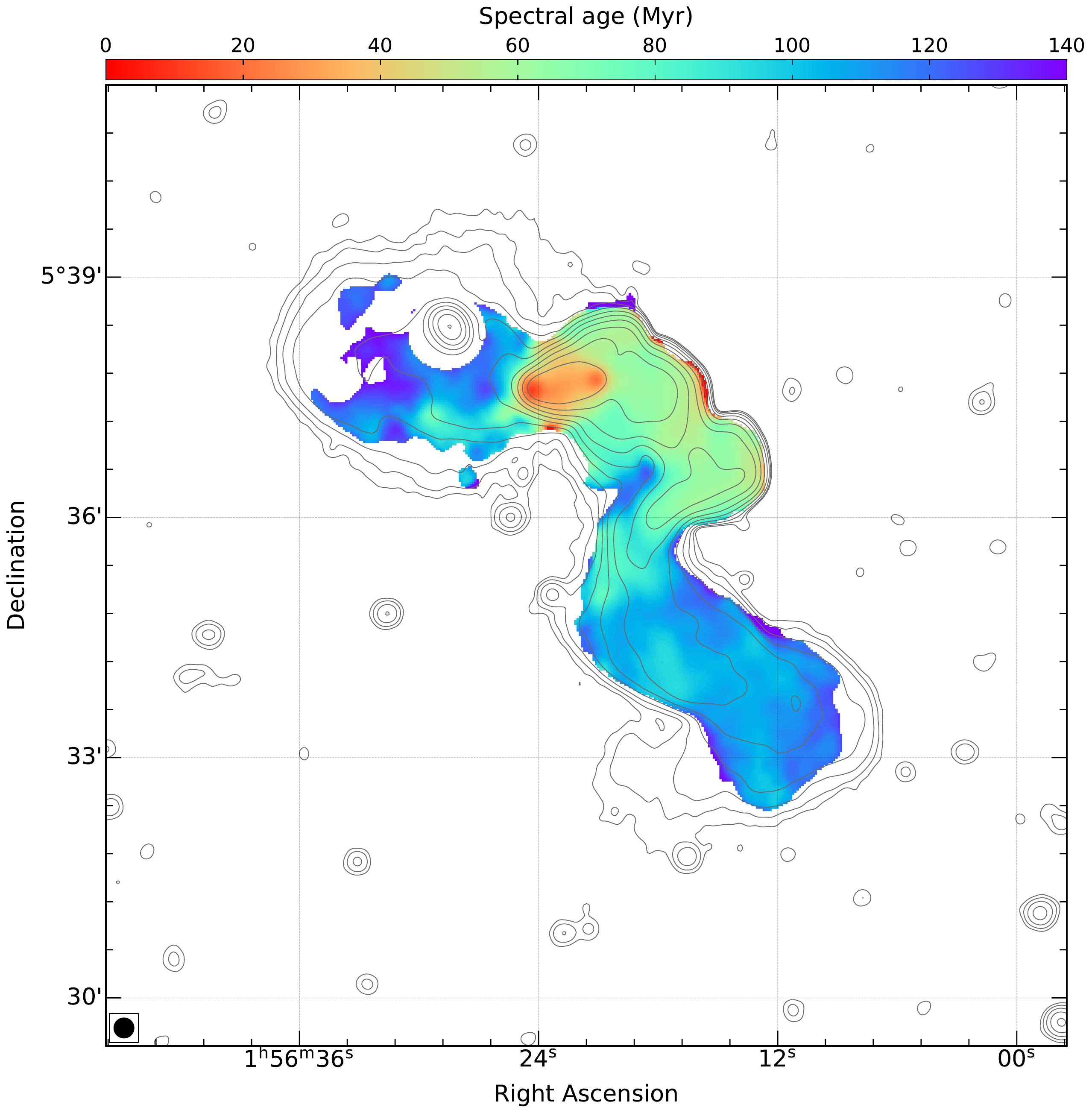} 
 \caption{Injection spectral index map (left) and the spectral age map (right) of NGC\,741~742 obtained  using the JP model. The best-fit injection index map has a resolution of 7\arcsec, created using 144~MHz, 400~MHz, 700~MHz, 1.28~GHz, 3~GHz, and 6~GHz. The injection map highlights the presence of two electron populations with different injection indices (i.e., region inside the dashed ellipse and the rest of the emitting area). The spectral age map is created using five 15\arcsec~ radio maps (excluding C-band image) and an injection index of $-0.5$, suggesting that the core region is the youngest and the tail and lobe regions are the oldest (see Table\,\ref{imaging} for image properties).}
      \label{age_maps}
\end{figure*}  

Figure\,\ref{global_spectrum_NE} shows the results. The plot is obtained by shifting the set of points for each region (see Figure\,\ref{spectra} left panel) in both frequency, $\log(\nu)$, and  intensity, $\log(I)$ using data from the 15\arcsec\, images. We used the model that best matches the observed data in the radio color-color plot, i.e., JP at $\alpha_{\rm inj}=-0.50$ and $\alpha_{\rm inj}=-0.67$. The JP model lines are used to determine the shifts in $\log(I)$ and $\log(\nu)$ to put the observed points onto the spectrum.  All of the multi-frequency data points within each region are shifted by the same amount. For the central region and the southern tail (excluding the radio blob region), we used the JP model with $\alpha_{\rm inj}=-0.50$ as a reference to shift in $\log(\nu)$ and $\log(I)$.  The data points from these two regions are nicely fitted with a single electron energy distribution or global spectrum, similar to the radio color-color plot. 

In Figure\,\ref{global_spectrum_NE} (middle panel), we show the global spectrum for the NE lobe (red circles). The data points from the NE lobe show deviations from the JP model line with $\alpha_{\rm inj}=-0.67$, as also observed in the radio color-color plot. In the central region, there is an area where the data points align more closely with the JP model with $\alpha_{\rm inj}=-0.67$ (yellow circles in Figure\,\ref{global_spectrum_NE}). This area overlaps with the X-ray cavity reported by \citet{Schellenberger2017}, see Figure\,\ref{age_maps}. To investigate this further, we created a global spectrum using high-resolution 7\arcsec\, radio images and, most importantly, including C-band data. The high resolution allowed us to separate that region better.  Evidently, the area to the south and southwest of NGC\,741 (dashed ellipse in Figure\,\ref{age_maps}) and the cavity region  exhibits a different injection index compared to the rest of the emitting region (Figure\,\ref{global_spectrum_NE} right panel). This finding suggests the presence of two separate populations of relativistic electrons in that region. It is plausible that one electron population gives rise to the NE lobe, radio blob, and areas to the south and north of NGC741-742 while the second population generates the core, bent tail, and the extended southern tail regions. The similar injection index across the NE lobe and radio blob points to a connected origin of radio emission.

\subsubsection{Spectral age analysis}
 We used the {\tt BRATS} package to estimate the age of radio emission from the NGC\,741-742 system on a pixel-by-pixel basis.  We refer to \cite{Harwood2013} for details about {\tt BRATS}  and its capabilities. The JP spectral aging model involve four free parameters: flux normalization, magnetic field ($B$), injection index ($\alpha_{\rm inj}$), and time since the last acceleration stage ($t_{\rm age}$). As an initial step, we searched for the best $\alpha_{\rm inj}$ value to be used to determine the final age by performing a series of fitting iterations in the range of $alpha_{\rm inj}=-0.5$ to $-0.9$ across the entire source using the highest common resolution images (i.e., 7\arcsec) at 144~MHz, 400~MHz, 700~MHz, 1.28~GHz, 3~GHz and 6~GHz data. Given that the NE lobe and the southern tail are not recovered at such a high resolution, we also ran the {\tt BRATS} on the 15\arcsec\, maps but excluding the C-band. We used the same resolution to obtain the final spectral age map. We note that the injection index values obtained from the 15\arcsec injection map (not shown) are in line with those derived from the 7\arcsec\, map. 

In Figure\,\ref{age_maps} left panel, we show the high resolution ($7\arcsec$) JP injection index map. Clearly the injection index is different in the cavity/radio blob region and some regions to the east of the cavity, namely by about $-0.6$.  These regions apparently form a lobe-like feature (dashed ellipse in Figure\,\ref{age_maps} left panel). It is worth noting that for the NE lobe, the color-color plot and global spectrum suggest an injection index of $-0.6$ and a distinct spectral shape. This hints a similar origin for the NE lobe and the region (including the cavity/radio blob) to the west of the NGC~742, which also exhibits an average injection index of about $-0.65$. 

The injection index in the southern tail is more or less consistent with $\alpha_{\rm inj}=-0.50$ (Figure\,\ref{age_maps} left panel). For the $15\arcsec$ resolution injection map (now shown), we find that the lobe displays variations in the injection index,  between $-0.5$ and $-0.9$. However, given the large uncertainties on the high-frequency spectral index and rather small emitting region at 3~GHz, it remains challenging to confirm whether the injection index really varies in this region. The average injection index is found to be about $-0.6$, similar to that observed in the cavity/radio blob region. 

To derive the spectral age map, we use a single injection index over the entire source equal to $\alpha_{\rm inj}=-0.50$, which represents the best fit value for most of the regions in the source. Although this marginally impacts the spectral age estimation for areas where the injection index differs from $-0.5$ (i.e., the cavity region and the NE lobe), the resulting discrepancies in age are anticipated to be less significant. We ran the final model fitting iteration using a magnetic field value of $1.95\,\rm\mu G$, estimated as $B=B_{cmb}/\sqrt{3}$ where $B_{\rm cmb}=3.25 (1+z)^{2} \mu \rm G$ is the equivalent magnetic field of the cosmic microwave background. This assumption of magnetic field minimizes the radiative losses and maximizes the lifetime of the source, resulting in a map of the maximum age of the source. However, for the NE lobe, our estimated age should be considered a lower limit because the most extended part of the lobe is not detected at 3~GHz. In Figure\,\ref{age_maps} right panel, we show the final age map. As expected, the spectral age distribution follows the observed spectral index distribution, with younger ages in the core region and older ages in the most distant regions.

Across the source, the age predominantly spans  11 to 130 Myrs. The cores of NGC\,741 and NGC~742 are the youngest. In the core and bent tail regions, the average age is about 40 Myr and 60 Myr, respectively. The NE lobe and the southern tail are the oldest with a mean age of 120~Myr and 105~Myr, respectively.  Our age estimates are more or less consistent with those reported by \citet{Schellenberger2017} despite their assumption of a very similar magnetic field strength, 2~$\mu$G and an injection index of $-0.76$. However, we were able to trace a maximum age of about 140\,Myr. This is expected, as our new observations provide much improved sensitivity and thus recover fainter (and in some cases older) regions of emission that were not previously detected.

\subsection{The radio ring}
One remarkable feature observed in the NGC\,741-742 system is the bright ring between NGC\,741 and the blob. Although the northern boundary of the V-shaped cone extends back to the northern edge of the ring, in the south the ring extends well beyond the boundary of the cone (Figure\,\ref{outer_structure}). This suggests that the ring is a separate structure, not a part of the cone. If we assume the ring is circular, its shape indicates it is likely slightly inclined to the line of sight, oriented closer to edge-on than to face-on. From the projected major and minor radii of the ring we estimate the inclination angle of $\sim68\degree$ relative to the plane of the sky.  A closed ring with a steep, constant spectral index (approximately $-2.0$ between 3 and 6\,GHz) and a weak polarization is reported in NGC~1265 \citep{Sijbring1998,deBruyn2005}. \cite{Pfrommer2011} argued that the ring is produced by the passage of the head-tail galaxy NGC~1265 and its older radio plasma bubble through an accretion shock front. The spectral index across the ring in NGC\,741-742 between 3 and 6\,GHz is flatter than the NGC~1265 ring, namely about $-1.3$. In addition, compared to the NGC~1265 ring, the NGC741-742 ring is significantly polarized.

The ring in the NGC~741-742 system may also be formed by the passing of a shock front through a radio plasma cocoon filled with turbulent magnetic fields  \citep{Ensslin2002,Pfrommer2011}. In this case, upon encountering the shock front, if the magnetic field within the radio cocoon is not high, it subsequently evolves into a vortex ring. The morphology and the fractional polarization of the ring in NGC~741-742 is consistent with MHD simulations of a shock passing through an under-dense bubble. As the shock would have to be associated with the passage of NGC~742 through the group, this suggests that the ring was formed by the interaction between the NGC~742 shock and a previously existing old radio structure, presumably another radio lobe associated with NGC\,741.

We obtained the Mach number of the shock using the ring's major radius and its width. 
Following \cite{Ensslin2002}, the compression ratio ($C$) of the shock can be estimated as:
\begin{equation}
C=\frac{2R^2}{3\pi r^2},
\label{compression_eq}
\end{equation}
where $R$ is the major radius of the ring and $r$ its width. Note that Equation\,\ref{compression_eq}  is valid only when the radio cocoon was spherical before shock crossing and that the major radius did not change. Assuming that the radio cocoon is in pressure equilibrium with its surroundings before and after shock
crossing and the magnetic field is not dynamically important, the pressure jump in the non-relativistic IGrM plasma at the shock front is given by $P_2/P_1=C^{\gamma_r}$, where $\gamma_r$ is the adiabatic index of the relativistic plasma in the radio cocoon. From our high resolution C-band image, we measure $R$ and $r$ of $13$~kpc and $2.6$~kpc, respectively. Using Equation\,\ref{compression_eq}, we obtained $C=5.3$. Assuming $\gamma_r=4/3$ \citep[suitable for the relativistic plasma in an old radio lobe][]{Ensslin2002}, this gives $P_2/P_1=9.2$. Applying standard Rankine Hugoniot jump conditions
\begin{equation}
\mathcal{M} = \sqrt{\frac{\frac{P_2}{P_1} (\gamma + 1) + (\gamma - 1)}{2 \gamma}},
\label{eq::P_jump}
\end{equation}
where $\gamma$ is the adiabatic index of the thermal plasma (assumed to be 5/3), we obtain a Mach number $\mathcal{M}=2.7$. This Mach number corresponds to a velocity of 1750 $\rm km\,s^{-1}$. 

\subsection{Velocity of NGC~742}
The velocity of NGC~742 can be estimated from the shock cone's opening angle using the relation
\begin{equation}
  \mathcal{M} =\frac{1}{ \sin\theta},
  \label{opening_angle}
\end{equation}
where $\theta$ is the half-angle of the shock cone. From our high resolution S-band image, we measure a  full opening angle (i.e, $2\theta$) of the shock cone of about $59\degree$. Using Equation\,\ref{opening_angle}, we obtained a Mach number of 2.1 which is equivalent to a velocity of 1358~km~s$^{-1}$. This value is in good agreement with the velocity reported by \citet{Schellenberger2017} (1300~km~s$^{-1}$) using the same method. Our estimate does not account for the component of motion of NGC~742 along the line of sight, which will cause the Mach cone to be foreshortened, widening its apparent, projected opening angle. However, if we take the inclination of the ring structure as an indicator of the angle of the galaxy's trajectory to the plane of the sky, we find only a small change in the estimated Mach number, to $\mathcal{M}$=2.15, equivalent to 1390~km~s$^{-1}$.

The radio tails represent the host galaxy's motion through the IGrM and the velocity can be also estimated from the spectral age and the physical size of the tails. The trajectory distance of the southern tail is about 200~kpc at 144~MHz. However, the very extended part of the tail is only detected at 144~MHz. In the southern tail, the synchrotron electrons are as old as 120 Myr, i.e., the maximum age in the tail. We note that the age of the tail can only be estimated out to 140~kpc, therefore we use this as a trajectory distance. Under this assumption, we estimate that the projected velocity component is about 1141 $\rm km\,s^{-1}$. The optical recession velocities of NGC\,741 and NGC~742 suggest a difference of 480 $\rm km\,s^{-1}$ . This results in a 3D velocity of about 1240 $\rm km\,s^{-1}$ ($\mathcal{M}=1.9$) which is reasonably consistent with the galaxy velocity based on the shock cone's opening angle. However, it is roughly 14\% lower than the velocity reported by \citet{Schellenberger2017} employing the same method. The difference is because they used a maximum age of 90~Myr to estimate the velocity, while our new observations have revealed previously undetected extended part of the southern tail, allowing us to recover regions as old as 120 Myr. Another alternative is to use the line of sight velocity difference between NGC~741 and NGC~741 and the inclination angle of the radio ring to estimate the 3D velocity. This suggests a true velocity of 1290~km~s$^{-1}$ (equivalent to $\mathcal{M}$=2), again in reasonable agreement with the estimates from the cone opening angle and age of the radio tail.

Lastly, the temperature jump across the shock suggests a velocity 1100$\pm$66~km~s$^{-1}$ which is equivalent to $\mathcal{M}=1.7$ \citep{Schellenberger2017}. In summary, the true velocity of NGC~742 estimated from aforementioned methods, is probably in the range of 1100 to 1390 $\rm km\,s^{-1}$ ($\mathcal{M}=1.7-2.15$). These value are significantly lower than $\sim$1750~km\,s$^{-1}$ obtained from the ring (i.e, $\mathcal{M}=2.7$). We emphasize that a velocity $\sim$1750~km\,s$^{-1}$ is most likely overestimated as the shock could be making an ellipse rather than a ring. 

\subsection{Effect of NGC~742 infall on the IGrM}
The supersonic infall of NGC~742 appears to be driving a bow shock through the core of the NGC~741 group, and this will deposit energy in the IGrM. The available X-ray data do not have sufficient depth to support detailed measurements of gas properties across the shock front, but we can estimate the energy input based on IGrM properties in the core  and our estimates of the Mach number of the shock. \citet{Schellenberger2017} provide profiles of gas temperature and entropy, and note that the cooling region (in which the cooling time is $\leq$3~Gyr) has a radius of $\sim$10~kpc and a luminosity L$_{\rm cool}$=2.59$\times$10$^{41}$~erg~s$^{-1}$.

As noted by \citet{Randalletal15} the amount of thermal energy deposited in gas by a shock can be determined from the change in its entropy, with the fractional energy gained by the gas being $\Delta$ln(P/$\rho^\gamma$), where P is the gass pressure, $\rho$ is its density, and $\gamma$ is its adiabatic index (5/3 for the IGrM). Assuming the shock has passed through the group core, we can estimate the pre-shock temperature and density for given Mach numbers, and thus estimate the change in entropy. We find that for $\mathcal{M}$=1.7-2.15, the shock would have deposited $\sim$8-22\% of the thermal energy in the IGrM gas. This is a significant contribution, comparable to that of shocks driven by AGN outbursts \citep[][found fractions of 0.4-12\% for the $\mathcal{M}$=1.2-1.8 shocks in NGC~5813]{Randalletal15} and equivalent to $\sim$2-5$\times$10$^{57}$~erg in the cooling region. \citet{Schellenberger2017} estimate the enthalpy of the western cavity to be 4.9$\times$10$^{57}$~erg. The energy injected by the shock may thus be comparable to that available from the last cycle of AGN feedback in NGC~741, and will have been injected on a much shorter timescale. At 1390~km~s$^{-1}$, NGC~742 would take $\sim$14~Myr to cross the cooling region, whereas the cavity must be significantly older, with much of its enthalpy still contained in the relativistic plasma and thus not yet available for heating the IGrM. It therefore seems that the infall of NGC~742 through the group core has had a significant impact on the IGrM, potentially injecting enough energy through shock heating to affect the cooling and therefore gas inflow into NGC~741. As a singular event, this kind of high-velocity merger cannot balance the radiative losses from the IGrM, but it can affect the cooling and feedback cycle of the group temporarily.

\subsection{Origin of the radio emission and dynamics of the group}
NGC\,741-742 has been known as a bright radio source for several decades, and its morphology and physical structure has been the subject of discussion in the literature \citep{Birkinshaw1985,Venkatesan1994,Giacintucci2011,Jetha2008,Schellenberger2017}. An early study  \citep{Birkinshaw1985} classified NGC\,741 as a classical double-lobed radio galaxy, with the second peak at NGC\,742 noted as a probable hotspot. However, \cite{Venkatesan1994} argued that the diffuse emission instead arises from a head-tail or narrow-angle tail associated with NGC~742, and that NGC\,741 only hosts a compact nuclear source. The interpretation of much of the emission as a radio tail was supported by \cite{Jetha2008,Schellenberger2017}, who observed jets emerging from NGC\,742. However, low frequency radio observations ($\leq 610\,\rm\,GHz$) revealed a fainter extended emission to the west of NGC~742, suggesting that the entire radio emission may not necessarily originate solely from NGC~742 \citep{Giacintucci2011,Schellenberger2017}. 

Based on relatively short {\xmm} and \chandra\ observations, \cite{Jetha2008} reported a ghost cavity without any radio emission to the west of the brightest group galaxy, potentially inflated by previous AGN activity originating from NGC\,741. This ghost cavity was not detected in deeper {\chandra} observations \citep{Schellenberger2017}, but a smaller cavity was identified, located along the line of sight to the radio structure we have referred to as the radio blob.

Our new images confirm the complex morphology found by past studies, while revealing several new features, including the ring, blob, the greater extent of the long southern tail and the filaments within it, and the filamentary structures within the NE lobe. However, the nature of the radio source, and even whether it is a single structure or a combination of structures originating from different hosts is still unclear. Of the possibilities suggested in previous work, the only one which we can immediately dismiss is that the whole source is an FR-I associated with NGC\,741. The jets originating in NGC~742 and their lack of connection to NGC\,741 rule this option out. Below we discuss possible explanations for the origin of the radio source, considering two main hypotheses:

\begin{enumerate}[label=(\roman*)]
\item The entire extended radio source is part of the head-tail radio galaxy NGC~742. In this hypothesis, NGC\,741 hosts a compact nuclear radio source which overlaps the extended emission but is not physically with the rest of the observed radio emission. In this scenario, the V-shaped structure detected in the X-ray and radio maps at the apex of a head-tail galaxy (NGC~742) traces a shock front. As NGC~742 interacts with the IGrM at supersonic speeds, it compresses and heats the surrounding gas, leading to the formation of a Mach cone (shock front). All observed spectral and polarization properties across the V-shaped cone are consistent with the fact it traces a shock front.  The location of NGC~742 close to NGC\,741 and the Mach cone implies that we are observing the infalling galaxy NGC~742 during its brief, maximum velocity passage through the group core. 

The NE lobe and the extended southern tail can be considered as lobes/tails linked to NGC~742. One of the tails is bent to the south, creating a 250 kpc long southern tail, while a second tail is bent along the line of sight away from us, it trails behind the core region, and then to the east, with its far end forming the apparent NE lobe. While the relatively similar radio brightnesses of the far end of the southern tail  and the NE lobe might support this possibility, the radio color-color and global spectrum analysis highlight the differences between the spectral properties of the NE lobe and the extended southern tail. In particular, our analysis suggests different injection indices for the southern tail and  NE lobe, and the presence of two different global spectra which are inconsistent with a single electron population in these two regions.  

Moreover, the radio morphologies of the NE lobe and the southern tail are also quite different. The braided appearance of the southern tail suggests that very likely contains both tails of NGC~742, wrapped around each other, while the embedded  filaments in it are aligned primarily along the line of the tail. By contrast, the filamentary structures in the NE lobe do not seem to be aligned from one side to the other, as might be expected in a tail, though projection effects could be an issue.

If the NE lobe is a second tail from NGC~742, a large fraction of the tail must be aligned along the line of sight and seen in projection behind the core region. In this case we would expect that the apparent emission from this region would be a combination of flat spectrum emission originating from jets and shock compression in the core region, and older emission from the aged plasma in the second tail. This combination would likely result in significantly steeper spectral indices outside the jets, not the flat indices we observe. However, the present data do not allow us to conclusively rule out this scenario.

\item The NE lobe originated from NGC\,741, while the remaining emission is associated with the tails of NGC\,742 which bend to the south. The integrated spectral index of the NE lobe differs significantly from that of the southern tail, suggesting that the emission in the NE lobe could be arising from an old radio lobe associated with NGC\,741. It is worth mentioning that the core of NGC\,741 is quite flat, therefore active, with no evidence for current large-scale jets. But, as the central galaxy of the group it is likely that it has undergone episodes of jet (and bubble) launching in the past. However, if the NE lobe is the product of the most recent episode, we must ask why there is no corresponding counter-lobe at an equivalent distance on the western side of NGC\,741.

One hypothesis is that the NE lobe comprises two distinct lobes associated with NGC\,741 that are seen in projection along our line of sight. In this case, NGC\,741 moved west because of its interaction  with NGC~742, as suggested by \citet{Schellenberger2017}. 

Another possibility is that there was a second lobe at the location of the small X-ray cavity from NGC~741, but that NGC\,742 passed through it, strongly disrupting it and causing the partial mixing of the old lobe plasma from the NGC~741 lobe with the younger plasma of the NGC~742 tail in the core and bent tail regions. It is worth emphasizing that the spectral index in the core and bent tail regions is indeed a mixture of both flatter and steeper spectral indices.  In this scenario the vortex ring would have been produced by the shock of NGC~742 interacting with the old lobe of NGC~741, with the blob perhaps also being partly composed of remnant plasma from the old lobe. 

Our spectral analysis suggests that the western X-ray cavity (radio blob) and the nearby regions, have an injection index of about $-0.6$, similar to the NE lobe. This suggests that the emission the NE lobe and radio blob could have originated from the NGC\,741 and that the radio blob is a counter lobe. This hypothesis offers an explanation for several of the observed structures, but we should note that it does not explain why the NE lobe is significantly more extended than the X-ray cavity nor does it fully explain the complex radio emission to the west of NGC\,741. Testing this scenario in detail is beyond the scope of this paper, and thus would require sophisticated simulations to test the effect of a high velocity interaction between a head-tail source and a passively aging lobe.
 
\end{enumerate}

\section{Summary and conclusions}
\label{summary}
We presented deep, wideband VLA ($2{-}8$\,GHz), MeerKAT ($0.9{-}1.7$\,GHz), uGMRT ($300{-}850$\,GHz), and LOFAR ($120{-}169$\,GHz) observations of the galaxy group NGC\,741. Our new images provide an unprecedented radio view of this system. We  summarize our main findings as follows:
\begin{enumerate}
    \item Our new images reveal complex radio emission and previously unseen features in the inner region, including a ring, a bright well-defined cloud (blob), several intricate filaments, radio edges, and ripples. On large scales, radio emission shows the great extent of a long ($\sim \rm 200~kpc$) southern tail and a lobe-like feature (NE lobe) to the northeast of NGC~742. 

    \item The radio morphologies of the southern tail and the NE lobe are different, and, in addition, the spectrum of the NE lobe is significantly steeper spectrum and shows greater curvature than that of the tail, indicating that the lobe contains older plasma. Moreover, we find evidence of different physical properties across the NE lobe and the southern tail. Specifically, we find different injection indices ($-0.50$ and $-0.65$) and two global spectra. The radio color-color plot and global spectrum analyses are inconsistent with a single electron population in these two regions. All the evidence suggests different origins for these two structures. We propose that the NE lobe is likely associated with NGC\,741, and the long braided southern tail consists of both tails associated with the head-tail radio galaxy NGC~742. The absence of the NGC~741 counter lobe to the west can be explained by assuming that it is completely disrupted through its interaction with the merging NGC~742.   
    
    \item We find that the blob coincides with the small X-ray cavity located to the west of NGC\,741 detected in {\chandra} observations. Its edges are found to be polarized. Our analysis of spectral index and curvature revealed an injection index of approximately $-0.67$ across the blob and surrounding areas, similar to the injection index of the NE lobe. The blob likely consists of remnant plasma from the NGC\,741 counter lobe.

    \item We propose that the ring forms as the NGC\,742 shock front passes through the remnant plasma of an old radio lobe previously inflated by the AGN in NGC\,741. The V-shaped cone originates from the interaction between NGC~742 and the IGrM.  Based on the major and minor radii of the ring, we estimate the shock Mach number to be about 2.7, implying the velocity of NGC~742 to be about 1750 ks$^{-1}$. From the opening angle of the shock cone, we measure a smaller Mach number of about 2.15 and a similar value of the Mach number of 1.9 from spectral aging across the southern tail. Including previous estimates of the shock velocity from the X-ray data, this suggests the true velocity of NGC\,742 is in the range $1100-1750$~$\rm km\,s^{-1}$, highly supersonic in the IGrM. We estimate the likely shock-heating effect of the infall of NGC~742 on the cooling core of the IGrM, and find that it has likely injected $\sim$8-22\% of the thermal energy in the central 10~kpc, equivalent to $\sim$2-5$\times$10$^{57}$~erg. This is comparable to the enthalpy of the previously identified western cavity, and suggests that NGC~742 could potentially have affected the cooling and feedback cycle of NGC~741, by slowing the IGrM cooling that fuels the group-central AGN.

    \item Between 2-4\,GHz, we find strong polarization (up to 48\%) along the boundaries of the V-shaped shock cone, with an average degree of polarization of 20\%. This is consistent with the compression of the magnetic field along the shock front. Additionally, the top and bottom ends of the ring also show a high degree of polarization. 

    \item  We find that some of the previously detected thermal X-ray filaments align with radio edges, suggesting compression of the IGrM as the relativistic plasma of the NGC\,742 tail interacts with the IGrM. A bright X-ray filament connecting NGC~742 and NGC\,741 is found to be correlated with highly linearly polarized radio emission. There are also X-ray filaments that have no radio counterpart. 

\end{enumerate}

We conclude that NGC\,741-742 represents the clearest observational example to date of the interaction between a head-tail radio galaxy and the hot X-ray emitting medium of a galaxy group or a cluster. On smaller scales, the supersonic passage of the head-tail galaxy through the IGrM has formed several complex features, including a V-shaped cone, ring, and blob. On large scales, the nonthermal plasma is presumably adiabatically compressed by the shock and may also be re-accelerated, generating sharp edges, ripples, and fine filaments. The wideband polarization and Faraday analysis will be ideal for investigating the origin of the detected features. Furthermore, highly sensitive observations at low frequencies, such as those that can be obtained with the LOFAR Low Band Antennas, may allow the detection of older electron populations.  The variety of radio features in the galaxy group NGC\,741 makes it a remarkable laboratory for studying the interactions between a shock front, a head-tail radio galaxy, and an aged lobe.

\section*{Acknowledgments}
{\small We thank Lawrence Rudnick for the helpful discussions. KR and EOS acknowledge support from the National Aeronautics and Space Administration (NASA) through  \textit{XMM-Newton} grant 80NSSC22K1641.  FV has been supported by Fondazione Cariplo and Fondazione CDP, through grant n. Rif: 2022-2088 CUP J33C22004310003 for ``BREAKTHRU" project. FL acknowledges financial support from the Italian Ministry of University and Research ? Project Proposal CIR01-00010. MB acknowledges support from the Next Generation EU funds within the National Recovery and Resilience Plan (PNRR), Mission 4 - Education and Research, Component 2 - From Research to Business (M4C2), Investment Line 3.1 - Strengthening and creation of Research Infrastructures, Project IR0000034 - `STILES - Strengthening the Italian Leadership in ELT and SKA'. Basic research in Radio Astronomy at the U.S. Naval Research Laboratory is supported by 6.1 Base funding. AB acknowledges financial support from the European Union - Next Generation EU. WF and CJ acknowledge support from the Smithsonian Institution and the Chandra High Resolution Camera project (through NASA contract NAS8-03060) and NASA ADP grant 80NSSC19K0116. AD acknowledges support by the BMBF Verbundforschung under the grant 05A20STA. The J\"ulich LOFAR Long Term Archive and the German LOFAR network are both coordinated and operated by the J\"ulich Supercomputing Centre (JSC), and computing resources on the supercomputer JUWELS at JSC were provided by the Gauss Centre for Supercomputing e.V. (grant CHTB00) through the John von Neumann Institute for Computing (NIC). This research has made use of VLA observations which is operated by the National Radio Astronomy Observatory. The National Radio Astronomy Observatory is a facility of the National Science Foundation operated under cooperative agreement by Associated Universities. The MeerKAT telescope is operated by the South African Radio Astronomy Observatory, which is a facility of the National Research Foundation, an agency of the Department of Science and Innovation. We thank the staff of the GMRT that made these observations possible. GMRT is run by the National Centre for Radio Astrophysics of the Tata Institute of Fundamental Research. LOFAR \citep{Haarlem2013} is the Low Frequency Array designed and constructed by ASTRON. It has observing, data processing, and data storage facilities in several countries, which are owned by various parties (each with their own funding sources), and that are collectively operated by the ILT foundation under a joint scientific policy. The ILT resources have benefited from the following recent major funding sources: CNRS-INSU, Observatoire de Paris and Universit\'{e} d'Orl\'{e}ans, France; BMBF, MIWF-NRW, MPG, Germany; Science Foundation Ireland (SFI), Department of Business, Enterprise and Innovation (DBEI), Ireland; NWO, The Netherlands; The Science and Technology Facilities Council, UK; Ministry of Science and Higher Education, Poland; The Istituto Nazionale di Astrofisica (INAF), Italy. This research made use of the LOFAR-UK computing facility located at the University of Hertfordshire and supported by STFC [ST/P000096/1], and of the LOFAR-IT computing infrastructure supported and operated by INAF, and by the Physics Dept. of Turin University (under the agreement with Consorzio Interuniversitario per la Fisica Spaziale) at the C3S Supercomputing Centre, Italy. The scientific results reported in this article are based in part on observations made by the {\chandra} X-ray Observatory and published previously in \cite{Schellenberger2017}. }

\facilities{VLA, MeerKAT, GMRT, LOFAR, Chandra}

\software{CARACal, AOflagger, WSClean, CASA, SPAM, DDF}

\bibliography{ref.bib}

\begin{thebibliography}{}
\expandafter\ifx\csname natexlab\endcsname\relax\def\natexlab#1{#1}\fi
\providecommand{\url}[1]{\href{#1}{#1}}
\providecommand{\dodoi}[1]{doi:~\href{http://doi.org/#1}{\nolinkurl{#1}}}
\providecommand{\doeprint}[1]{\href{http://ascl.net/#1}{\nolinkurl{http://ascl.net/#1}}}
\providecommand{\doarXiv}[1]{\href{https://arxiv.org/abs/#1}{\nolinkurl{https://arxiv.org/abs/#1}}}

\bibitem[{{Andreon} {et~al.}(2019){Andreon}, {Moretti}, {Trinchieri}, \&
  {Ishwara-Chandra}}]{Andreon2019}
{Andreon}, S., {Moretti}, A., {Trinchieri}, G., \& {Ishwara-Chandra}, C.~H.
  2019, \aap, 630, A78, \dodoi{10.1051/0004-6361/201935702}

\bibitem[{{Beattie} {et~al.}(2024){Beattie}, {Federrath}, {Klessen}, {Cielo},
  \& {Bhattacharjee}}]{Beattie2024}
{Beattie}, J.~R., {Federrath}, C., {Klessen}, R.~S., {Cielo}, S., \&
  {Bhattacharjee}, A. 2024, arXiv e-prints, arXiv:2405.16626,
  \dodoi{10.48550/arXiv.2405.16626}

\bibitem[{{Bennett} {et~al.}(2014){Bennett}, {Larson}, {Weiland}, \&
  {Hinshaw}}]{Bennett2014}
{Bennett}, C.~L., {Larson}, D., {Weiland}, J.~L., \& {Hinshaw}, G. 2014, \apj,
  794, 135, \dodoi{10.1088/0004-637X/794/2/135}

\bibitem[{{Birkinshaw} \& {Davies}(1985)}]{Birkinshaw1985}
{Birkinshaw}, M., \& {Davies}, R.~L. 1985, \apj, 291, 32,
  \dodoi{10.1086/163038}

\bibitem[{Botteon {et~al.}(2018)Botteon, Gastaldello, \&
  Brunetti}]{Botteon2018}
Botteon, A., Gastaldello, F., \& Brunetti, G. 2018, Monthly Notices of the
  Royal Astronomical Society, 476, 5591, \dodoi{10.1093/mnras/sty598}

\bibitem[{{Botteon} {et~al.}(2022){Botteon}, {Shimwell}, {Cassano}, {Cuciti},
  {Zhang}, {Bruno}, {Camillini}, {Natale}, {Jones}, {Gastaldello},
  {Simionescu}, {Rossetti}, {Akamatsu}, {van Weeren}, {Brunetti},
  {Br{\"u}ggen}, {Groeneveld}, {Hoang}, {Hardcastle}, {Ignesti}, {Di Gennaro},
  {Bonafede}, {Drabent}, {R{\"o}ttgering}, {Hoeft}, \& {de
  Gasperin}}]{Botteon2022a}
{Botteon}, A., {Shimwell}, T.~W., {Cassano}, R., {et~al.} 2022, \aap, 660, A78,
  \dodoi{10.1051/0004-6361/202143020}

\bibitem[{{Bourdin} {et~al.}(2013){Bourdin}, {Mazzotta}, {Markevitch},
  {Giacintucci}, \& {Brunetti}}]{Bourdin2013}
{Bourdin}, H., {Mazzotta}, P., {Markevitch}, M., {Giacintucci}, S., \&
  {Brunetti}, G. 2013, \apj, 764, 82, \dodoi{10.1088/0004-637X/764/1/82}

\bibitem[{{Brienza} {et~al.}(2021){Brienza}, {Shimwell}, {de Gasperin},
  {Bikmaev}, {Bonafede}, {Botteon}, {Br{\"u}ggen}, {Brunetti}, {Burenin},
  {Capetti}, {Churazov}, {Hardcastle}, {Khabibullin}, {Lyskova},
  {R{\"o}ttgering}, {Sunyaev}, {van Weeren}, {Gastaldello}, {Mandal}, {Purser},
  {Simionescu}, \& {Tasse}}]{Brienza2021}
{Brienza}, M., {Shimwell}, T.~W., {de Gasperin}, F., {et~al.} 2021, Nature
  Astronomy, 5, 1261, \dodoi{10.1038/s41550-021-01491-0}

\bibitem[{{Candini} {et~al.}(2023){Candini}, {Brienza}, {Bonafede},
  {Rajpurohit}, {Biava}, {Murgia}, {Loi}, {van Weeren}, \&
  {Vazza}}]{Candini2023}
{Candini}, S., {Brienza}, M., {Bonafede}, A., {et~al.} 2023, \aap, 677, A4,
  \dodoi{10.1051/0004-6361/202347036}

\bibitem[{{CASA Team} {et~al.}(2022){CASA Team}, {Bean}, {Bhatnagar}, {Castro},
  {Donovan Meyer}, {Emonts}, {Garcia}, {Garwood}, {Golap}, {Gonzalez Villalba},
  {Harris}, {Hayashi}, {Hoskins}, {Hsieh}, {Jagannathan}, {Kawasaki},
  {Keimpema}, {Kettenis}, {Lopez}, {Marvil}, {Masters}, {McNichols},
  {Mehringer}, {Miel}, {Moellenbrock}, {Montesino}, {Nakazato}, {Ott}, {Petry},
  {Pokorny}, {Raba}, {Rau}, {Schiebel}, {Schweighart}, {Sekhar}, {Shimada},
  {Small}, {Steeb}, {Sugimoto}, {Suoranta}, {Tsutsumi}, {van Bemmel},
  {Verkouter}, {Wells}, {Xiong}, {Szomoru}, {Griffith}, {Glendenning}, \&
  {Kern}}]{casa2022}
{CASA Team}, {Bean}, B., {Bhatnagar}, S., {et~al.} 2022, \pasp, 134, 114501,
  \dodoi{10.1088/1538-3873/ac9642}

\bibitem[{{Chandra} \& {Kanekar}(2017)}]{Chandra2017}
{Chandra}, P., \& {Kanekar}, N. 2017, \apj, 846, 111,
  \dodoi{10.3847/1538-4357/aa85a2}

\bibitem[{{Condon} {et~al.}(2021){Condon}, {Cotton}, {White}, {Legodi},
  {Goedhart}, {McAlpine}, {Ratcliffe}, \& {Camilo}}]{Condon2021}
{Condon}, J.~J., {Cotton}, W.~D., {White}, S.~V., {et~al.} 2021, arXiv
  e-prints, arXiv:2106.05340.
\newblock \doarXiv{2106.05340}

\bibitem[{{Cornwell} {et~al.}(2008){Cornwell}, {Golap}, \&
  {Bhatnagar}}]{Cornwell2008}
{Cornwell}, T.~J., {Golap}, K., \& {Bhatnagar}, S. 2008, IEEE Journal of
  Selected Topics in Signal Processing, 2, 647,
  \dodoi{10.1109/JSTSP.2008.2005290}

\bibitem[{{de Bruyn} \& {Brentjens}(2005)}]{deBruyn2005}
{de Bruyn}, A.~G., \& {Brentjens}, M.~A. 2005, \aap, 441, 931,
  \dodoi{10.1051/0004-6361:20052992}

\bibitem[{{En{\ss}lin} \& {Br{\"u}ggen}(2002)}]{Ensslin2002}
{En{\ss}lin}, T.~A., \& {Br{\"u}ggen}, M. 2002, \mnras, 331, 1011,
  \dodoi{10.1046/j.1365-8711.2002.05261.x}

\bibitem[{{Gendron-Marsolais} {et~al.}(2020){Gendron-Marsolais},
  {Hlavacek-Larrondo}, {van Weeren}, {Rudnick}, {Clarke}, {Sebastian},
  {Mroczkowski}, {Fabian}, {Blundell}, {Sheldahl}, {Nyland}, {Sanders},
  {Peters}, \& {Intema}}]{Gendron2020}
{Gendron-Marsolais}, M., {Hlavacek-Larrondo}, J., {van Weeren}, R.~J., {et~al.}
  2020, \mnras, 499, 5791, \dodoi{10.1093/mnras/staa2003}

\bibitem[{{Giacintucci} {et~al.}(2011){Giacintucci}, {O'Sullivan}, {Vrtilek},
  {David}, {Raychaudhury}, {Venturi}, {Athreya}, {Clarke}, {Murgia},
  {Mazzotta}, {Gitti}, {Ponman}, {Ishwara-Chandra}, {Jones}, \&
  {Forman}}]{Giacintucci2011}
{Giacintucci}, S., {O'Sullivan}, E., {Vrtilek}, J., {et~al.} 2011, \apj, 732,
  95, \dodoi{10.1088/0004-637X/732/2/95}

\bibitem[{{Giacintucci} {et~al.}(2022){Giacintucci}, {Venturi}, {Markevitch},
  {Bourdin}, {Mazzotta}, {Merluzzi}, {Dallacasa}, {Bardelli}, {Sikhosana},
  {Smirnov}, \& {Bernardi}}]{Giacintucci2022}
{Giacintucci}, S., {Venturi}, T., {Markevitch}, M., {et~al.} 2022, \apj, 934,
  49, \dodoi{10.3847/1538-4357/ac7805}

\bibitem[{{Hardcastle} {et~al.}(2016){Hardcastle}, {G{\"u}rkan}, {van Weeren},
  {Williams}, {Best}, {de Gasperin}, {Rafferty}, {Read}, {Sabater}, {Shimwell},
  {Smith}, {Tasse}, {Bourne}, {Brienza}, {Br{\"u}ggen}, {Brunetti},
  {Chy{\.z}y}, {Conway}, {Dunne}, {Eales}, {Maddox}, {Jarvis}, {Mahony},
  {Morganti}, {Prandoni}, {R{\"o}ttgering}, {Valiante}, \&
  {White}}]{Hardcastle2016}
{Hardcastle}, M.~J., {G{\"u}rkan}, G., {van Weeren}, R.~J., {et~al.} 2016,
  \mnras, 462, 1910, \dodoi{10.1093/mnras/stw1763}

\bibitem[{{Harwood} {et~al.}(2013){Harwood}, {Hardcastle}, {Croston}, \&
  {Goodger}}]{Harwood2013}
{Harwood}, J.~J., {Hardcastle}, M.~J., {Croston}, J.~H., \& {Goodger}, J.~L.
  2013, \mnras, 435, 3353, \dodoi{10.1093/mnras/stt1526}

\bibitem[{{Intema} {et~al.}(2009){Intema}, {van der Tol}, {Cotton}, {Cohen},
  {van Bemmel}, \& {R{\"o}ttgering}}]{Intema2009}
{Intema}, H.~T., {van der Tol}, S., {Cotton}, W.~D., {et~al.} 2009, \aap, 501,
  1185, \dodoi{10.1051/0004-6361/200811094}

\bibitem[{{Jaffe} \& {Perola}(1973)}]{Jaffe1973}
{Jaffe}, W.~J., \& {Perola}, G.~C. 1973, \aap, 26, 423

\bibitem[{{Jetha} {et~al.}(2008){Jetha}, {Hardcastle}, {Babul}, {O'Sullivan},
  {Ponman}, {Raychaudhury}, \& {Vrtilek}}]{Jetha2008}
{Jetha}, N.~N., {Hardcastle}, M.~J., {Babul}, A., {et~al.} 2008, \mnras, 384,
  1344, \dodoi{10.1111/j.1365-2966.2007.12829.x}

\bibitem[{{Jonas} \& {MeerKAT Team}(2016)}]{Jonas2016}
{Jonas}, J., \& {MeerKAT Team}. 2016, in MeerKAT Science: On the Pathway to the
  SKA, 1, \dodoi{10.22323/1.277.0001}

\bibitem[{{J{\'o}zsa} {et~al.}(2020){J{\'o}zsa}, {White}, {Thorat}, {Smirnov},
  {Serra}, {Ramatsoku}, {Ramaila}, {Perkins}, {Moln{\'a}r}, {Makhathini},
  {Maccagni}, {Kleiner}, {Kamphuis}, {Hugo}, {de Blok}, \&
  {Andati}}]{caracal2020}
{J{\'o}zsa}, G. I.~G., {White}, S.~V., {Thorat}, K., {et~al.} 2020, {CARACal:
  Containerized Automated Radio Astronomy Calibration pipeline}, Astrophysics
  Source Code Library, record ascl:2006.014.
\newblock \doeprint{2006.014}

\bibitem[{{Kardashev}(1962)}]{Kardashev1962}
{Kardashev}, N.~S. 1962, \sovast, 6, 317

\bibitem[{{Katz-Stone} {et~al.}(1993){Katz-Stone}, {Rudnick}, \&
  {Anderson}}]{Katz1993}
{Katz-Stone}, D.~M., {Rudnick}, L., \& {Anderson}, M.~C. 1993, \apj, 407, 549,
  \dodoi{10.1086/172536}

\bibitem[{{Knowles} {et~al.}(2022){Knowles}, {Cotton}, {Rudnick}, {Camilo},
  {Goedhart}, {Deane}, {Ramatsoku}, {Bietenholz}, {Br{\"u}ggen}, {Button},
  {Chen}, {Chibueze}, {Clarke}, {de Gasperin}, {Ianjamasimanana}, {J{\'o}zsa},
  {Hilton}, {Kesebonye}, {Kolokythas}, {Kraan-Korteweg}, {Lawrie}, {Lochner},
  {Loubser}, {Marchegiani}, {Mhlahlo}, {Moodley}, {Murphy}, {Namumba},
  {Oozeer}, {Parekh}, {Pillay}, {Passmoor}, {Ramaila}, {Ranchod},
  {Retana-Montenegro}, {Sebokolodi}, {Sikhosana}, {Smirnov}, {Thorat},
  {Venturi}, {Abbott}, {Adam}, {Adams}, {Aldera}, {Bauermeister}, {Bennett},
  {Bode}, {Botha}, {Botha}, {Brederode}, {Buchner}, {Burger}, {Cheetham}, {de
  Villiers}, {Dikgale-Mahlakoana}, {du Toit}, {Esterhuyse}, {Fadana},
  {Fanaroff}, {Fataar}, {Foley}, {Fourie}, {Frank}, {Gamatham}, {Gatsi},
  {Geyer}, {Gouws}, {Gumede}, {Heywood}, {Hlakola}, {Hokwana}, {Hoosen},
  {Horn}, {Horrell}, {Hugo}, {Isaacson}, {Jonas}, {Jordaan}, {Joubert},
  {Julie}, {Kapp}, {Kasper}, {Kenyon}, {Kotz{\'e}}, {Kotze}, {Kriek}, {Kriel},
  {Krishnan}, {Kusel}, {Legodi}, {Lehmensiek}, {Liebenberg}, {Lord}, {Lunsky},
  {Madisa}, {Magnus}, {Main}, {Makhaba}, {Makhathini}, {Malan}, {Manley},
  {Marais}, {Maree}, {Martens}, {Mauch}, {McAlpine}, {Merry}, {Millenaar},
  {Mokone}, {Monama}, {Mphego}, {New}, {Ngcebetsha}, {Ngoasheng}, {Ockards},
  {Otto}, {Patel}, {Peens-Hough}, {Perkins}, {Ramanujam}, {Ramudzuli},
  {Ratcliffe}, {Renil}, {Robyntjies}, {Rust}, {Salie}, {Sambu}, {Schollar},
  {Schwardt}, {Schwartz}, {Serylak}, {Siebrits}, {Sirothia}, {Slabber},
  {Sofeya}, {Taljaard}, {Tasse}, {Tiplady}, {Toruvanda}, {Twum}, {van Balla},
  {van der Byl}, {van der Merwe}, {van Dyk}, {Van Tonder}, {Van Wyk}, {Venter},
  {Venter}, {Welz}, {Williams}, \& {Xaia}}]{Knowles2022}
{Knowles}, K., {Cotton}, W.~D., {Rudnick}, L., {et~al.} 2022, \aap, 657, A56,
  \dodoi{10.1051/0004-6361/202141488}

\bibitem[{{Komissarov} \& {Gubanov}(1994)}]{Komissarov1994}
{Komissarov}, S.~S., \& {Gubanov}, A.~G. 1994, \aap, 285, 27

\bibitem[{{Koribalski} {et~al.}(2024){Koribalski}, {Duchesne}, {Lenc},
  {Venturi}, {Botteon}, {Shabala}, {Vernstrom}, {Carretti}, {Norris},
  {Anderson}, {Hopkins}, {Riseley}, {Gupta}, {Velovi{\'c}}, \&
  {-}}]{Koribalski2024}
{Koribalski}, B.~S., {Duchesne}, S.~W., {Lenc}, E., {et~al.} 2024, arXiv
  e-prints, arXiv:2405.04374, \dodoi{10.48550/arXiv.2405.04374}

\bibitem[{{Liu} {et~al.}(2019){Liu}, {Sun}, {Nulsen}, {Clarke}, {Sarazin},
  {Forman}, {Gaspari}, {Giacintucci}, {Lal}, \& {Edge}}]{Liu2019}
{Liu}, W., {Sun}, M., {Nulsen}, P., {et~al.} 2019, \mnras, 484, 3376,
  \dodoi{10.1093/mnras/stz229}

\bibitem[{{Mahdavi} \& {Geller}(2004)}]{MahdaviGeller04}
{Mahdavi}, A., \& {Geller}, M.~J. 2004, \apj, 607, 202, \dodoi{10.1086/383458}

\bibitem[{{McMullin} {et~al.}(2007){McMullin}, {Waters}, {Schiebel}, {Young},
  \& {Golap}}]{McMullin2007}
{McMullin}, J.~P., {Waters}, B., {Schiebel}, D., {Young}, W., \& {Golap}, K.
  2007, in Astronomical Society of the Pacific Conference Series, Vol. 376,
  Astronomical Data Analysis Software and Systems XVI, ed. R.~A. {Shaw},
  F.~{Hill}, \& D.~J. {Bell}, 127

\bibitem[{{McNamara} {et~al.}(2005){McNamara}, {Nulsen}, {Wise}, {Rafferty},
  {Carilli}, {Sarazin}, \& {Blanton}}]{McNamara2004}
{McNamara}, B.~R., {Nulsen}, P.~E.~J., {Wise}, M.~W., {et~al.} 2005, Nature,
  433, 45, \dodoi{10.1038/nature03202}

\bibitem[{{Mohan} \& {Rafferty}(2015)}]{Mohan2015}
{Mohan}, N., \& {Rafferty}, D. 2015, {PyBDSM: Python Blob Detection and Source
  Measurement}, Astrophysics Source Code Library.
\newblock \doeprint{1502.007}

\bibitem[{{Nolting} {et~al.}(2019){Nolting}, {Jones}, {O'Neill}, \&
  {Mendygral}}]{Nolting2019}
{Nolting}, C., {Jones}, T.~W., {O'Neill}, B.~J., \& {Mendygral}, P.~J. 2019,
  \apj, 876, 154, \dodoi{10.3847/1538-4357/ab16d6}

\bibitem[{{Offringa} {et~al.}(2010){Offringa}, {de Bruyn}, {Biehl}, {Zaroubi},
  {Bernardi}, \& {Pandey}}]{Offringa2010}
{Offringa}, A.~R., {de Bruyn}, A.~G., {Biehl}, M., {et~al.} 2010, \mnras, 405,
  155, \dodoi{10.1111/j.1365-2966.2010.16471.x}

\bibitem[{{Offringa} {et~al.}(2014){Offringa}, {McKinley}, {Hurley-Walker},
  {Briggs}, {Wayth}, {Kaplan}, {Bell}, {Feng}, {Neben}, {Hughes}, {Rhee},
  {Murphy}, {Bhat}, {Bernardi}, {Bowman}, {Cappallo}, {Corey}, {Deshpand e},
  {Emrich}, {Ewall-Wice}, {Gaensler}, {Goeke}, {Greenhill}, {Hazelton},
  {Hindson}, {Johnston-Hollitt}, {Jacobs}, {Kasper}, {Kratzenberg}, {Lenc},
  {Lonsdale}, {Lynch}, {McWhirter}, {Mitchell}, {Morales}, {Morgan},
  {Kudryavtseva}, {Oberoi}, {Ord}, {Pindor}, {Procopio}, {Prabu}, {Riding},
  {Roshi}, {Shankar}, {Srivani}, {Subrahmanyan}, {Tingay}, {Waterson},
  {Webster}, {Whitney}, {Williams}, \& {Williams}}]{Offringa2014}
{Offringa}, A.~R., {McKinley}, B., {Hurley-Walker}, N., {et~al.} 2014, \mnras,
  444, 606, \dodoi{10.1093/mnras/stu1368}

\bibitem[{{O'Sullivan} {et~al.}(2011){O'Sullivan}, {Giacintucci}, {David},
  {Gitti}, {Vrtilek}, {Raychaudhury}, \& {Ponman}}]{O'Sullivan2011}
{O'Sullivan}, E., {Giacintucci}, S., {David}, L.~P., {et~al.} 2011, \apj, 735,
  11, \dodoi{10.1088/0004-637X/735/1/11}

\bibitem[{{O'Sullivan} {et~al.}(2019){O'Sullivan}, {Schellenberger}, {Burke},
  {Sun}, {Vrtilek}, {David}, \& {Sarazin}}]{O'Sullivan2019}
{O'Sullivan}, E., {Schellenberger}, G., {Burke}, D.~J., {et~al.} 2019, \mnras,
  488, 2925, \dodoi{10.1093/mnras/stz1711}

\bibitem[{{Pacholczyk}(1970)}]{Pacholczyk1970}
{Pacholczyk}, A.~G. 1970, {Radio astrophysics. Nonthermal processes in galactic
  and extragalactic sources}

\bibitem[{{Perley} \& {Butler}(2013)}]{Perley2013}
{Perley}, R.~A., \& {Butler}, B.~J. 2013, \apjs, 204, 19,
  \dodoi{10.1088/0067-0049/204/2/19}

\bibitem[{{Pfrommer} \& {Jones}(2011)}]{Pfrommer2011}
{Pfrommer}, C., \& {Jones}, T.~W. 2011, \apj, 730, 22,
  \dodoi{10.1088/0004-637X/730/1/22}

\bibitem[{{Rajpurohit} {et~al.}(2020){Rajpurohit}, {Hoeft}, {Vazza}, {Rudnick},
  {van Weeren}, {Wittor}, {Drabent}, {Brienza}, {Bonnassieux}, {Locatelli},
  {Kale}, \& {Dumba}}]{Rajpurohit2020a}
{Rajpurohit}, K., {Hoeft}, M., {Vazza}, F., {et~al.} 2020, \aap, 636, A30,
  \dodoi{10.1051/0004-6361/201937139}

\bibitem[{{Rajpurohit} {et~al.}(2021{\natexlab{a}}){Rajpurohit}, {Vazza}, {van
  Weeren}, {Hoeft}, {Brienza}, {Bonnassieux}, {Riseley}, {Brunetti},
  {Bonafede}, {Br{\"u}ggen}, {Formann}, {Rajpurohit}, {R{\"o}ttgering},
  {Drabent}, {Dom{\'\i}nguez-Fern{\'a}ndez}, {Wittor}, \&
  {Andrade-Santos}}]{Rajpurohit2021c}
{Rajpurohit}, K., {Vazza}, F., {van Weeren}, R.~J., {et~al.}
  2021{\natexlab{a}}, \aap, 654, A41, \dodoi{10.1051/0004-6361/202141060}

\bibitem[{{Rajpurohit} {et~al.}(2021{\natexlab{b}}){Rajpurohit}, {Wittor}, {van
  Weeren}, {Vazza}, {Hoeft}, {Rudnick}, {Locatelli}, {Eilek}, {Forman},
  {Bonafede}, {Bonnassieux}, {Riseley}, {Brienza}, {Brunetti}, {Br{\"u}ggen},
  {Loi}, {Rajpurohit}, {R{\"o}ttgering}, {Botteon}, {Clarke}, {Drabent},
  {Dom{\'\i}nguez-Fern{\'a}ndez}, {Di Gennaro}, \&
  {Gastaldello}}]{Rajpurohit2021a}
{Rajpurohit}, K., {Wittor}, D., {van Weeren}, R.~J., {et~al.}
  2021{\natexlab{b}}, \aap, 646, A56, \dodoi{10.1051/0004-6361/202039428}

\bibitem[{{Rajpurohit} {et~al.}(2022){Rajpurohit}, {van Weeren}, {Hoeft},
  {Vazza}, {Brienza}, {Forman}, {Wittor}, {Dom{\'\i}nguez-Fern{\'a}ndez},
  {Rajpurohit}, {Riseley}, {Botteon}, {Osinga}, {Brunetti}, {Bonnassieux},
  {Bonafede}, {Rajpurohit}, {Stuardi}, {Drabent}, {Br{\"u}ggen}, {Dallacasa},
  {Shimwell}, {R{\"o}ttgering}, {Gasperin}, {Miley}, \&
  {Rossetti}}]{Rajpurohit2022b}
{Rajpurohit}, K., {van Weeren}, R.~J., {Hoeft}, M., {et~al.} 2022, \apj, 927,
  80, \dodoi{10.3847/1538-4357/ac4708}

\bibitem[{{Ramatsoku} {et~al.}(2020){Ramatsoku}, {Murgia}, {Vacca}, {Serra},
  {Makhathini}, {Govoni}, {Smirnov}, {Andati}, {de Blok}, {J{\'o}zsa},
  {Kamphuis}, {Kleiner}, {Maccagni}, {Moln{\'a}r}, {Ramaila}, {Thorat}, \&
  {White}}]{Ramatsoku2020}
{Ramatsoku}, M., {Murgia}, M., {Vacca}, V., {et~al.} 2020, \aap, 636, L1,
  \dodoi{10.1051/0004-6361/202037800}

\bibitem[{{Randall} {et~al.}(2009){Randall}, {Jones}, {Kraft}, {Forman}, \&
  {O'Sullivan}}]{Randalletal09}
{Randall}, S.~W., {Jones}, C., {Kraft}, R., {Forman}, W.~R., \& {O'Sullivan},
  E. 2009, {ApJ}, 696, 1431, \dodoi{10.1088/0004-637X/696/2/1431}

\bibitem[{{Randall} {et~al.}(2015){Randall}, {Nulsen}, {Jones}, {Forman},
  {Bulbul}, {Clarke}, {Kraft}, {Blanton}, {David}, {Werner}, {Sun}, {Donahue},
  {Giacintucci}, \& {Simionescu}}]{Randalletal15}
{Randall}, S.~W., {Nulsen}, P.~E.~J., {Jones}, C., {et~al.} 2015, \apj, 805,
  112, \dodoi{10.1088/0004-637X/805/2/112}

\bibitem[{{Rau} \& {Cornwell}(2011)}]{Rau2011}
{Rau}, U., \& {Cornwell}, T.~J. 2011, \aap, 532, A71,
  \dodoi{10.1051/0004-6361/201117104}

\bibitem[{{Rudnick} {et~al.}(1994){Rudnick}, {Katz-Stone}, \&
  {Anderson}}]{Rudnick1994}
{Rudnick}, L., {Katz-Stone}, D.~M., \& {Anderson}, M.~C. 1994, \apjs, 90, 955,
  \dodoi{10.1086/191931}

\bibitem[{{Rudnick} {et~al.}(2022){Rudnick}, {Br{\"u}ggen}, {Brunetti},
  {Cotton}, {Forman}, {Jones}, {Nolting}, {Schellenberger}, \& {van
  Weeren}}]{Rudnick2022}
{Rudnick}, L., {Br{\"u}ggen}, M., {Brunetti}, G., {et~al.} 2022, \apj, 935,
  168, \dodoi{10.3847/1538-4357/ac7c76}

\bibitem[{{Russell} {et~al.}(2014){Russell}, {Fabian}, {McNamara}, {Edge},
  {Sanders}, {Nulsen}, {Baum}, {Donahue}, \& {O'Dea}}]{Russelletal14}
{Russell}, H.~R., {Fabian}, A.~C., {McNamara}, B.~R., {et~al.} 2014, {MNRAS},
  444, 629, \dodoi{10.1093/mnras/stu1469}

\bibitem[{{Sanders} {et~al.}(2016){Sanders}, {Fabian}, {Russell}, {Walker}, \&
  {Blundell}}]{Sanders2016}
{Sanders}, J.~S., {Fabian}, A.~C., {Russell}, H.~R., {Walker}, S.~A., \&
  {Blundell}, K.~M. 2016, \mnras, 460, 1898, \dodoi{10.1093/mnras/stw1119}

\bibitem[{{Scaife} \& {Heald}(2012)}]{Scaife2012}
{Scaife}, A. M.~M., \& {Heald}, G.~H. 2012, \mnras, 423, L30,
  \dodoi{10.1111/j.1745-3933.2012.01251.x}

\bibitem[{{Schellenberger} {et~al.}(2017){Schellenberger}, {Vrtilek}, {David},
  {O'Sullivan}, {Giacintucci}, {Johnston-Hollitt}, {Duchesne}, \&
  {Raychaudhury}}]{Schellenberger2017}
{Schellenberger}, G., {Vrtilek}, J.~M., {David}, L., {et~al.} 2017, \apj, 845,
  84, \dodoi{10.3847/1538-4357/aa7f2e}

\bibitem[{{Shimwell} {et~al.}(2017){Shimwell}, {R{\"o}ttgering}, {Best},
  {Williams}, {Dijkema}, {de Gasperin}, {Hardcastle}, {Heald}, {Hoang},
  {Horneffer}, {Intema}, {Mahony}, {Mandal}, {Mechev}, {Morabito}, {Oonk},
  {Rafferty}, {Retana-Montenegro}, {Sabater}, {Tasse}, {van Weeren},
  {Br{\"u}ggen}, {Brunetti}, {Chy{\.z}y}, {Conway}, {Haverkorn}, {Jackson},
  {Jarvis}, {McKean}, {Miley}, {Morganti}, {White}, {Wise}, {van Bemmel},
  {Beck}, {Brienza}, {Bonafede}, {Calistro Rivera}, {Cassano}, {Clarke},
  {Cseh}, {Deller}, {Drabent}, {van Driel}, {Engels}, {Falcke}, {Ferrari},
  {Fr{\"o}hlich}, {Garrett}, {Harwood}, {Heesen}, {Hoeft}, {Horellou},
  {Israel}, {Kapi{\'n}ska}, {Kunert-Bajraszewska}, {McKay}, {Mohan},
  {Orr{\'u}}, {Pizzo}, {Prandoni}, {Schwarz}, {Shulevski}, {Sipior}, {Smith},
  {Sridhar}, {Steinmetz}, {Stroe}, {Varenius}, {van der Werf}, {Zensus}, \&
  {Zwart}}]{Shimwell2017}
{Shimwell}, T.~W., {R{\"o}ttgering}, H.~J.~A., {Best}, P.~N., {et~al.} 2017,
  \aap, 598, A104, \dodoi{10.1051/0004-6361/201629313}

\bibitem[{{Shimwell} {et~al.}(2019){Shimwell}, {Tasse}, {Hardcastle}, {Mechev},
  {Williams}, {Best}, {R{\"o}ttgering}, {Callingham}, {Dijkema}, {de Gasperin},
  {Hoang}, {Hugo}, {Mirmont}, {Oonk}, {Prandoni}, {Rafferty}, {Sabater},
  {Smirnov}, {van Weeren}, {White}, {Atemkeng}, {Bester}, {Bonnassieux},
  {Br{\"u}ggen}, {Brunetti}, {Chy{\.z}y}, {Cochrane}, {Conway}, {Croston},
  {Danezi}, {Duncan}, {Haverkorn}, {Heald}, {Iacobelli}, {Intema}, {Jackson},
  {Jamrozy}, {Jarvis}, {Lakhoo}, {Mevius}, {Miley}, {Morabito}, {Morganti},
  {Nisbet}, {Orr{\'u}}, {Perkins}, {Pizzo}, {Schrijvers}, {Smith}, {Vermeulen},
  {Wise}, {Alegre}, {Bacon}, {van Bemmel}, {Beswick}, {Bonafede}, {Botteon},
  {Bourke}, {Brienza}, {Calistro Rivera}, {Cassano}, {Clarke}, {Conselice},
  {Dettmar}, {Drabent}, {Dumba}, {Emig}, {En{\ss}lin}, {Ferrari}, {Garrett},
  {G{\'e}nova-Santos}, {Goyal}, {G{\"u}rkan}, {Hale}, {Harwood}, {Heesen},
  {Hoeft}, {Horellou}, {Jackson}, {Kokotanekov}, {Kondapally},
  {Kunert-Bajraszewska}, {Mahatma}, {Mahony}, {Mandal}, {McKean}, {Merloni},
  {Mingo}, {Miskolczi}, {Mooney}, {Nikiel-Wroczy{\'n}ski}, {O'Sullivan},
  {Quinn}, {Reich}, {Roskowi{\'n}ski}, {Rowlinson}, {Savini}, {Saxena},
  {Schwarz}, {Shulevski}, {Sridhar}, {Stacey}, {Urquhart}, {van der Wiel},
  {Varenius}, {Webster}, \& {Wilber}}]{Shimwell2019}
{Shimwell}, T.~W., {Tasse}, C., {Hardcastle}, M.~J., {et~al.} 2019, \aap, 622,
  A1, \dodoi{10.1051/0004-6361/201833559}

\bibitem[{{Shimwell} {et~al.}(2022){Shimwell}, {Hardcastle}, {Tasse}, {Best},
  {R{\"o}ttgering}, {Williams}, {Botteon}, {Drabent}, {Mechev}, {Shulevski},
  {van Weeren}, {Bester}, {Br{\"u}ggen}, {Brunetti}, {Callingham}, {Chy{\.z}y},
  {Conway}, {Dijkema}, {Duncan}, {de Gasperin}, {Hale}, {Haverkorn}, {Hugo},
  {Jackson}, {Mevius}, {Miley}, {Morabito}, {Morganti}, {Offringa}, {Oonk},
  {Rafferty}, {Sabater}, {Smith}, {Schwarz}, {Smirnov}, {O'Sullivan},
  {Vedantham}, {White}, {Albert}, {Alegre}, {Asabere}, {Bacon}, {Bonafede},
  {Bonnassieux}, {Brienza}, {Bilicki}, {Bonato}, {Calistro Rivera}, {Cassano},
  {Cochrane}, {Croston}, {Cuciti}, {Dallacasa}, {Danezi}, {Dettmar}, {Di
  Gennaro}, {Edler}, {En{\ss}lin}, {Emig}, {Franzen}, {Garc{\'\i}a-Vergara},
  {Grange}, {G{\"u}rkan}, {Hajduk}, {Heald}, {Heesen}, {Hoang}, {Hoeft},
  {Horellou}, {Iacobelli}, {Jamrozy}, {Jeli{\'c}}, {Kondapally}, {Kukreti},
  {Kunert-Bajraszewska}, {Magliocchetti}, {Mahatma}, {Ma{\l}ek}, {Mandal},
  {Massaro}, {Meyer-Zhao}, {Mingo}, {Mostert}, {Nair}, {Nakoneczny},
  {Nikiel-Wroczy{\'n}ski}, {Orr{\'u}}, {Pajdosz-{\'S}mierciak}, {Pasini},
  {Prandoni}, {van Piggelen}, {Rajpurohit}, {Retana-Montenegro}, {Riseley},
  {Rowlinson}, {Saxena}, {Schrijvers}, {Sweijen}, {Siewert}, {Timmerman},
  {Vaccari}, {Vink}, {West}, {Wo{\l}owska}, {Zhang}, \& {Zheng}}]{Shimwell2022}
{Shimwell}, T.~W., {Hardcastle}, M.~J., {Tasse}, C., {et~al.} 2022, \aap, 659,
  A1, \dodoi{10.1051/0004-6361/202142484}

\bibitem[{{Sijbring} \& {de Bruyn}(1998)}]{Sijbring1998}
{Sijbring}, D., \& {de Bruyn}, A.~G. 1998, \aap, 331, 901

\bibitem[{{Tasse} {et~al.}(2021){Tasse}, {Shimwell}, {Hardcastle},
  {O'Sullivan}, {van Weeren}, {Best}, {Bester}, {Hugo}, {Smirnov}, {Sabater},
  {Calistro-Rivera}, {de Gasperin}, {Morabito}, {R{\"o}ttgering}, {Williams},
  {Bonato}, {Bondi}, {Botteon}, {Br{\"u}ggen}, {Brunetti}, {Chy{\.z}y},
  {Garrett}, {G{\"u}rkan}, {Jarvis}, {Kondapally}, {Mandal}, {Prandoni},
  {Repetti}, {Retana-Montenegro}, {Schwarz}, {Shulevski}, \&
  {Wiaux}}]{Tasse2020}
{Tasse}, C., {Shimwell}, T., {Hardcastle}, M.~J., {et~al.} 2021, \aap, 648, A1,
  \dodoi{10.1051/0004-6361/202038804}

\bibitem[{{Ubertosi} {et~al.}(2023){Ubertosi}, {Gitti}, {Brighenti},
  {McDonald}, {Nulsen}, {Donahue}, {Brunetti}, {Randall}, {Gaspari}, {Ettori},
  {Calzadilla}, {Ignesti}, {Feretti}, \& {Blanton}}]{Ubertosi2023}
{Ubertosi}, F., {Gitti}, M., {Brighenti}, F., {et~al.} 2023, \apj, 944, 216,
  \dodoi{10.3847/1538-4357/acacf9}

\bibitem[{{van den Bosch} {et~al.}(2015){van den Bosch}, {Gebhardt},
  {G{\"u}ltekin}, {Y{\i}ld{\i}r{\i}m}, \& {Walsh}}]{VandenBoschetal15}
{van den Bosch}, R. C.~E., {Gebhardt}, K., {G{\"u}ltekin}, K.,
  {Y{\i}ld{\i}r{\i}m}, A., \& {Walsh}, J.~L. 2015, \apjs, 218, 10,
  \dodoi{10.1088/0067-0049/218/1/10}

\bibitem[{{van Haarlem} {et~al.}(2013){van Haarlem}, {Wise}, {Gunst}, {Heald},
  {McKean}, {Hessels}, {de Bruyn}, {Nijboer}, {Swinbank}, {Fallows},
  {Brentjens}, {Nelles}, {Beck}, {Falcke}, {Fender}, {H{\"o}randel},
  {Koopmans}, {Mann}, {Miley}, {R{\"o}ttgering}, {Stappers}, {Wijers},
  {Zaroubi}, {van den Akker}, {Alexov}, {Anderson}, {Anderson}, {van Ardenne},
  {Arts}, {Asgekar}, {Avruch}, {Batejat}, {B{\"a}hren}, {Bell}, {Bell}, {van
  Bemmel}, {Bennema}, {Bentum}, {Bernardi}, {Best}, {B{\^i}rzan}, {Bonafede},
  {Boonstra}, {Braun}, {Bregman}, {Breitling}, {van de Brink}, {Broderick},
  {Broekema}, {Brouw}, {Br{\"u}ggen}, {Butcher}, {van Cappellen}, {Ciardi},
  {Coenen}, {Conway}, {Coolen}, {Corstanje}, {Damstra}, {Davies}, {Deller},
  {Dettmar}, {van Diepen}, {Dijkstra}, {Donker}, {Doorduin}, {Dromer}, {Drost},
  {van Duin}, {Eisl{\"o}ffel}, {van Enst}, {Ferrari}, {Frieswijk}, {Gankema},
  {Garrett}, {de Gasperin}, {Gerbers}, {de Geus}, {Grie{\ss}meier}, {Grit},
  {Gruppen}, {Hamaker}, {Hassall}, {Hoeft}, {Holties}, {Horneffer}, {van der
  Horst}, {van Houwelingen}, {Huijgen}, {Iacobelli}, {Intema}, {Jackson},
  {Jelic}, {de Jong}, {Juette}, {Kant}, {Karastergiou}, {Koers}, {Kollen},
  {Kondratiev}, {Kooistra}, {Koopman}, {Koster}, {Kuniyoshi}, {Kramer},
  {Kuper}, {Lambropoulos}, {Law}, {van Leeuwen}, {Lemaitre}, {Loose}, {Maat},
  {Macario}, {Markoff}, {Masters}, {McFadden}, {McKay-Bukowski}, {Meijering},
  {Meulman}, {Mevius}, {Middelberg}, {Millenaar}, {Miller-Jones}, {Mohan},
  {Mol}, {Morawietz}, {Morganti}, {Mulcahy}, {Mulder}, {Munk}, {Nieuwenhuis},
  {van Nieuwpoort}, {Noordam}, {Norden}, {Noutsos}, {Offringa}, {Olofsson},
  {Omar}, {Orr{\'u}}, {Overeem}, {Paas}, {Pandey-Pommier}, {Pandey}, {Pizzo},
  {Polatidis}, {Rafferty}, {Rawlings}, {Reich}, {de Reijer}, {Reitsma},
  {Renting}, {Riemers}, {Rol}, {Romein}, {Roosjen}, {Ruiter}, {Scaife}, {van
  der Schaaf}, {Scheers}, {Schellart}, {Schoenmakers}, {Schoonderbeek},
  {Serylak}, {Shulevski}, {Sluman}, {Smirnov}, {Sobey}, {Spreeuw}, {Steinmetz},
  {Sterks}, {Stiepel}, {Stuurwold}, {Tagger}, {Tang}, {Tasse}, {Thomas},
  {Thoudam}, {Toribio}, {van der Tol}, {Usov}, {van Veelen}, {van der Veen},
  {ter Veen}, {Verbiest}, {Vermeulen}, {Vermaas}, {Vocks}, {Vogt}, {de Vos},
  {van der Wal}, {van Weeren}, {Weggemans}, {Weltevrede}, {White}, {Wijnholds},
  {Wilhelmsson}, {Wucknitz}, {Yatawatta}, {Zarka}, {Zensus}, \& {van
  Zwieten}}]{Haarlem2013}
{van Haarlem}, M.~P., {Wise}, M.~W., {Gunst}, A.~W., {et~al.} 2013, \aap, 556,
  A2, \dodoi{10.1051/0004-6361/201220873}

\bibitem[{{van Weeren} {et~al.}(2012){van Weeren}, {R{\"o}ttgering}, {Intema},
  {Rudnick}, {Br{\"u}ggen}, {Hoeft}, \& {Oonk}}]{vanWeeren2012a}
{van Weeren}, R.~J., {R{\"o}ttgering}, H.~J.~A., {Intema}, H.~T., {et~al.}
  2012, \aap, 546, A124, \dodoi{10.1051/0004-6361/201219000}

\bibitem[{{van Weeren} {et~al.}(2016){van Weeren}, {Williams}, {Hardcastle},
  {Shimwell}, {Rafferty}, {Sabater}, {Heald}, {Sridhar}, {Dijkema}, {Brunetti},
  {Br{\"u}ggen}, {Andrade-Santos}, {Ogrean}, {R{\"o}ttgering}, {Dawson},
  {Forman}, {de Gasperin}, {Jones}, {Miley}, {Rudnick}, {Sarazin}, {Bonafede},
  {Best}, {B{\^\i}rzan}, {Cassano}, {Chy{\.z}y}, {Croston}, {Ensslin},
  {Ferrari}, {Hoeft}, {Horellou}, {Jarvis}, {Kraft}, {Mevius}, {Intema},
  {Murray}, {Orr{\'u}}, {Pizzo}, {Simionescu}, {Stroe}, {van der Tol}, \&
  {White}}]{vanWeeren2016c}
{van Weeren}, R.~J., {Williams}, W.~L., {Hardcastle}, M.~J., {et~al.} 2016,
  \apjs, 223, 2, \dodoi{10.3847/0067-0049/223/1/2}

\bibitem[{{van Weeren} {et~al.}(2017){van Weeren}, {Andrade-Santos}, {Dawson},
  {Golovich}, {Lal}, {Kang}, {Ryu}, {Br{\`i}ggen}, {Ogrean}, {Forman}, {Jones},
  {Placco}, {Santucci}, {Wittman}, {Jee}, {Kraft}, {Sobral}, {Stroe}, \&
  {Fogarty}}]{vanWeeren2017a}
{van Weeren}, R.~J., {Andrade-Santos}, F., {Dawson}, W.~A., {et~al.} 2017,
  Nature Astronomy, 1, 0005, \dodoi{10.1038/s41550-016-0005}

\bibitem[{{van Weeren} {et~al.}(2021){van Weeren}, {Shimwell}, {Botteon},
  {Brunetti}, {Br{\"u}ggen}, {Boxelaar}, {Cassano}, {Di Gennaro},
  {Andrade-Santos}, {Bonnassieux}, {Bonafede}, {Cuciti}, {Dallacasa}, {de
  Gasperin}, {Gastaldello}, {Hardcastle}, {Hoeft}, {Kraft}, {Mandal},
  {Rossetti}, {R{\"o}ttgering}, {Tasse}, \& {Wilber}}]{vanWeeren2020}
{van Weeren}, R.~J., {Shimwell}, T.~W., {Botteon}, A., {et~al.} 2021, \aap,
  651, A115, \dodoi{10.1051/0004-6361/202039826}

\bibitem[{{Venkatesan} {et~al.}(1994){Venkatesan}, {Batuski}, {Hanisch}, \&
  {Burns}}]{Venkatesan1994}
{Venkatesan}, T.~C.~A., {Batuski}, D.~J., {Hanisch}, R.~J., \& {Burns}, J.~O.
  1994, \apj, 436, 67, \dodoi{10.1086/174881}

\bibitem[{{Wilber} {et~al.}(2018){Wilber}, {Br{\"u}ggen}, {Bonafede}, {Savini},
  {Shimwell}, {van Weeren}, {Rafferty}, {Mechev}, {Intema}, {Andrade-Santos},
  {Clarke}, {Mahony}, {Morganti}, {Prand oni}, {Brunetti}, {R{\"o}ttgering},
  {Mandal}, {de Gasperin}, \& {Hoeft}}]{Wilber2018}
{Wilber}, A., {Br{\"u}ggen}, M., {Bonafede}, A., {et~al.} 2018, \mnras, 473,
  3536, \dodoi{10.1093/mnras/stx2568}

\bibitem[{{Williams} {et~al.}(2016){Williams}, {van Weeren}, {R{\"o}ttgering},
  {Best}, {Dijkema}, {de Gasperin}, {Hardcastle}, {Heald}, {Prand oni},
  {Sabater}, {Shimwell}, {Tasse}, {van Bemmel}, {Br{\"u}ggen}, {Brunetti},
  {Conway}, {En{\ss}lin}, {Engels}, {Falcke}, {Ferrari}, {Haverkorn},
  {Jackson}, {Jarvis}, {Kapi{\'n}ska}, {Mahony}, {Miley}, {Morabito},
  {Morganti}, {Orr{\'u}}, {Retana-Montenegro}, {Sridhar}, {Toribio}, {White},
  {Wise}, \& {Zwart}}]{Williams2016}
{Williams}, W.~L., {van Weeren}, R.~J., {R{\"o}ttgering}, H.~J.~A., {et~al.}
  2016, \mnras, 460, 2385, \dodoi{10.1093/mnras/stw1056}

\bibitem[{{Xu} {et~al.}(2000){Xu}, {Baum}, {O'Dea}, {Wrobel}, \&
  {Condon}}]{Chun2000}
{Xu}, C., {Baum}, S.~A., {O'Dea}, C.~P., {Wrobel}, J.~M., \& {Condon}, J.~J.
  2000, \aj, 120, 2950, \dodoi{10.1086/316842}

\bibitem[{{Zabludoff} \& {Mulchaey}(1998)}]{ZabludoffMulchaey98}
{Zabludoff}, A.~I., \& {Mulchaey}, J.~S. 1998, {ApJ}, 496, 39

\end{thebibliography}

\end{document}